\documentclass[twoside,12pt]{article}
\usepackage{graphicx,amsbsy,amsmath,color}

\def\Ref{\operatorname{\Re}\!e}
\def\Imf{\operatorname{\Im}\!m}
\def\be{\begin{eqnarray} &&}
\def\nonu{\nonumber \\ &&}
\def\ee{\end{eqnarray}}
\def\psla{ \slash \! \! \!}
\def\Psla{ \slash \! \! \! \!}

\newcommand{\mbf}[1]{\boldsymbol{#1}}
\newcommand{\specialcell}[2][c]{%
  \begin{tabular}[#1]{@{}c@{}}#2\end{tabular}}
\begin{document}

\title{ \vspace{1cm} Nucleon Electromagnetic Form Factors \\
in the Timelike Region}
\author{Achim Denig$^1$ and Giovanni Salm\`e$^2$
\\
$^1$ Institute of Nuclear Physics of Johannes Gutenberg University Mainz, 
\\
Helmholtz Institute Mainz, and Cluster of Excellence PRISMA,
\\
Johann-Joachim-Becher-Weg 45, 55099 Mainz,
Germany\\
$^2$ Istituto Nazionale di Fisica Nucleare, Sezione di Roma, \\Piazzale Aldo Moro 2,
00185 Rome, Italy}

\maketitle
\begin{abstract} The electromagnetic form factors of the proton and the neutron 
in the timelike
region are reviewed. In view of the forthcoming experimental projects devoted to
investigate these observables,  we present 
the current status of the field and we emphasize the relevant role, 
that accurate measurements, refined phenomenological
analyses, as well as microscopic models will play with the goal of 
achieving deeper insights into the structure of the nucleon and its inner dynamics.
\end{abstract}


\section{Introduction}

The investigation of the nucleon structure through electromagnetic (em) probes
has  played a central role in our present understanding of strong
interactions. In particular, the 
measurements of em nucleon form factors (FF's) in the spacelike (SL) region,
which started with the pioneering work of Hofstadter \cite{Hofst},
have given the indisputable evidence of the non-elementary nature of the 
proton. The possibility, offered by SL photons,
of achieving an accurate description of the three-dimensional structure of 
the nucleon has
motivated many and successful experimental efforts, which resulted in an 
accurate data base of the SL nucleon FF's (see Ref.~\cite{SLrev} for a list of recent
reviews). With the advent of a new experimental technique -- the so-called 
polarization-transfer method -- 
a striking fall-off of the proton electric FF relative to the magnetic one, which
largely follows a dipole behavior, has been discovered. This relatively recent
achievement impressively demonstrates the relevance of the polarization degrees of
freedom in the studies of the nucleon structure.

The transition to the timelike (TL) region, widening in such a way the 
investigation of the nucleon FF's,
opens unique possibilities for studying 
   peculiar features of the nucleon state.
In particular, in the TL region, one can exploit the hadronic components of the
virtual photon state, that become physical, according to the proper energy
thresholds.  A very popular model of those hadronic components is the vector
meson dominance model  (see Ref. \cite{VMD} for a recent review), that allows i) to  develop a simple picture of the 
  production mechanism of a physical hadron-antihadron pair, and ii) to  accurately 
  identify   quantities,  which are containing relevant details on the nucleon/antinucleon
  states.
   The information accessed in the TL region could appear 
less
intuitive than the electric and magnetic distribution densities of the nucleon, but indeed 
they  
play a relevant role for shedding  light on  the long-range behavior of
the strong interactions. TL FF's are indeed more sensitive to the discrete
hadron
spectrum (given the pole structure that occurs) and to the presence of many transition
amplitudes of physical processes.  For instance, though the SL nucleon FF's can be described by models, that take into 
account
only  few vector mesons, the pole pattern shown by the experimental TL FF's cannot be reproduced without considering {\em all} the
 relevant vector mesons.
 To better understand  the appealing feature of the TL region, let us go back
to the fifties, when the theoretical studies on this issue began.
In particular, in a paper by  Federbush, Goldberger and  Treiman \cite{Feder58},
it was pointed out that a bridge can
be established between the studies  of the nucleon em FF's 
and the transition amplitudes of physical processes, like em decays of vector mesons (VM's) 
as well as the production of hadronic states with well-defined quantum
numbers. This link was found through the dispersion relations that allow to
express the FF's in terms of an integral over  their imaginary parts, that
 are non-vanishing only in the TL region. At that time, the mesonic and nucleonic degrees of
 freedom were the relevant ones in the field of the strong interactions, and therefore
 the mentioned em VM decays and  hadronic transition amplitudes were
 considered as the relevant inputs for a phenomenological analysis. Our modern paradigm of
 strong interactions, Quantum ChromoDynamics (QCD), 
 allows us  to understand  
  those physical
 quantities, through  a microscopical description in terms of quark and gluon degrees of
freedom. { It should be noticed that   microscopic models  could represent an
 effective
tool for evaluating the above mentioned quantities (see, e.g., the analyses in Refs. \cite{AdS,CQMpion,Demelo}), since   
Lattice QCD  is
greatly challenged in the TL region,   differently from  the SL
case where electric and magnetic  density distributions  begin to be addressed~\cite{LQCD}.}

Summarizing what has been illustrated above, it is clear that  a measurement 
of the imaginary part of the TL nucleon FF's, that can be achieved by
exploiting  the polarization degrees of freedom,  
is highly desirable, with a final goal of
constructing a more detailed database of the nucleon FF's, that does  contain
not only
information on the electric and magnetic distribution densities, but also 
on other physical hadronic processes.

At present, the experimental knowledge  of the TL nucleon FF's unfortunately 
does not yet allow  to achieve
this goal. Compared to the SL sector, where precision measurements of FF's 
have been achieved on the percent level, the data base of TL FF's is rather scarce.
Recent measurements of the BaBar collaboration~\cite{BABAR} have somewhat improved the knowledge of the proton FF.
The BaBar result concerning the ratio of the electric to the magnetic FF is however still largely limited by 
statistical uncertainties, and it is furthermore in conflict
with a previous measurement at LEAR~\cite{PS170_94}. This only demonstrates the need for a new
generation of high-precision measurements of the em FF of the proton and neutron.
Indeed, new facilities at Novosibirsk, Beijing, and 
FAIR/Darmstadt recently came into operation or are about to start in the coming years.
From these facilities we can expect significantly improved results.

The aim of the work is therefore to give an updated (maybe not complete) view to the present
status of both the experimental and theoretical investigations in the field.
Moreover, we hope that this review
could represent an opportunity, allowing to share the common expertise, and to compare 
the different experimental as well as theoretical techniques, which have been worked
out in the past. But, after all, the main goal is to  show the
wealth of  information which is provided by  the TL nucleon FF's for
understanding in depth 
the non-perturbative regime of QCD. In view of this,  we will  emphasize that only 
  measurements of 
  the relative phases will  fully
 determine the
nucleon FF's in the TL region, so that the maximal phenomenological knowledge can be reached.  
 Therefore, 
  this review appears to be timely,  given  
  the near- and mid-term future experimental programs.

The review is organized as follows. 
Sect.\ 2 is devoted to the experimental studies, in particular in Subsect.\ 2.1 an overview
of the facilities as well as the experimental techniques is given.
In Subsect.\ 2.2, the present experimental
data base is presented, and in Subsect.\ 2.3 the future facilities and perspectives are discussed.  
In Sect.\ 3 we present the theoretical background.
We start in Subsect.~\ref{SLFF} with a brief review of the SL em FF's;
in Subsect.~\ref{TLFF} the general formalism for investigating 
the TL nucleon FF's  is presented. It follows in Subsect.~\ref{thresh} a discussion
of the threshold energy region; in Subsect. \ref{DR} the path how to
obtain the dispersion relations between the real and the
imaginary part of the nucleon FF's is shown; in Subsects. 3.5 - 3.9, several theoretical approaches are finally
reviewed.
In Sect.~\ref{concl} some conclusions are drawn and future perspectives are presented.

\newpage

\section{The experimental investigation of TL nucleon form factors}
In the year 1972 the first measurement of a TL nucleon FF  
was performed at the $e^+e^-$ collider ADONE in Frascati using
the process $e^+e^- \rightarrow p \bar{p}$~\cite{Adone73}. 
This historically first result was obtained with an optical spark chamber setup 
at a center-of-mass energy (c.m.\ energy) of $\sqrt{s}=2.1$  GeV/c. In the following years a series
of measurement campaigns were performed at the electron-positron colliders ADONE  with the FENICE experiment~\cite{Fenice1,Fenice}, as well as at the Orsay colliding beam facility (DCI) with the detectors DM1~\cite{DM179}
and DM2~\cite{DM283,DM290}.
The em FF of the proton was explored by these facilities 
from nearly production threshold ($\sqrt{s}=2M_p\cdot { c}$, where $M_p$ is the mass of the proton) 
up to c.m. energies of $2.4$ { GeV/c}. 
Precision measurements were also obtained with the BES-II experiment~\cite{BES05} at BEPC, 
and with CLEO~\cite{CLEO05} at CESR.  
The information concerning the TL  FF of the neutron is especially scarce, with only 
the FENICE experiment so far having been able to perform this measurement, cf.\ Refs.~\cite{Fenice,Fenice2}.   

First attempts to measure the proton FF using the inverse reaction $p \bar{p}\rightarrow e^+e^-$ 
date back to the mid 1960's. Indeed, first upper limits stem from antiproton beam experiments at BNL~\cite{BNL65} and CERN~\cite{PS69}. The discovery of this reaction was finally possible using an antiproton beam at  PS/CERN in 1976~\cite{PS7677}.  Antiproton experiments were later continued with great success at 
LEAR/CERN with the PS170 experiment~\cite{PS170_94,PS170_91,PS170_91b} and at 
FNAL~\cite{E760_93,E835_99,E835_03}. 

In the $e^+e^-$ experiments mentioned above, em
FF's have been measured using the so-called energy scan
method, i.e.\ by systematically varying the c.m.\ energy of the $e^+e^-$ collider.
Around the beginning of the 21$^{\rm st}$ century, $e^+e^-$ particle factories came into 
operation, such as 
the B-factory PEP-II at SLAC, which was operated at a 
c.m.\ energy corresponding to the mass of the $\Upsilon$(4S) resonance of $10.6$ GeV. It was 
realized that the use of events with
photon radiation from the initial state (ISR) appears to be a copious source of hadronic final states
with invariant masses {\it below} the actual c.m.\ energy of the collider. Competitive results of the proton~\cite{BABAR} 
as well as of the $\Lambda \bar{\Lambda}$,
$\Lambda \bar{\Sigma^0}$, and $\Sigma^0 \bar{\Sigma^0}$ FF's~\cite{BABAR2} have 
been achieved at the BaBar experiment in the course of the years. 
The BaBar data set does not only feature the best statistical and systematic precision achieved to date, 
but it is also spanning the entire energy range of interest from threshold up to $4.5$ GeV. The BELLE 
collaboration at the Japanese B-factory project KEK-B has been using the ISR-technique in the
charmonium energy region and has measured the process $e^+e^- \rightarrow \Lambda_c^+ \Lambda_c^-$~\cite{BELLE}. 

\subsection{Experimental techniques and facilities}
\label{exptec}

We report in this Subsect. on
the various experimental techniques (see also Fig.~(\ref{fig:feynman})) and facilities, 
as outlined above. A compilation of the experimental results
follows in Subsect.~\ref{expris}. An outlook concerning future perspectives 
in the field will finally be discussed in Subsect.~\ref{expoutlook}.

\subsubsection{Energy scan experiments in $e^+e^-$ annihilation}
 \label{enscan}
\begin{figure}[t]
\begin{center}
\includegraphics[width=18.cm]{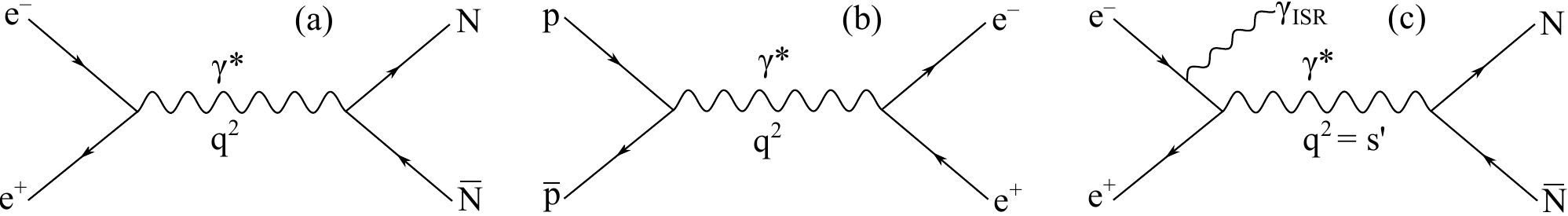}
\caption{Diagrammatic representation of the experimental processes used for the measurement of timelike nucleon FF's:
(a) $e^+e^-$ annihilation scan experiments; (b) $p\bar{p}$ annihilation; 
(c) the Initial State Radiation technique at $e^+e^-$ colliders; in all cases the
form factor is measured as a function of the square four-momentum transfer $q^2$ of the virtual
photon coupling to the baryon pair.}
\label{fig:feynman}
\end{center}
\end{figure}

The standard technique for measurements of hadron production in electron-positron annihilation is
the so-called energy scan, in which the c.m.\ energy of the collider, $\sqrt{s}$, is varied 
systematically, see Fig.~(\ref{fig:feynman},a). 
At each energy point a measurement of the associated cross section is carried out. 
In the case of nucleon pair production, $e^+e^- \rightarrow N \bar{N}$, the associated 
FF's are 
related to the total cross section in  one-photon approximation  
 (see Subsect.~\ref{TLFF} for a more detailed  discussion) according to:
\be
\sigma_{e^+e^- \rightarrow N \bar{N}} = \frac{4 \pi \alpha^2 \beta}{3s} ~C_N(s)~
\left [ \left| G_M^N(q^2) \right|^2 + \frac{2M_N^2}{s} \left| G_E^N(q^2) \right|^2 \right ]{\rm ,} 
\label{eq:ff_general}
\ee
where  $\alpha$ is the electromagnetic fine structure constant, $\beta$ 
the nucleon velocity, $C_N(s)$ 
is the s-wave 
Sommerfeld-Gamow factor, 
that takes into account the Coulomb effects at threshold (see \ Eq. 
(\ref{sommer}) and below for more details),  and $M_N$ the nucleon mass.  
$G_E^N$ and $G_M^N$ are the electric and magnetic Sachs 
FF's, respectively,
which depend on the square four-momentum transfer of the virtual photon, $q^2$. 
Neglecting radiative corrections, the  four-momentum transfer and the  c.m.\ energy
of the collider are identical, cf.\ Eq.~(\ref{eq:ff_general}): 
$$
q^2 = s.
$$ 
In none of the energy scan experiments performed so far, the Sachs FF's could be 
disentangled by means of a measurement of the differential cross section (see below),
and the data are therefore shown in terms of an {\em effective} FF
defined as follows:
\be
|G_{\rm eff}^N (q^2)|=\sqrt{\sigma_{e^+e^-\rightarrow N \bar{N}}(q^2)\over {4\pi\alpha^2 
\beta \over 3q^2} ~C_N(q^2)   \left [1
+\frac{2M^2_N}{q^2}\right ]}=\sqrt{q^2~|G_{M}^N(q^2)  |^2+ 2 M^2_N|G_{E}^N(q^2)|^2
\over q^2 +2M^2_N },
\label{geffa}\ee
where the rightmost-hand side shows the relation with the Sachs FF's 
in one-photon approximation. 

We want to stress that any {\it experimental} measurement 
of $G_{\rm eff}^N$ is independent of any assumption on 
$\left| G_E \right|$ or $ \left| G_M \right|$~\footnote{ In literature, it is often
stated that the experimental value of $|G_{\rm eff}^N (q^2)|$ is obtained assuming
$\left| G_E \right| = \left| G_M \right|$, which is incorrect.}, 
and it quantitatively indicates
 how much the  experimental cross section differs from a point-like 
 one (cf.\ Eq.~(\ref{plsigma})).
Summarizing,  the cross section 
$\sigma_{e^+e^- \rightarrow N \bar{N}}$ can be written as a product of  a 
point-like cross
section  and $|G_{\rm eff}^N (q^2)|^2$, 
that contains information on the  transition
amplitude from a virtual photon state to a $N \bar N$ state (i.e, in a naive
picture,
$|\gamma^*\rangle \to
|qqq \bar q \bar q \bar q \rangle \to | N\bar N\rangle$, through  all the 
allowed  paths, see also Subsect. \ref{TLFF}).

The cross section itself is obtained by measuring the total number of nucleon pairs, 
$\mathcal{N}_{N\bar{N}}$, after 
background subtraction ($\mathcal{N}_{\rm bkg}$) and by normalizing to the total integrated luminosity 
($\mathcal{L}_{\rm int}$). Since detectors cannot cover the full solid angle, the geometrical acceptance ($\epsilon_{\rm geom}$) and
detection efficiencies ($\epsilon_{\rm detect}$), as well as radiative corrections ($\delta_{\rm rad}$) need to be 
considered:
\be
\sigma_{e^+e^- \rightarrow N \bar{N}} 
= \frac{\mathcal{N}_{N\bar{N}} - \mathcal{N}_{\rm bkg}}{\mathcal{L}_{\rm int}}
\cdot \frac{1}{\epsilon_{\rm geom}} \cdot \frac{1}{\epsilon_{\rm detect}} \cdot \frac{1}{1+\delta_{\rm rad}}.
\label{eq:ff_exp}
\ee
Table~(\ref{tab:scan}) shows a compilation of nucleon FF measurements, which have been obtained
via the energy scan technique. While the $p \bar{p}$ channel has been measured by
several experiments (Refs.~\cite{Adone73,Fenice1,Fenice,DM179,DM283,DM290}), data on the $n \bar{n}$ final state so far exist only from FENICE~\cite{Fenice1,Fenice}. The experiment DM2 also measured
at one single energy point the process $\Lambda \bar{\Lambda}$~\cite{DM290}. As 
can be seen from Table~(\ref{tab:scan}), the energy range
from production threshold
up to $\sim3$ GeV has been covered via energy scan experiments. 
Typical energy steps in the individual scanning campaigns vary from $40$ MeV -- in the case of 
the Frascati and Orsay experiments -- up to $100$ MeV in the case of BEPC~\cite{BES05}. 
At 3.67 GeV, one single energy point has been published by CLEO at Cornell~\cite{CLEO05}. 
With integrated luminosities well below $1$ pb$^{-1}$ per scan point (in almost all cases) and given the
small cross section for the process $e^+e^- \rightarrow p\bar{p}$, 
the collected statistics typically was very low. Even close to
threshold, where the cross section for baryon-pair production is highest (but still below $1$ nb), 
the statistics per scan point did not exceed $25$ events.
We briefly discuss in the following
the methodology used at
the experiments DM1 / DM2 (Orsay), the FENICE experiment (Frascati), and the BES experiment (Beijing).
%
%
%


%
%
\begin{figure}[t]
\begin{minipage}[t]{0.485\linewidth}
\centering
\includegraphics[width=9.cm]{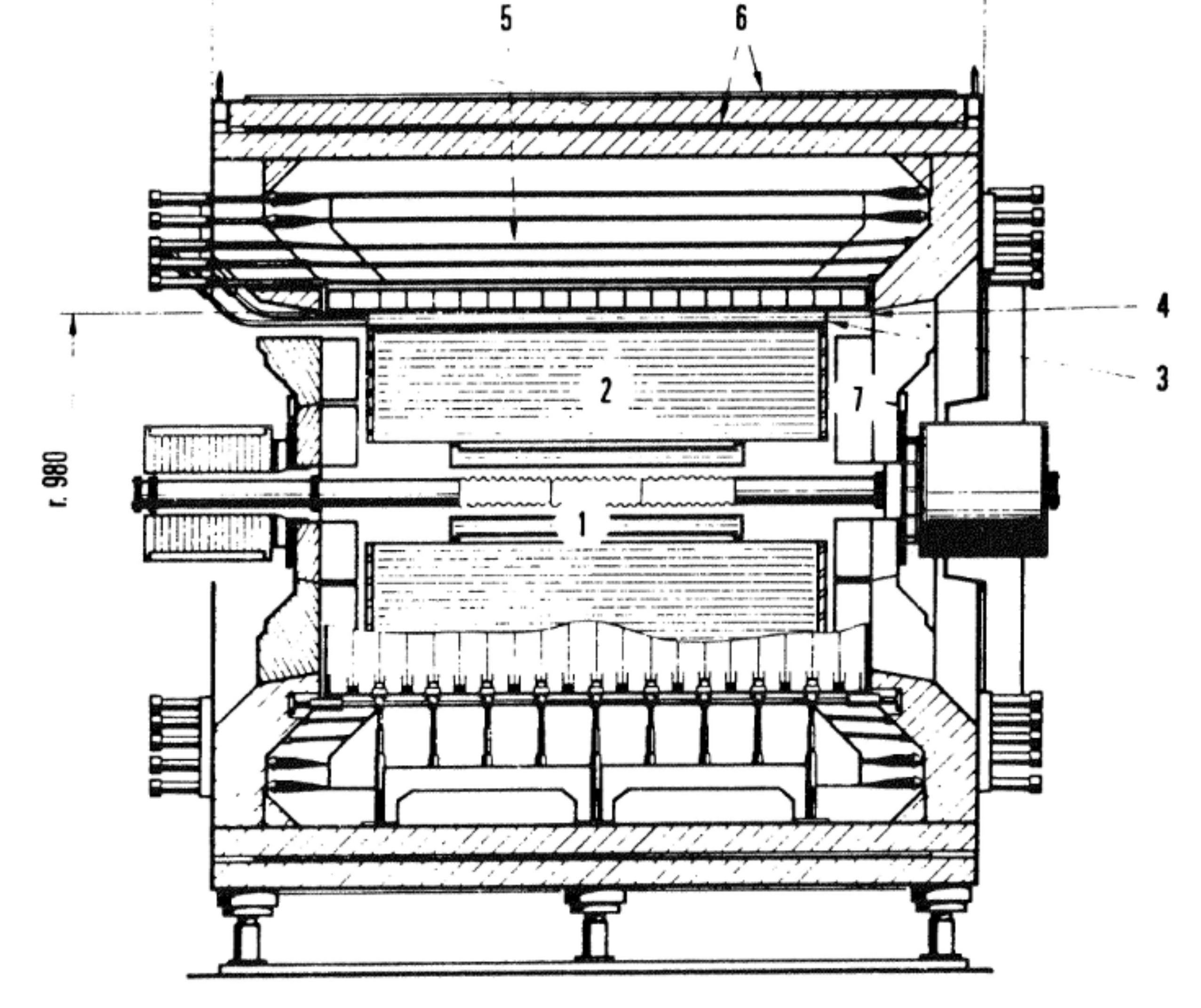}
\caption{Detector DM2: (1) Proportional wire chamber, (2) Drift chamber,  (3) Cerenkov counters, (4) Scintillation counters, (5) Photon detector, (6) Muon detector, (7) End cap detector.
Figure taken from Ref.~\cite{DM2plot}.}
\label{fig:dm2fig}
\end{minipage}
\hspace{0.3cm}
\begin{minipage}[t]{0.485\linewidth}
\centering
\includegraphics[width=8.3cm]{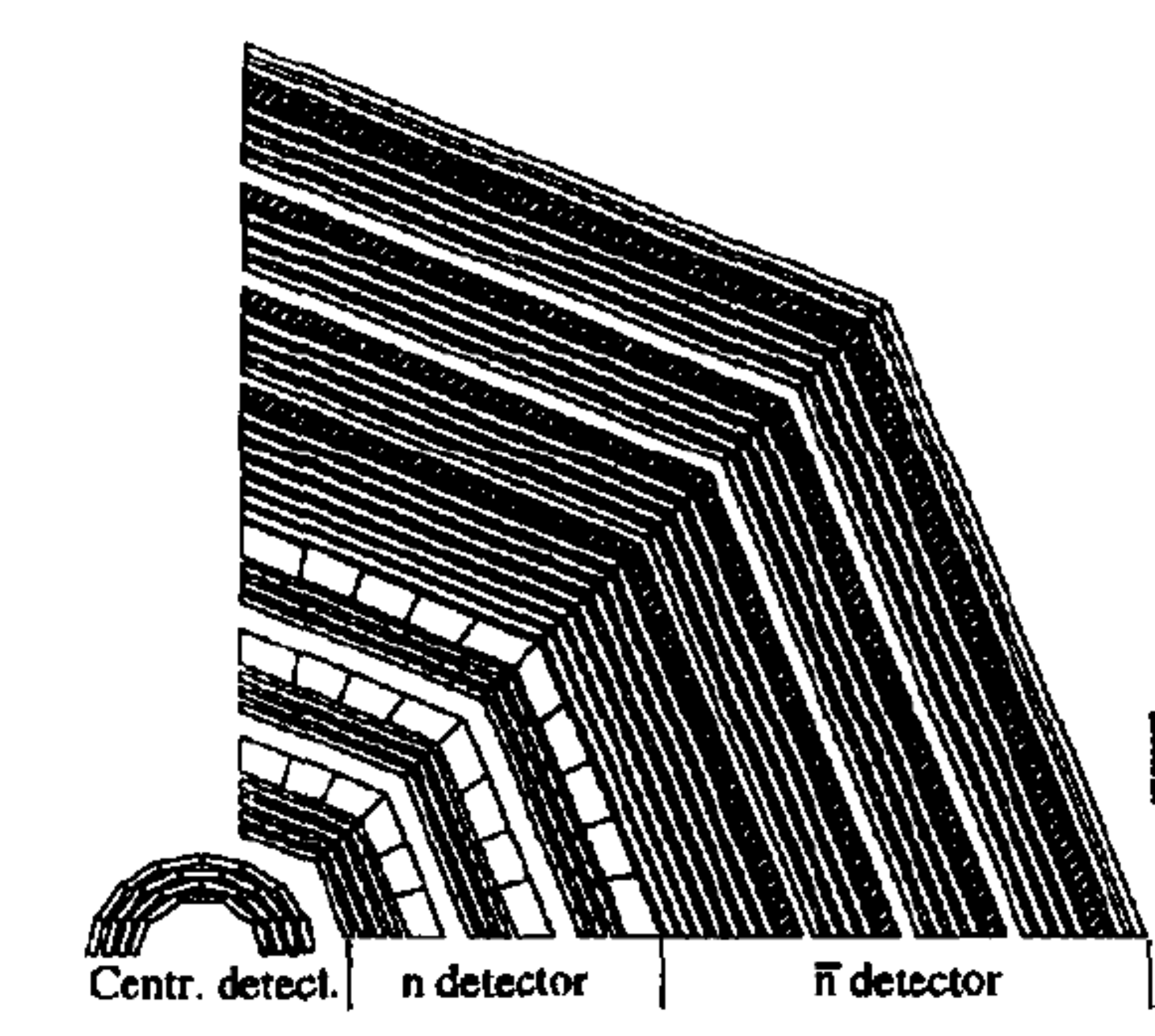}
\caption{Detector FENICE in a view
transversal to the beam direction. Figure taken from Ref.~\cite{Fenice}. }
\label{fig:fenicefig}
\end{minipage}
\end{figure}

\begin{table}[htbp]
\begin{center}
\begin{tabular}{|l|c|c|c|c|c|c|c|} \hline
\textbf{Exp.} &	\multicolumn{1}{c|}{\textbf{Reaction}} & 	\textbf{\specialcell{Year \\ Publ.}} & \textbf{\specialcell{Scan \\ Points}} & \textbf{Range [GeV]} & \textbf{\specialcell{$\mathcal{L}_\mathrm{int}$ \\ ~[pb${}^{-1}$]~}} & \multicolumn{1}{c|}{\textbf{Events}} & \textbf{Ref.} \\ \hline
DM1 &  			$e^+e^- \rightarrow p\bar{p}$ & 			1979 & 	4 & $1.925 - 2.180$ & 		0.4 & $\sim 70$ &~\cite{DM179} \\
DM2 &  			$e^+e^- \rightarrow p\bar{p}$ & 			1983 & 	6 & $1.975 - 2.25$ & 		0.5 & $\sim 100$ &~\cite{DM283} \\
DM2 &  			$e^+e^- \rightarrow p\bar{p}$ & 			1990 & 	1 & 2.4 & 			0.2 & 7 &~\cite{DM290} \\
\rule{0pt}{11pt}DM2 & 	$e^+e^- \rightarrow \Lambda\bar{\Lambda}$ & 		1990 & 	1 & 2.4 & 			0.2 & 4 &~\cite{DM290} \\ \hline
ADONE 73 &  		$e^+e^- \rightarrow p\bar{p}$ & 			1973 & 	1 & 2.1 & 			0.2 & 25 &~\cite{Adone73} \\
FENICE &  		$e^+e^- \rightarrow n\bar{n}$ & 			1993 & 	2 & $2.0 - 2.1$ & 		$<\,$0.1 & 27 &~\cite{Fenice2} \\
FENICE &  		$e^+e^- \rightarrow p\bar{p}$ & 			1993 & 	1 & 2.1 & 			0.1 & 28 &~\cite{Fenice2} \\
FENICE &  		$e^+e^- \rightarrow p\bar{p}$ & 			1994 & 	4 & $1.9 - 2.4$ & 		0.3 & 70 &~\cite{Fenice1} \\
FENICE &  		$e^+e^- \rightarrow n\bar{n}$ & 			1998 & 	5 & $1.9 - 2.44$ & 		0.4 & 74 &~\cite{Fenice} \\
FENICE &  		$e^+e^- \rightarrow p\bar{p}$ & 			1998 & 	1 & 2.1 & 			$<\,$0.1 & 7 &~\cite{Fenice} \\ \hline
BES-II &  		$e^+e^- \rightarrow p\bar{p}$ & 			2005 & 	10 & $2.0 - 3.07$ & 		5 & 80 &~\cite{BES05} \\
CLEO &  		$e^+e^- \rightarrow p\bar{p}$ & 			2005 & 	1 & 3.671 & 			21 & 16 &~\cite{CLEO05} \\ \hline
\end{tabular}
\end{center}
\caption{Summary of experimental results obtained at $e^+e^-$ colliders via the energy scan technique.}
\label{tab:scan}
\end{table}

\begin{itemize}

\item 
At the Orsay colliding beam facility DCI the nucleon FF measurements were performed with the detectors DM1 and DM2.
A sketch of the DM2 detector is shown in Fig.~(\ref{fig:dm2fig}). 
It already resembles the typical structure of modern collider experiments with
large geometrical acceptance. 
DM2 consisted of several layers of cylindrical multiwire proportional chambers (MWPCs) 
inside a magnetic field of solenoidal shape. 
In two scan campaigns (four and six scanning points each) the energy range between $1.925$ GeV and 
$2.25$ GeV could be  covered; in total, approximately $170$ $p\bar{p}$ events were selected. 

\item 
The non-magnetic FENICE detector (cf.\ Fig.~(\ref{fig:fenicefig})) at the ADONE collider
was designed with the primary goal to measure the neutron FF. The main components of
the experiments were Limited-Streamer-Tube (LST) modules as tracking devices, scintillation counters as triggering and timing 
devices and thin iron plates as distributed converters for the antineutron detection~\cite{Fenice}. The specific 
signature of an antineutron annihilation 'star' together with 
the long TOF between production and annihilation, were mainly used to detect anti-neutrons. 
No dedicated neutron detection was performed due to the relatively low neutron detection efficiency.
The neutron FF has been measured from values close to threshold up to $2.44$ GeV. 
A precise determination of the proton FF in the same energy range has been possible as well.

\item 
From an energy scan (10 scanning points) of the BEPC collider 
in the { range $2.0$ GeV/c $< \sqrt{s} < 3.07$ GeV/c}, the BES-II experiment could extract the proton 
FF, extending significantly the energy range covered before at
Frascati and Orsay. BES-II was a typical multiple-purpose detector with large acceptance, 
consisting of a vertex chamber, a large cylindrical drift chamber, a TOF system as well as a photon detector setup. 
The whole apparatus
was embedded in a solenoidal coil. 
The precision of the proton FF measurement at BEPC was 
mostly limited by statistics. In total, $80$ $e^+e^- \rightarrow p \bar{p}$ events have been
detected. The systematic uncertainty was close to $10\%$ in all scan points.

\end{itemize}
\begin{figure}[h]
\begin{center}
\includegraphics[width=8.5cm]{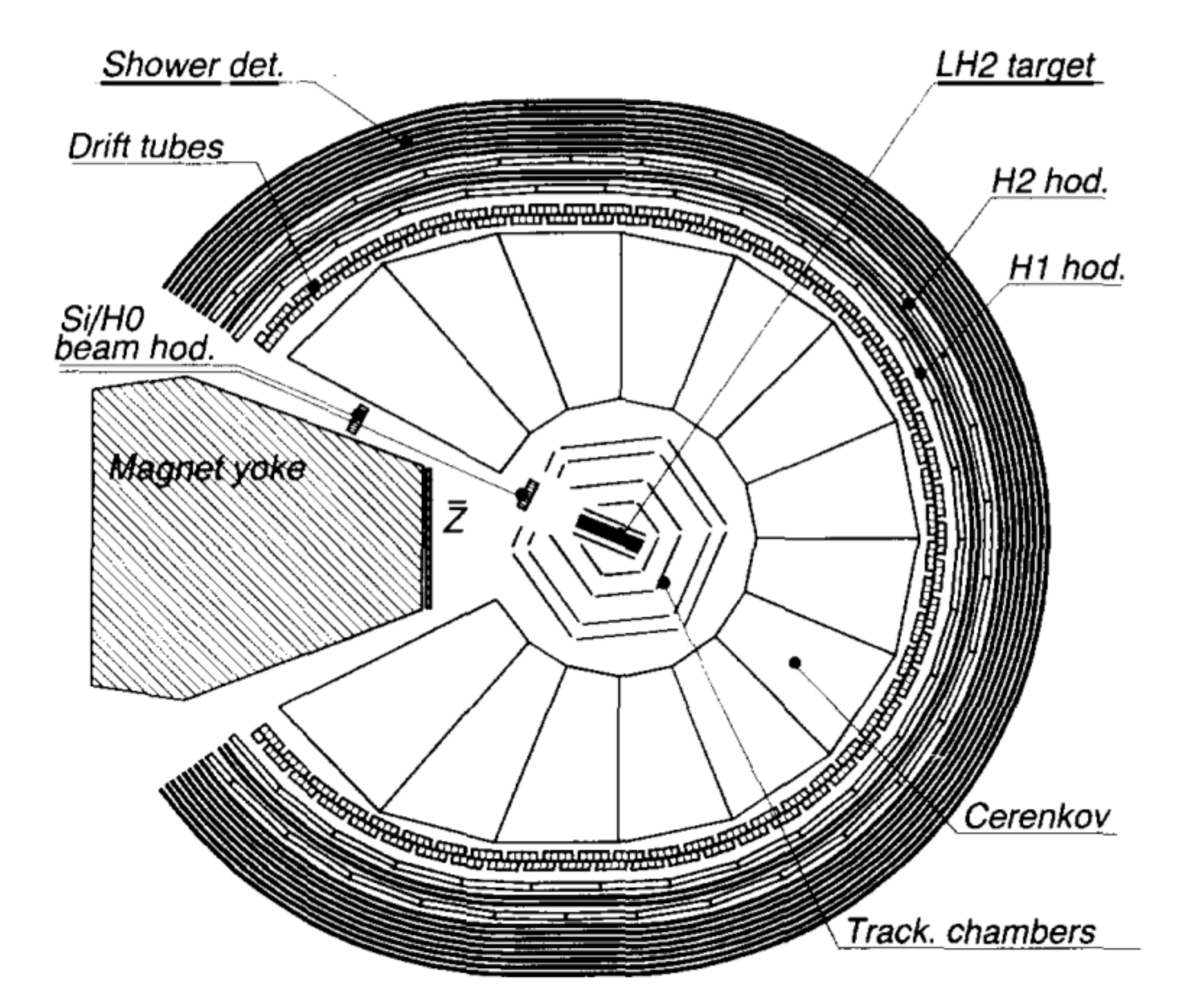}
\caption{Experiment PS170 at the antiproton storage ring LEAR. Figure taken from Ref.~\cite{PS170_94}.}
\label{fig:ps170}
\end{center}
\end{figure}

\subsubsection{$p\bar{p}$ annihilation}

Starting in the 1960's, antiproton beams became available, which were also used to search for the process 
$p \bar{p} \rightarrow e^+e^-$, see Fig.~(\ref{fig:feynman},b). 
The antiproton beam is scattered on a hydrogen target in a fixed-target
configuration. The momentum transfer squared, $q^2$, of the em FF in this case is accessible by
measuring the invariant mass of the lepton pair, or, alternatively, it is
calculated for a given antiproton beam momentum assuming the target protons to be at rest. 
First upper limits for the TL proton FF have indeed been
obtained at relatively high energies at BNL~\cite{BNL65} and CERN~\cite{PS69} 
in $q^2$ ranges between $5$ (GeV/c)$^2$ and $8$ (GeV/c)$^2$. For historical reasons we report FF results from
$e^+e^-$ experiments as a function of the c.m.\ energy of the collider, 
$\sqrt{s}$, while results
from $p\bar{p}$ annihilation experiments are quoted as a function of $q^2$. 
\begin{figure}[t]
\begin{center}
\includegraphics[width=14.5cm]{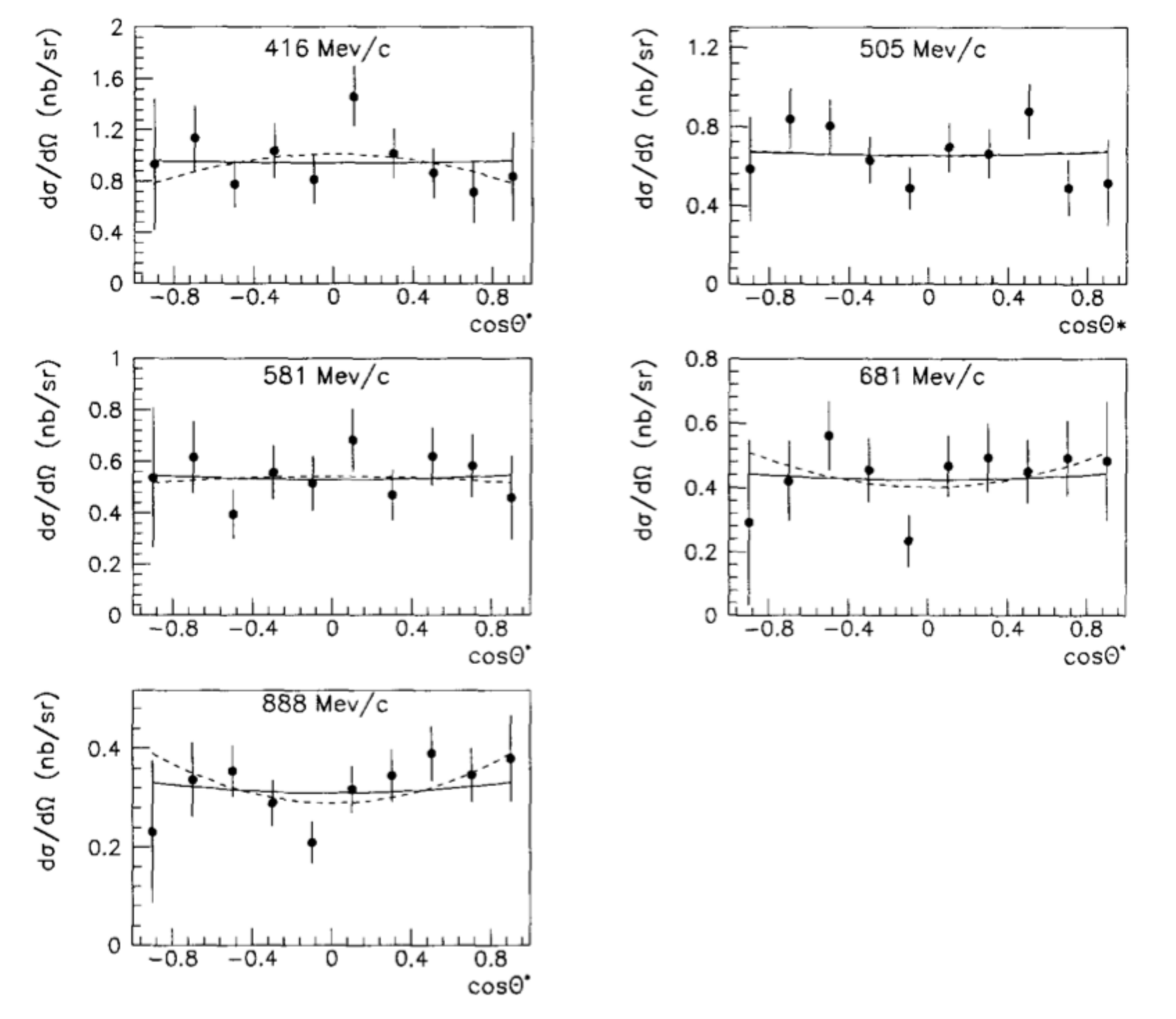}
\caption{Differential cross section for the reaction $p \bar{p} \rightarrow e^+e^-$. The angle
$\theta^*$ denotes the angle between the outgoing lepton and the antiproton 
beam direction in the c.m.\ reference frame. The differential cross section has been measured
by PS170 for five different antiproton momenta. The full line is a fit to the differential 
cross section assuming 
$\left| G_E^p \right| = \left| G_M^p \right|$; the dashed line shows a fit to data with 
$\left| G_M^p \right|$ and $\left| G_E^p / G_M^p \right|$ used as free parameters. Figure taken 
from Ref.~\cite{PS170_94}.}
\label{fig:ps170dsdO}
\end{center}
\end{figure}
%


In the year 1976 -- 3 years after the first Frascati measurement of a TL nucleon 
FF in $e^+e^-$ scattering -- the first positive result was achieved by the Mulhouse-Strasbourg-Torino collaboration~\cite{PS7677} with an antiproton beam at PS/CERN. In total,
$25$ candidate events, which were produced by stopped antiprotons in a liquid hydrogen target, as well as $5$ additional events, in which the annihilation process took place 'in-flight', 
had been recorded.  The experimental setup consisted of several optical spark chambers surrounded by an
array of scintillators.
Due to the low antiproton beam momenta, for the first time the proton 
FF could be measured at $q^2$ values almost at the production threshold.

Indeed, $p \bar{p}$ annihilation experiments allow for a complementary access to the TL
proton FF. Differently to $e^+e^-$ colliders, the production threshold does not feature a  limited phase space for the outgoing particles, and furthermore, one does not have to deal with the experimental issues related to low-momenta protons in the target region. On the other hand, no baryon 
FF's different from protons can be accessed in $p \bar{p}$ annihilation experiments, and the rejection of background with 
hadronic final states is an experimental challenge.

With the advent of a dedicated storage ring for low-momenta antiprotons at CERN, the 
{ LEAR ring}, 
significant progress could be achieved also for measurements of the TL proton FF.
A dedicated experiment PS170 was able extract the FF at $9$ different $q^2$ values from threshold 
up to $4.2$ (GeV/c)$^2$~\cite{PS170_94,PS170_91,PS170_91b}. Not only the beam intensity was increased with respect to the previous CERN experiment,
but also the instrumentation was more advanced. The PS170 detector consisted of several 
layers of MWPCs as tracking detectors, 
Cerenkov detectors, and hodoscopes inside a $1.2$ T magnetic field. It was possible to 
separate signal events  from two-body processes involving hadrons. Fig.~(\ref{fig:ps170}) 
shows a sketch of the detector. In a sophisticated analysis,
the yield of the $e^+e^-$ final state was normalized to the yield of the two-body hadronic reactions 
$p \bar{p} \rightarrow \pi^+\pi^-$ and $p \bar{p} \rightarrow K^+K^-$. The cross section
for the process $p\bar{p} \rightarrow e^+e^-$ was then
computed using the known ratio of hadronic two-body final states to the total $p\bar{p}$ annihilation 
cross section, cf.\ Ref.\cite{PS170_94} for more details.

For antiprotons stopped in
the liquid hydrogen target -- i.e.\ for events very close to threshold -- a large statistics of almost 
$2,000$ events could be
collected, exceeding the previously  available statistics by more than one order of magnitude. 
The extracted FF was however limited by systematic 
uncertainties. Due to the fact that the stopped antiproton is forming a $p\bar{p}$
atom, only the configuration with $J^{PC}=1^{--}$ quantum numbers is relevant for the
em FF. Hence, 
the fraction of $L=1$ quantum states, as well as the ratio of the spin singlet to 
triplet probabilities need to be known. Overall, an uncertainty
of approximately $10$\% was achieved at threshold. 

The large angular acceptance 
together with the relatively high statistics did also allow for a first measurement of the 
{differential cross section using 'in-flight' 
events. In one-photon approximation, this differential cross section is given by}
\be
\frac{d\sigma}{d{\rm cos}{\theta}} = \frac{\pi \alpha^2 }{4 k \sqrt{q^2}}
~C_p(q^2)~
\left[\left| G_M^p(q^2) \right|^2 (1+{\rm cos}^2\theta) + 
\frac{4M_p^2}{q^2} \left| G_E^p(q^2) \right|^2 {\rm sin}^2\theta \right]{\rm ,} 
\label{eq:dsdo}
\ee
where $\theta$ is the angle between the proton and the beam direction in the c.m. system, and
$k$ the antiproton momentum in the CM system ({ see} also Eq. (\ref{tlcross}). From the measurement of the 
differential cross section (cf.\ Fig.~(\ref{fig:ps170dsdO})) and using Eq.~(\ref{eq:dsdo}),
the ratio $\left| G_E^p / G_M^p \right|$ could be extracted. The experiment PS170 has measured
this ratio for five different $q^2$ values, although with relatively
large statistical and systematic uncertainties of $\sim 30$\%. The result for the ratio will be
presented in the next Subsect..

In the 1990's and at the beginning of the new century, measurements
of the TL proton FF have been obtained also at FNAL, where a high-intensity, stochastically-cooled
antiproton beam used to be in operation. 
Using a hydrogen gas target, the experiments E761 and E835 covered the
high-$q^2$ range between $8.2$ (GeV/c)$^2$ and $18.2$ (GeV/c)$^2$. Those were the highest momentum
transfers achieved at the time, exceeding the measurements at BES-II in electron-positron annihilation.
At the highest energy point, $q^2=18.22$ (GeV/c)$^2$, only an upper limit could be established. 
In contrast to the low-momenta antiproton experiments performed at LEAR, 
at FNAL the cross section measurement
could be normalized to the integrated luminosity, which was monitored by a dedicated monitoring system
using elastically scattered antiprotons as a reference process. The cross section is determined 
using Eq.~(\ref{geffa}) -- with the only difference that a different kinematic flux factor has to be considered -- , and Eq.~(\ref{eq:ff_exp}) with $\mathcal{N}_{N\bar{N}}$ to be replaced with 
$\mathcal{N}_{ee}$,  where $\mathcal{N}_{ee}$ corresponds to the
number of $p\bar{p} \rightarrow e^+e^-$ events.

Table~(\ref{tab:ppbar}) shows a summary of all results from $p\bar{p}$ annihilation 
experiments for extracting  TL nucleon FF's.

\begin{table}[h]
\begin{center}
\begin{tabular}{|l|c|c|c|c|c|c|} \hline
\textbf{Exp.} &	\multicolumn{1}{c|}{\textbf{Reaction}} & 	\textbf{\specialcell{Year \\ Publ.}} & \textbf{\specialcell{Scan \\ Points}} & \textbf{Range [GeV]} & \multicolumn{1}{c|}{\textbf{Events}} & \textbf{Ref.} \\ \hline
M.S.T. Coll. &  	$p\bar{p} \rightarrow e^+e^-$ & 			1976/77 & 2 & near threshold & 34 &~\cite{PS7677} \\ \hline
PS170 &  		$p\bar{p} \rightarrow e^+e^-$ & 			1991 & 	4 & near threshold & $\sim 2000$ &~\cite{PS170_91} \\
PS170 &  		$p\bar{p} \rightarrow e^+e^-$ & 			1991 & 	4 & $1.94 - 2.05$ & $\sim 1300$ &~\cite{PS170_91b} \\
PS170 &  		$p\bar{p} \rightarrow e^+e^-$ & 			1994 & 	9 & $\text{threshold} - 2.05$ & $\sim 2000$ &~\cite{PS170_94} \\ \hline
E760 &  		$p\bar{p} \rightarrow e^+e^-$ & 			1993 & 	3 & $3.0 - 3.6$ & 29 &~\cite{E760_93} \\
E835 &  		$p\bar{p} \rightarrow e^+e^-$ & 			1999 & 	4 & $3.0 - 3.8$ & 144 &~\cite{E835_99} \\
E835 &  		$p\bar{p} \rightarrow e^+e^-$ & 			2003 & 	2 & $3.4 - 3.5$ & 66 &~\cite{E835_03} \\ \hline
\end{tabular}
\end{center}
\caption{Summary of experimental results using the reaction $p\bar{p} \to e^+e^-$.}
\label{tab:ppbar}
\end{table}

\subsubsection{Initial State Radiation}
\label{isr}
Around the year 2000, a new generation of high-luminosity electron-positron colliders came into operation, which were explicitly designed to operate at fixed c.m.\ energies, $\sqrt{s}$, corresponding to the mass of either the 
$\phi(1020)$-resonance -- in the case of the $\phi$-factory DA$\Phi$NE in 
Frascati -- or the mass of the 
$\Upsilon(4S)$-resonance -- in the case of the B-factories PEP-II at SLAC and 
KEK-B in Tsukuba. As those resonances provide the highest statistics for 
coherent production of  pairs of K- and B-mesons, the main physics motivation 
for the above mentioned facilities was the measurement of CP-violation in 
 K- and  B-meson systems. 
%

Energy scans over wide energy ranges were { technically} not possible at the first generation $\phi$- and B-factories. It was realized, however (see Refs.~\cite{zerwas, eva, isrbfactories}), 
that the very high luminosities did allow for a complementary approach to the standard energy 
scan. 
Events with photon radiation from the 
initial state (ISR), cf.\ Fig.~(\ref{fig:feynman},c), lead to a reduction of the invariant mass of the virtual 
photon, $s^\prime$, and hence allow for a measurement of the hadronic 
FF's at energies 
below $\sqrt{s}$. 
By measuring the invariant mass of the hadronic system, $M_{\rm hadr}$, the entire energy range
below the c.m.\ energy of the collider becomes accessible: $\sqrt{s^\prime} < \sqrt{s}$. By neglecting
effects of final state radiation (FSR), the following relation holds for the 
four-momentum
transfer $q^2$: 
$$
q^2 = M_{\rm hadr}^2\cdot c^2 = s^\prime
. $$
Comprehensive reviews of the ISR method, which is also called {\it Radiative Return} technique,
can be found in Refs.~\cite{actis} and \cite{isrreview}.
The method was both used for measurements of hadronic reactions
with mesons in the final state, as for the measurement of baryon pair production:
$$
e^+e^- \rightarrow B \bar{B} \gamma_{\rm ISR}.
$$
The BaBar experiment at the SLAC B-factory PEP-II has measured the TL proton FF below $4.5$ GeV, and
has furthermore investigated the following three hyperon final states: $\Lambda\bar{\Lambda}$, $\Lambda\bar{\Sigma^0}$, and 
$\Sigma^0\bar{\Sigma^0}$. Before BaBar no experiment had the sensitivity to
measure the latter two channels.
The full BaBar data set comprises an integrated luminosity of $\sim 500$ fb$^{-1}$,
collected between 1999 and 2008. BaBar results of
mesonic final states, i.e.\ of final states containing two pions, three pions, four pions etc.\, 
are very valuable input for evaluating the hadronic contributions to 
$(g-2)_\mu$ and  the running fine structure constant 
$\Delta \alpha_{\rm had}(M_Z^2)$. Notably, a measurement of the TL pion FF from threshold up to $3$ GeV 
with $<1$\% precision in the $\rho$ peak region was achieved~\cite{babar_2pi}.
Furthermore, a series of interesting 
structures have been observed in the mass spectra, 
e.g.\ a destructive interference effect in the $6\pi$ final states~\cite{babar_6pi} at an energy being
very close to the $p\bar{p}$ threshold.  Among the major results 
of the ISR programme at BaBar, we want to mention also  
the discoveries of the $Y(4260)$~\cite{babar_y4260} and $Y(2175)$~\cite{babar_y2175} resonances.

\begin{figure}[t]
\begin{center}
\includegraphics[width=10.5cm]{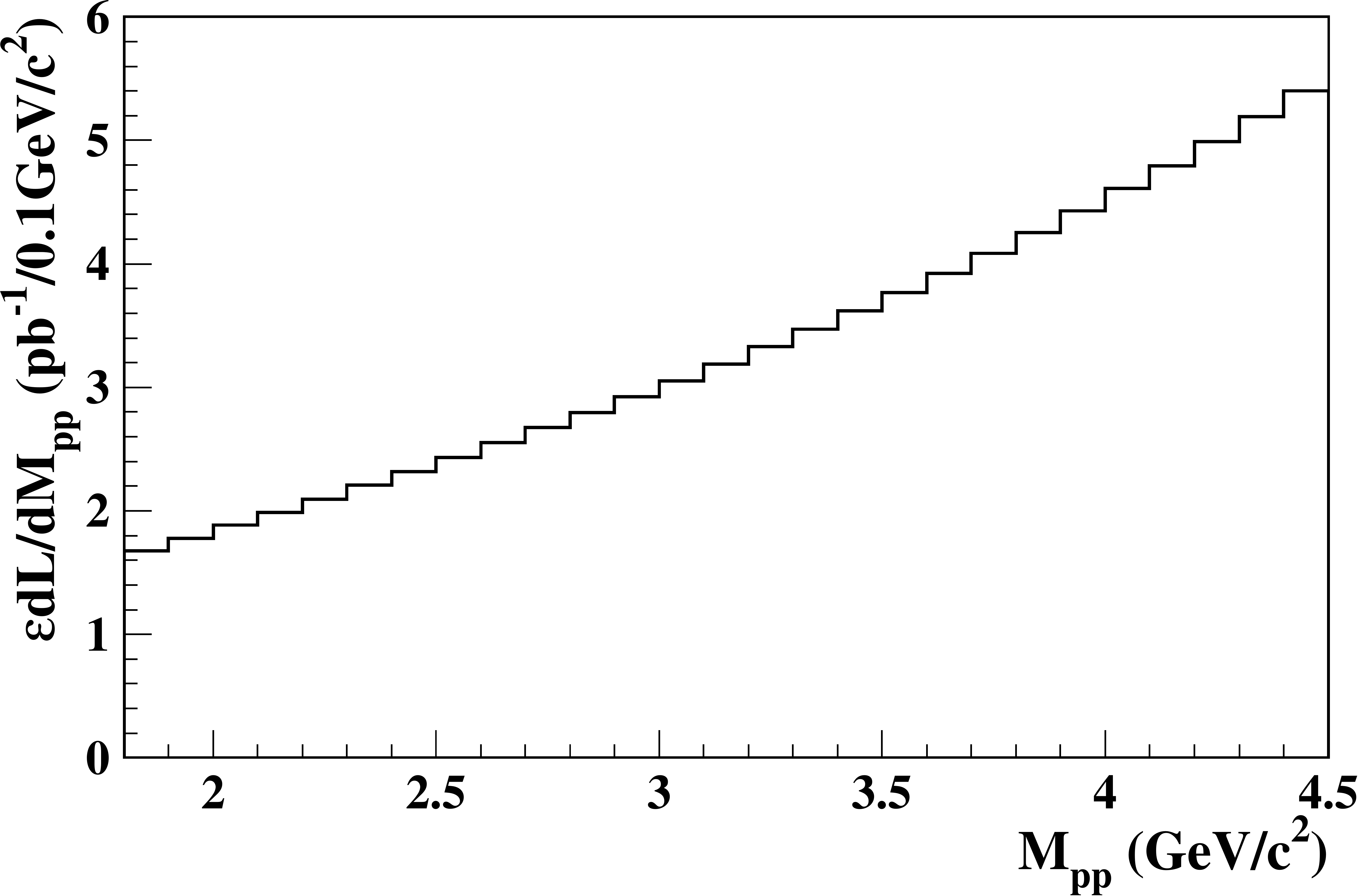}
\caption{ISR luminosity in the case of BaBar kinematics for $\mathcal{L_{\rm int}}=500$ fb$^{-1}$.
}
\label{fig:isrlumi}
\end{center}
\end{figure}
On the theoretical side, the ISR process was calculated within QED up to Next-to-Leading Order (NLO) 
and the probability for ISR photon emission
can be expressed by means of a radiator function $H_s(s^\prime)$~\cite{eva, isrbfactories, baier}. 
Notice that high energies of the ISR photon correspond to small values of $s^\prime$. 
Neglecting again effects of FSR (see Ref.~\cite{egle_fsr} concerning this issue), 
i.e.\ assuming { $M_{\rm hadr} \cdot c=\sqrt{s^\prime}$}, 
the {\it non-radiative} cross section $\sigma_{\rm hadr}=\sigma(e^+e^- \rightarrow {\rm hadrons})$
can be extracted from the measured {\it radiative} cross section 
$\sigma_{{\rm hadr +} \gamma}=\sigma(e^+e^- \rightarrow {\rm hadrons}+\gamma_{\rm ISR})$ 
by using this radiator function:
\be
\frac{d\sigma_{{\rm hadr} +\gamma}}{ds^\prime} = \sigma_{\rm hadr}(s^\prime) \cdot H_s(s^\prime) 
\cdot \frac{1}{s^\prime}.
\label{eq:hfunct}
\ee
The calculation of the radiator function is usually performed with the Monte Carlo method; a dedicated
Monte Carlo generator for ISR processes has been developed within the PHOKHARA package, which simulates
a large number of hadronic reactions, including also the baryonic final states~\cite{phokhara_ppbar} 
$p\bar{p}\gamma$, $n\bar{n}\gamma$, and 
$\Lambda \bar{\Lambda}\gamma$~\cite{phokhara_LLbar}. 
PHOKHARA simulates the full NLO-ISR radiative corrections~\cite{phokhara}. 
Within the BaBar collaboration, 
besides PHOKHARA, also the AFKQED Monte Carlo package has been developed, which is 
simulating FSR corrections by means of PHOTOS~\cite{photos} and which uses the 
structure function approach for higher order ISR corrections.
The AFKQED generator is based on an early version of PHOKHARA, called EVA, cf.\ Refs.~\cite{eva,eva4pi}. 

Due to the specific energy dependence of the ISR photon, the radiator function $H_s(s^\prime)$ 
decreases steeply with $s^\prime$. In Fig.~(\ref{fig:isrlumi}), 
the so-called ISR luminosity is shown, which is the
product of the radiator function times the integrated luminosity: 
$\mathcal{L}_{\rm ISR}(s^\prime)=H_s(s^\prime)\cdot \mathcal{L}_{\rm int}$. The specific
values for BaBar ($\sqrt{s}=10.6$ GeV, $\mathcal{L}_{\rm int}=500$ fb$^{-1}$) have been used
in Fig.~(\ref{fig:isrlumi}). The geometrical acceptance ($\epsilon_{\rm geom}$) and 
detection efficiency ($\epsilon_{\rm detect}$) 
have been considered as well ($\epsilon = \epsilon_{\rm geom} \cdot \epsilon_{\rm detect}$).

The ISR luminosity allows to compare the number of ISR events produced for a given final state and for a 
given mass value $s^\prime$ with the direct production in an energy scan; 
$\mathcal{L}_{\rm ISR}$ depends of course on the chosen bin size and decreases with finer binning.
At the $p\bar{p}$ production threshold, the ISR luminosity 
is found to be $1.7$ pb$^{-1}$ (assuming a { bin width of $100$ MeV/c$^2$}) and therefore 
exceeds the statistics collected at ADONE and DCI. 
Apart from the large statistics, the ISR method offers some additional advantages over conventional 
energy scan measurements:

\begin{figure}[t]
\begin{minipage}[t]{0.485\linewidth}
\centering
\includegraphics[width=\textwidth]{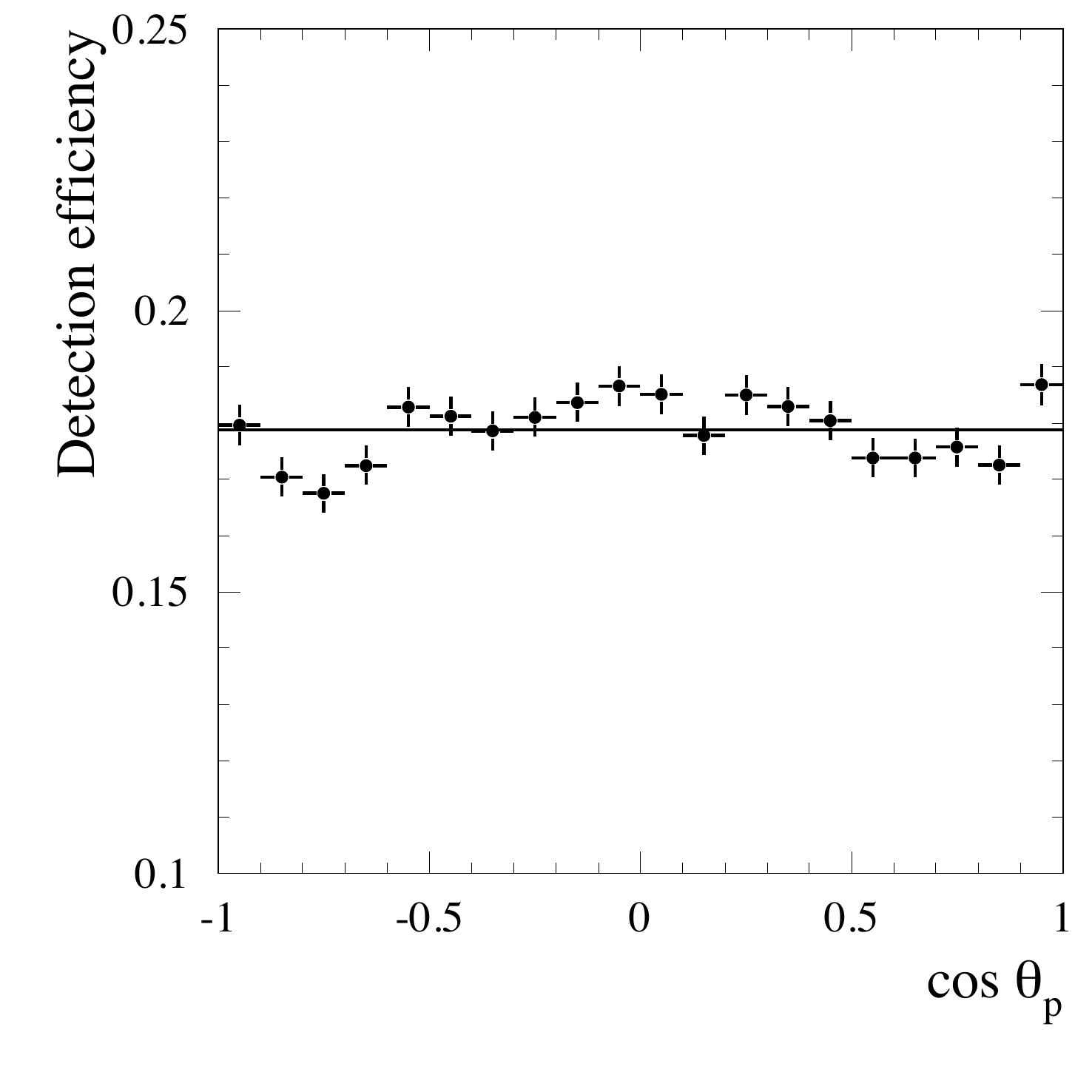}
\caption{BaBar measurement of the process $e^+e^- \rightarrow p\bar{p}\gamma$.
Dependence of the detection efficiency on $\cos{\theta_p}$, 
where $\theta_p$ is the proton
angle measured in the $p\bar{p}$ rest frame with respect to the ISR photon 
direction. 
Right: Dependence of the $M_{p\bar{p}}$ mass resolution on $M_{p\bar{p}}$. 
Figure reprinted with permission from 
Ref.~\cite{isrreview}, \copyright 2011 American Physical Society.
}
\label{fig:babar_analysis1}
\end{minipage}
\hspace{0.3cm}
\begin{minipage}[t]{0.485\linewidth}
\centering
\includegraphics[width=\textwidth]{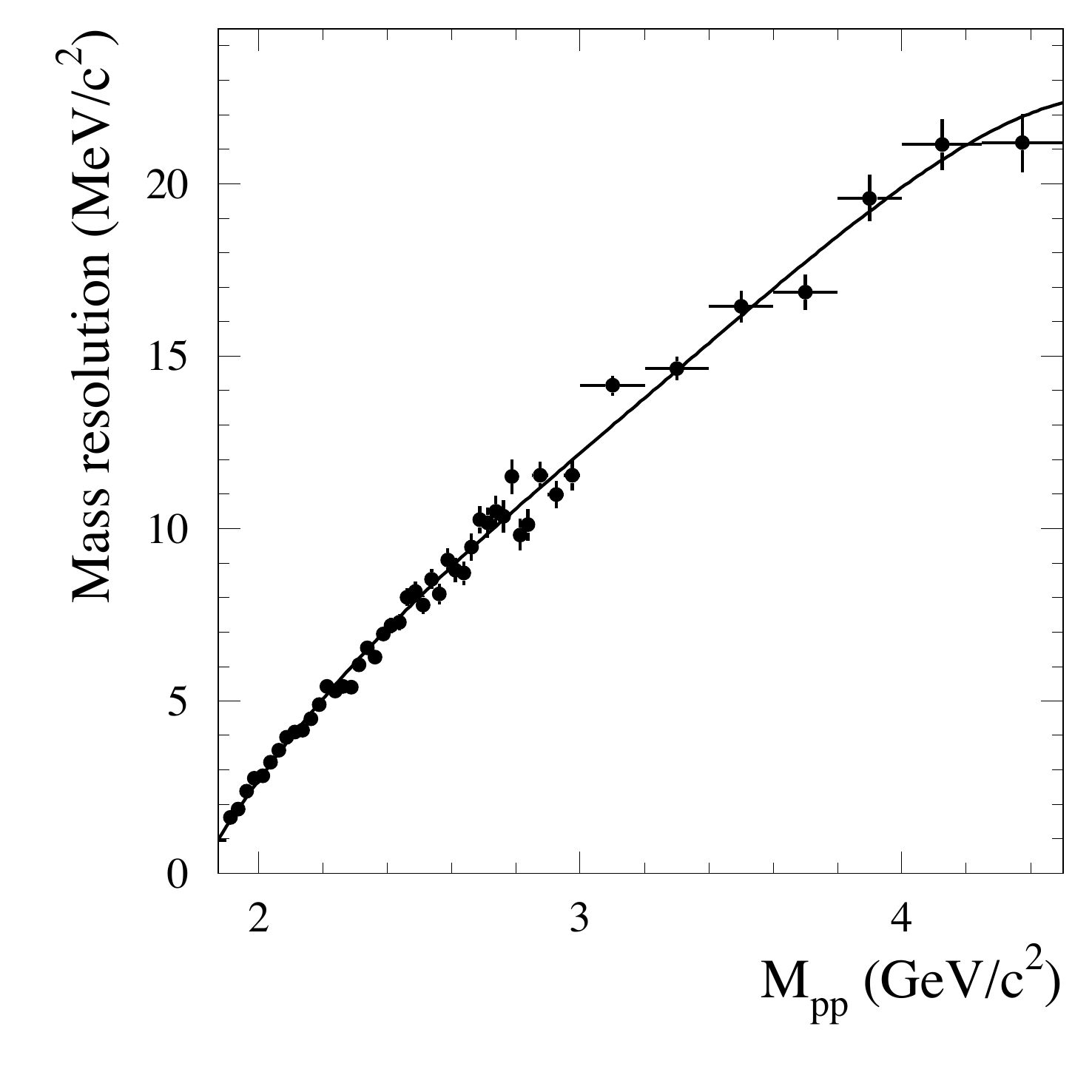}
\caption{
BaBar measurement of the process $e^+e^- \rightarrow p\bar{p}\gamma$.
Dependence of the $M_{p\bar{p}}$ mass resolution on $M_{p\bar{p}}$. 
Figure reprinted with permission from 
Ref.~\cite{isrreview}, \copyright 2011 American Physical Society.
}
\label{fig:babar_analysis2}
\end{minipage}
\end{figure}

\begin{figure}
\begin{center}
\includegraphics[width=0.35\linewidth]{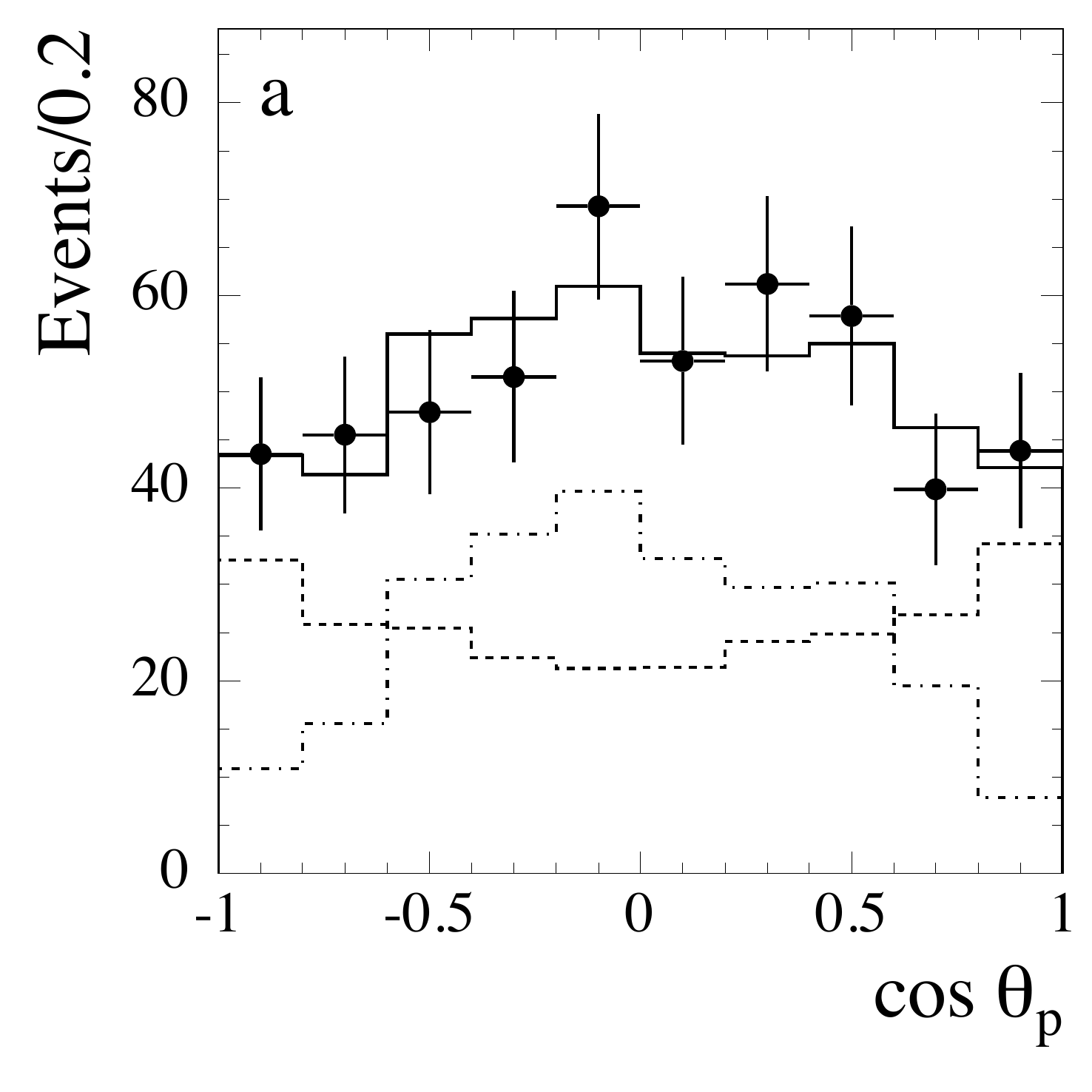}
\hspace{0.7cm}
\includegraphics[width=0.35\linewidth]{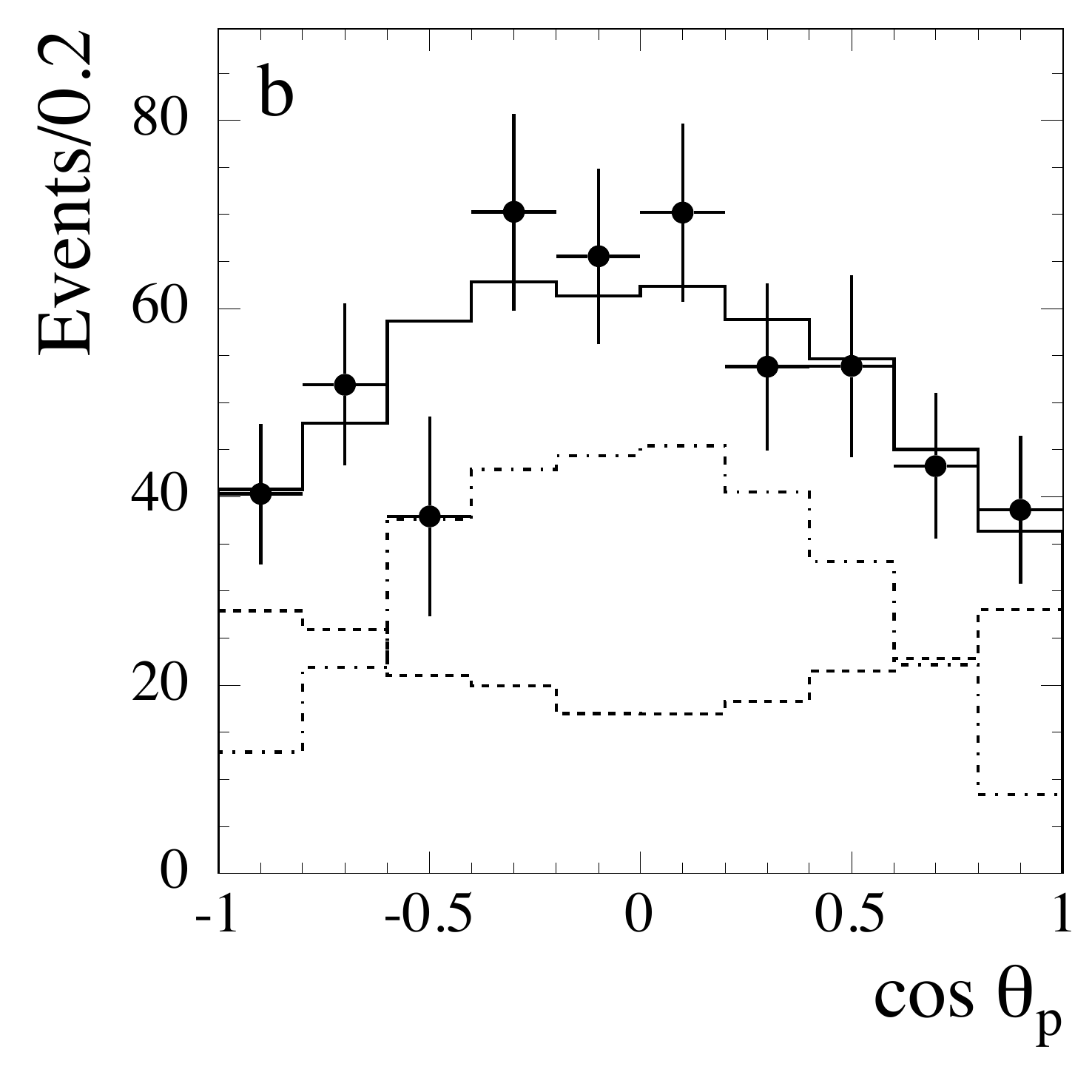}
\vspace{2mm}\\
\includegraphics[width=0.35\linewidth]{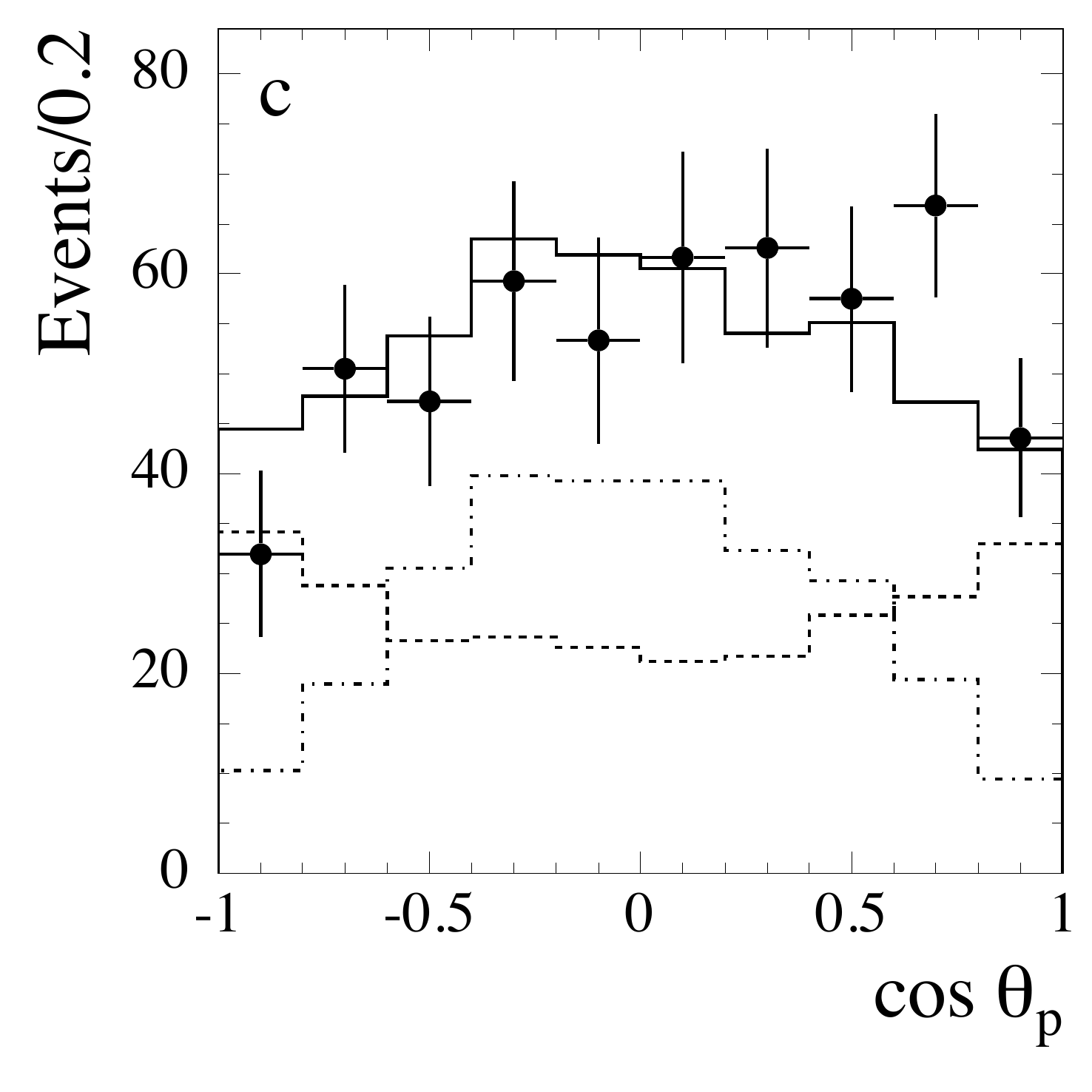}
\hspace{0.7cm}
\includegraphics[width=0.35\linewidth]{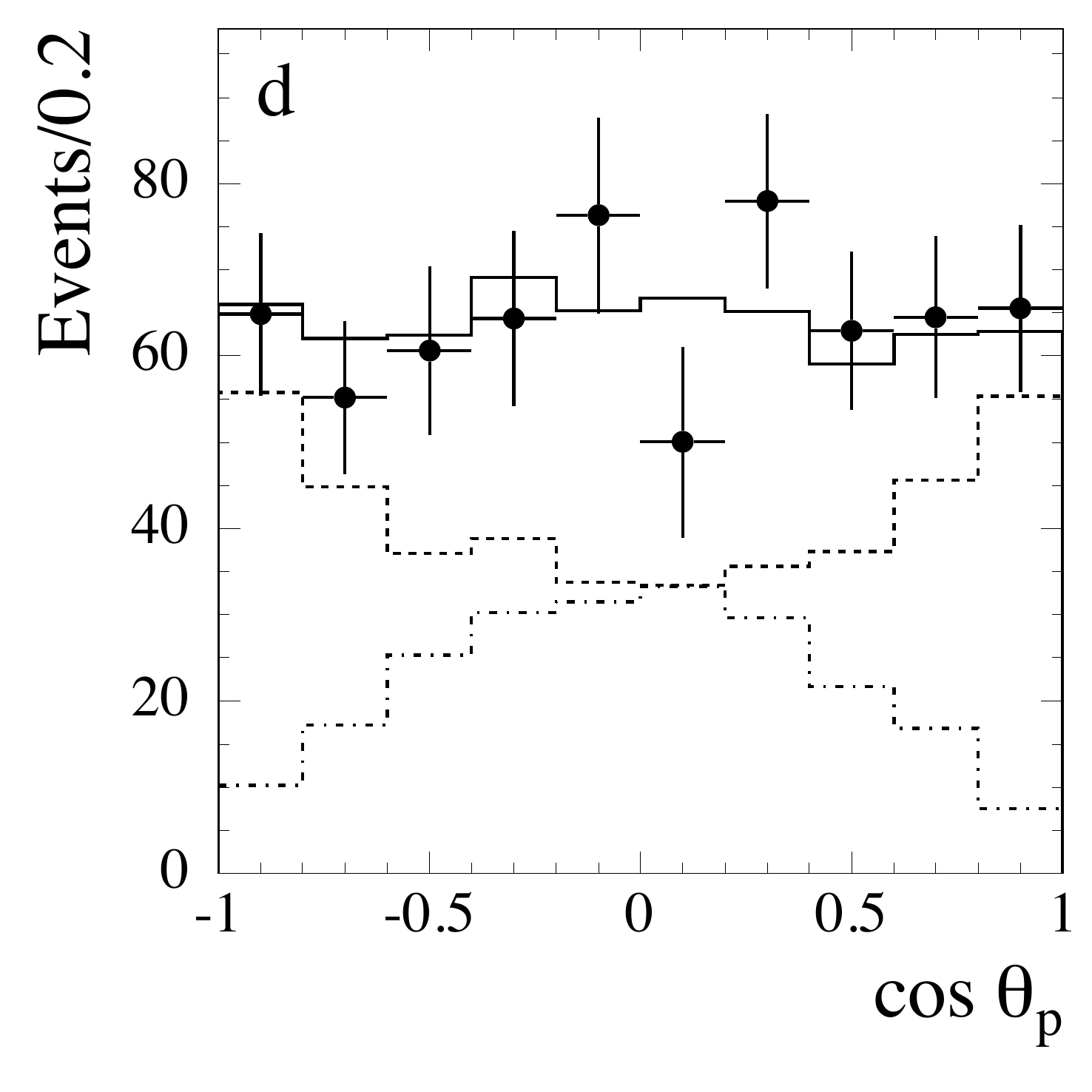}
\vspace{2mm}\\
\includegraphics[width=0.35\linewidth]{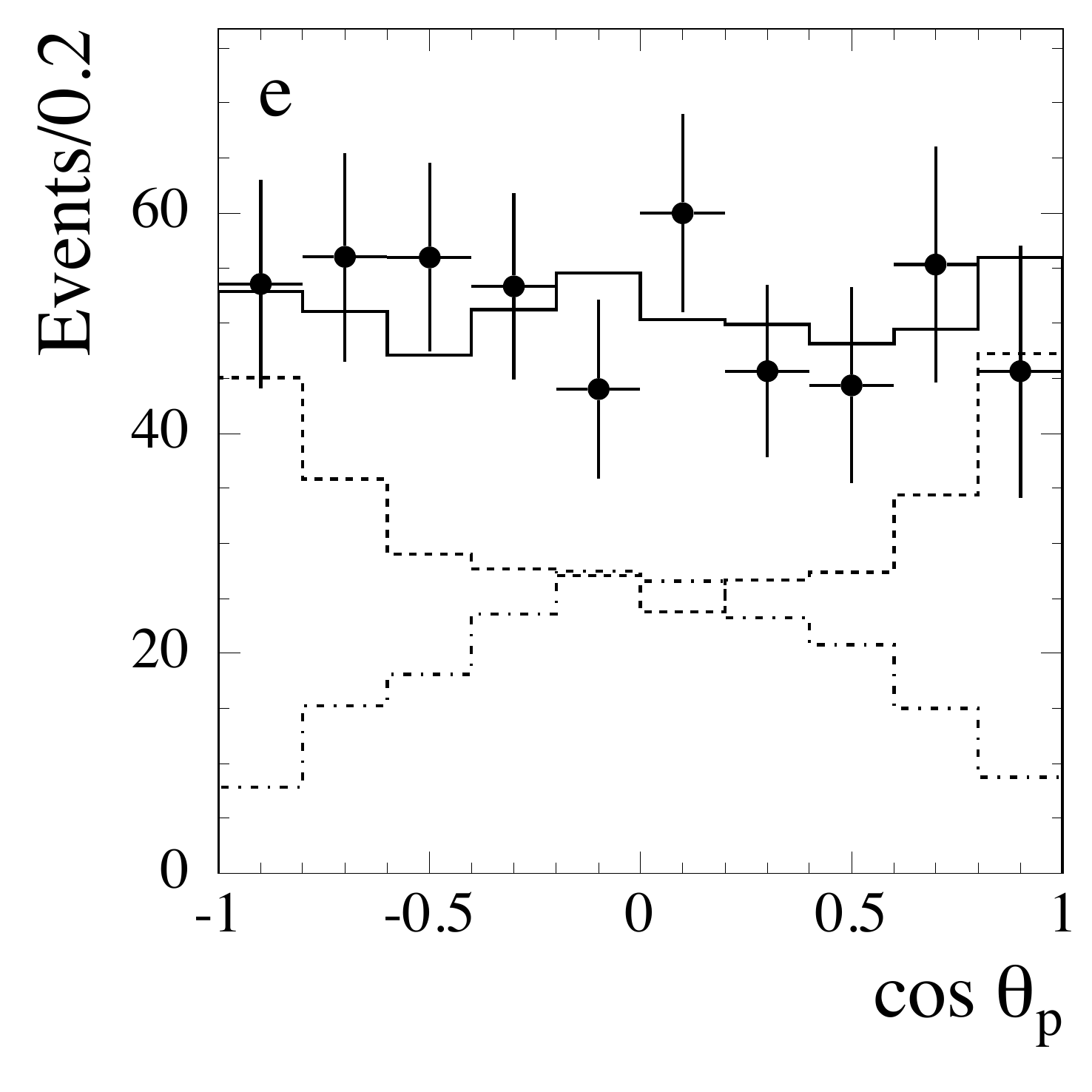}
\hspace{0.7cm}
\includegraphics[width=0.35\linewidth]{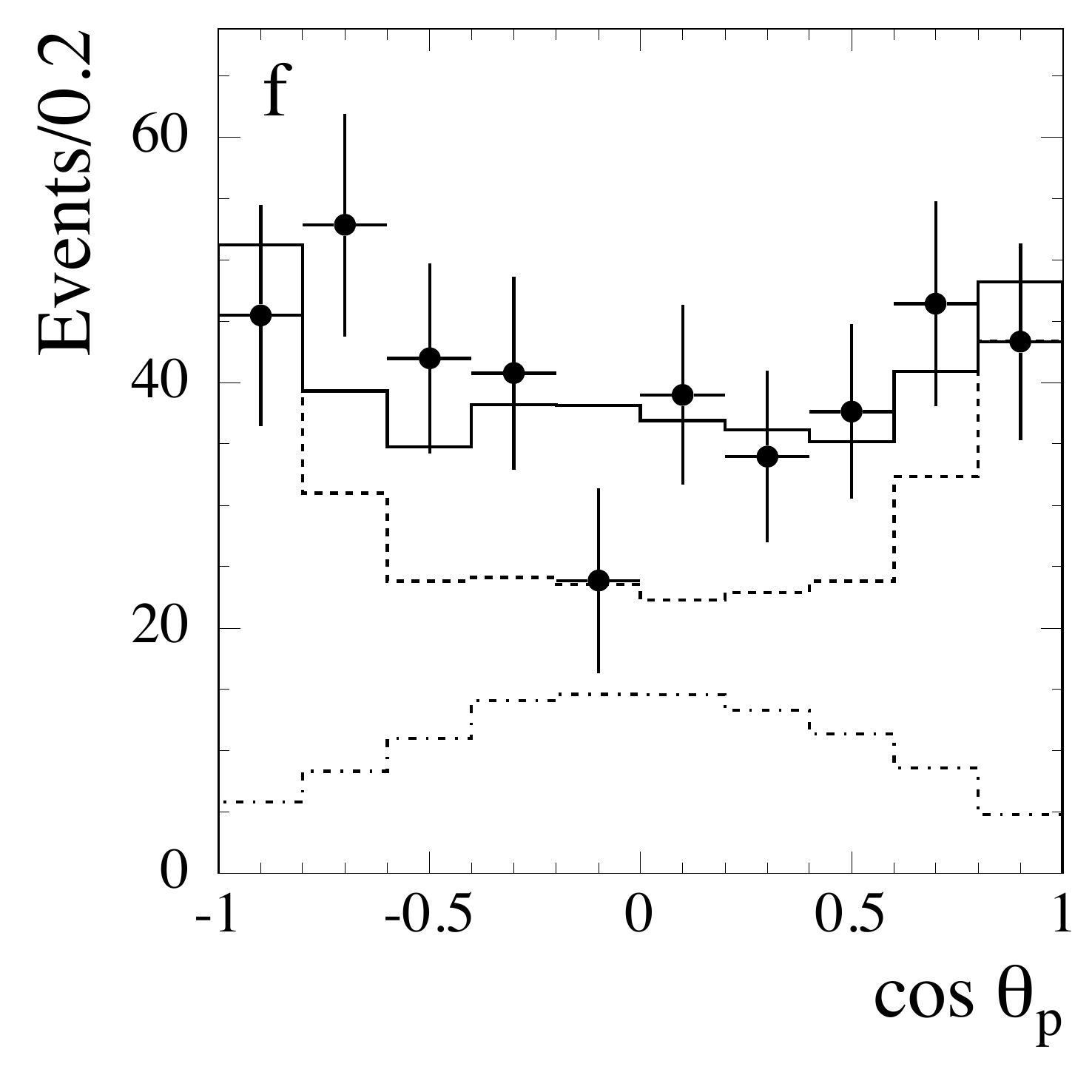}
\caption{ISR measurement of the process $e^+e^- \rightarrow p\bar{p}$ with
the BaBar detector.
The $\cos{\theta}$ distributions are presented for
different $p \bar{p}$ mass regions:
(a) 1.877--1.950~GeV/$c^2$,
(b) 1.950--2.025~GeV/$c^2$,
(c) 2.025--2.100~GeV/$c^2$,
(d) 2.100--2.200~GeV/$c^2$,
(e) 2.200--2.400~GeV/$c^2$,
(f) 2.400--3.000~GeV/$c^2$.
The points with error bars show data distributions after
background subtraction.  
The histograms are fit results: the dashed histograms
show the contributions  corresponding
to the magnetic form factor; the dash-dotted histograms
show the contributions from the electric
      form factor. 
Figure reprinted with permission from 
Ref.~\cite{isrreview}, \copyright 2006 American Physical Society.
\label{as}}
\end{center}
\end{figure}
\begin{figure}[h]
\begin{center}
\includegraphics[width=11.cm]{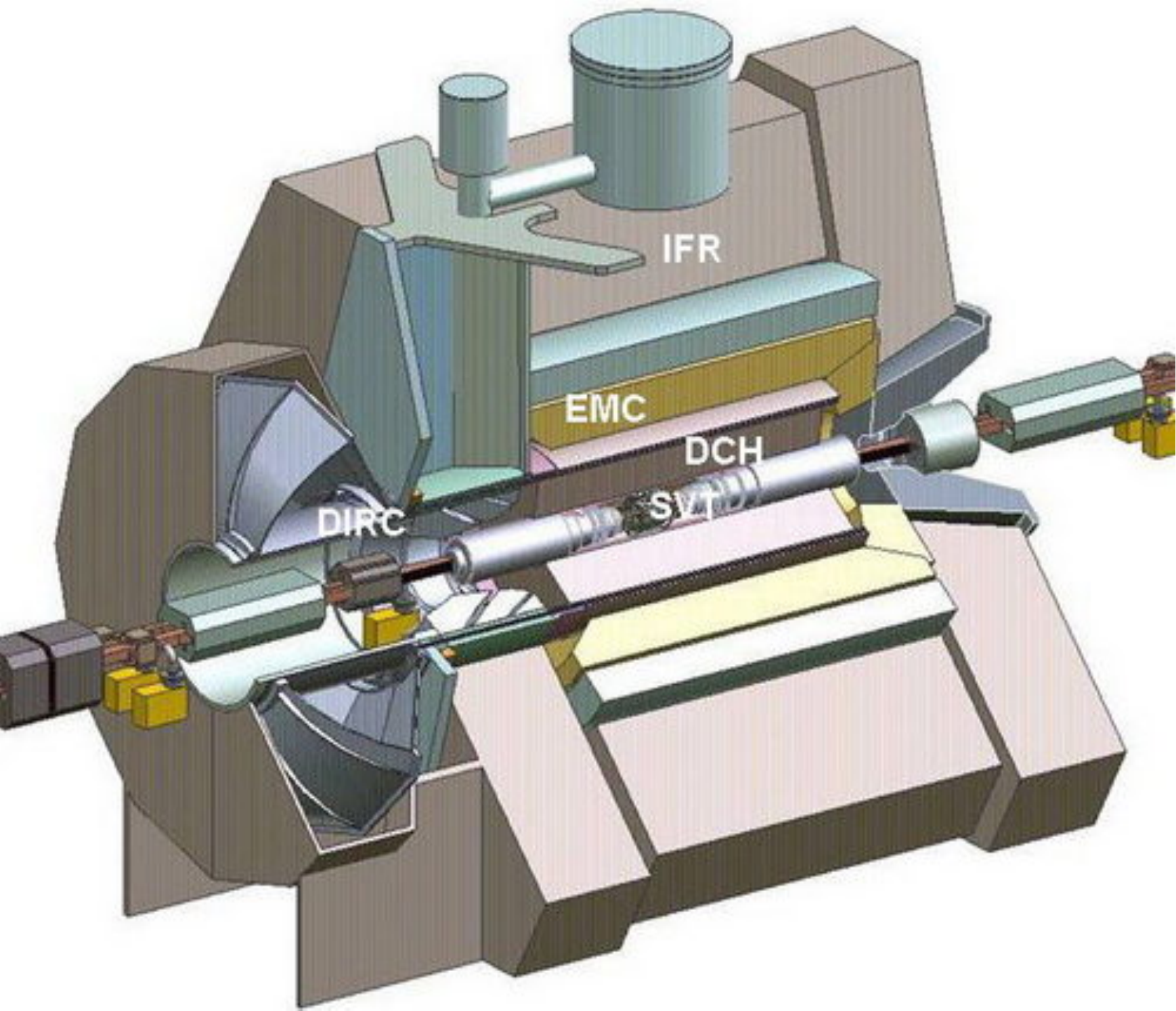}
\caption{The BaBar detector at the B-factory PEP-II at SLAC: (SVT) Silicon 
Vertex Tracker, (DCH) Drift Chamber, (EMC) Electromagnetic Calorimeter,
(DIRC) Detector of internally reflected Cerenkov Light, (IFR) Instrumented
Flux Return.}
\label{fig:babar_det}
\end{center}
\end{figure}
\begin{figure}[t]
\begin{minipage}[b]{0.485\linewidth}
\centering
\includegraphics[width=\textwidth]{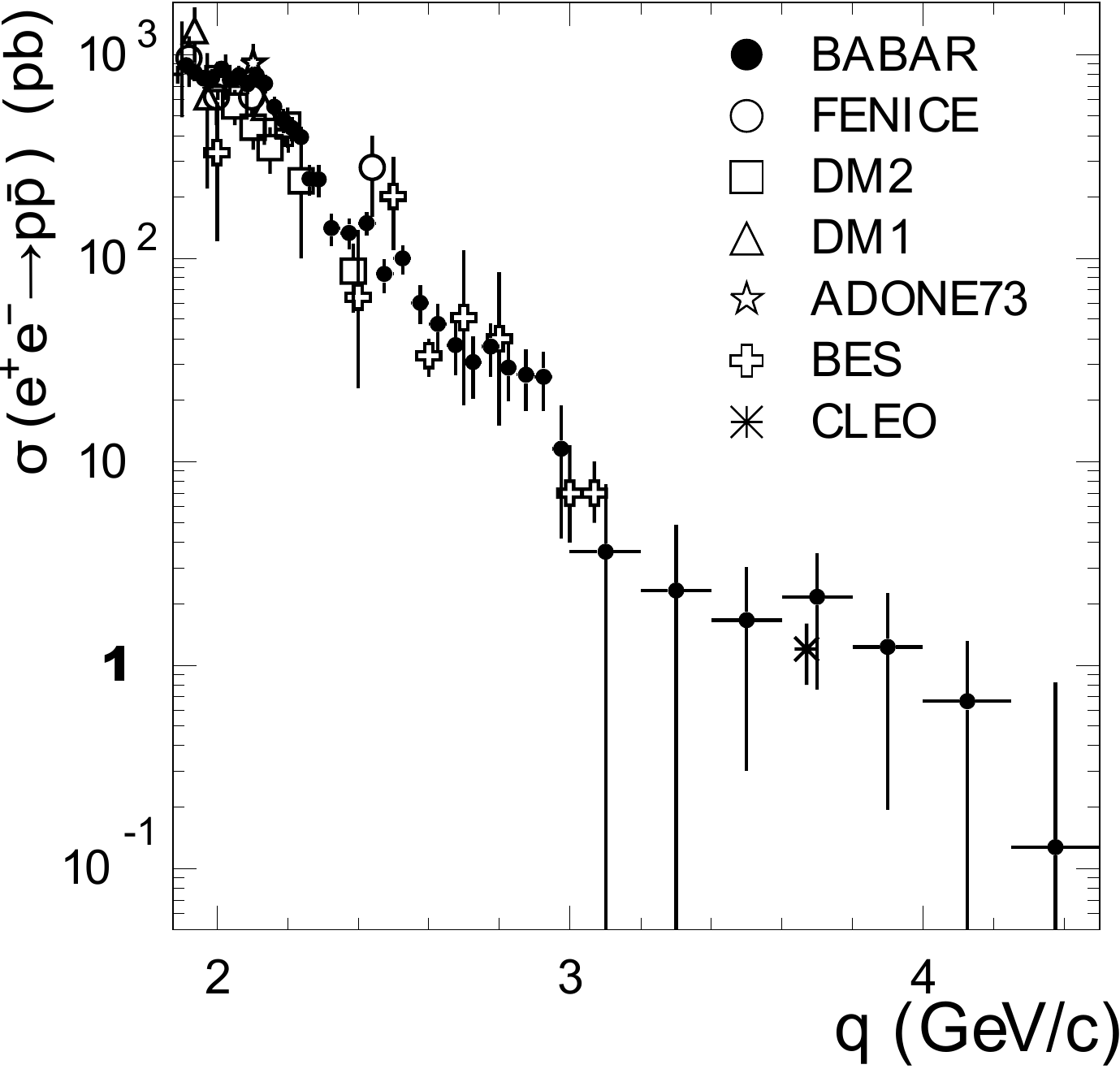}
\caption{Cross section measurements of the process 
$e^+e^- \rightarrow p \bar{p}$ obtained at $e^+e^-$ colliders: 
BaBar~\cite{BABAR}, FENICE~\cite{Fenice}, DM2~\cite{DM283,DM290}, DM1~\cite{DM179}, ADONE73~\cite{Adone73}, BES~\cite{BES05}, CLEO~\cite{CLEO05}.}
\label{fig:xsect_p_log}
\end{minipage}
\hspace{0.3cm}
\begin{minipage}[b]{0.485\linewidth}
\centering
\includegraphics[width=\textwidth]{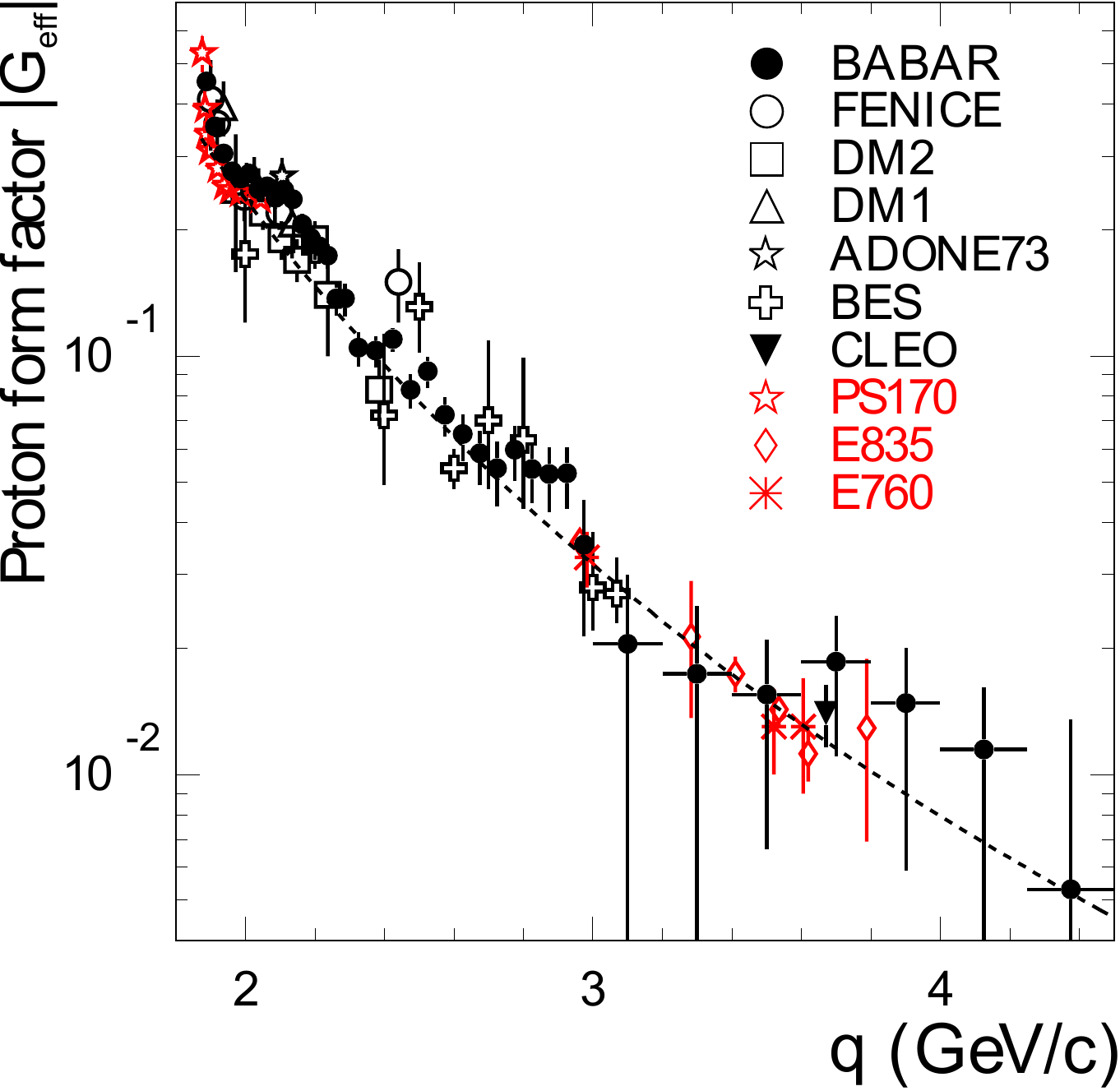}
\caption{The proton {\it effective} form factor 
extracted from measurements at $e^+e^-$ colliders 
(Refs. as in Fig.~(\ref{fig:xsect_p_log})) and at
antiproton accelerators: PS170~\cite{PS170_94}, E835~\cite{E835_99,E835_03}, E760~\cite{E760_93}. }
\label{fig:ff_p_log}
\end{minipage}
\end{figure}
\begin{figure}[h]
\begin{center}
\includegraphics[width=9.cm]{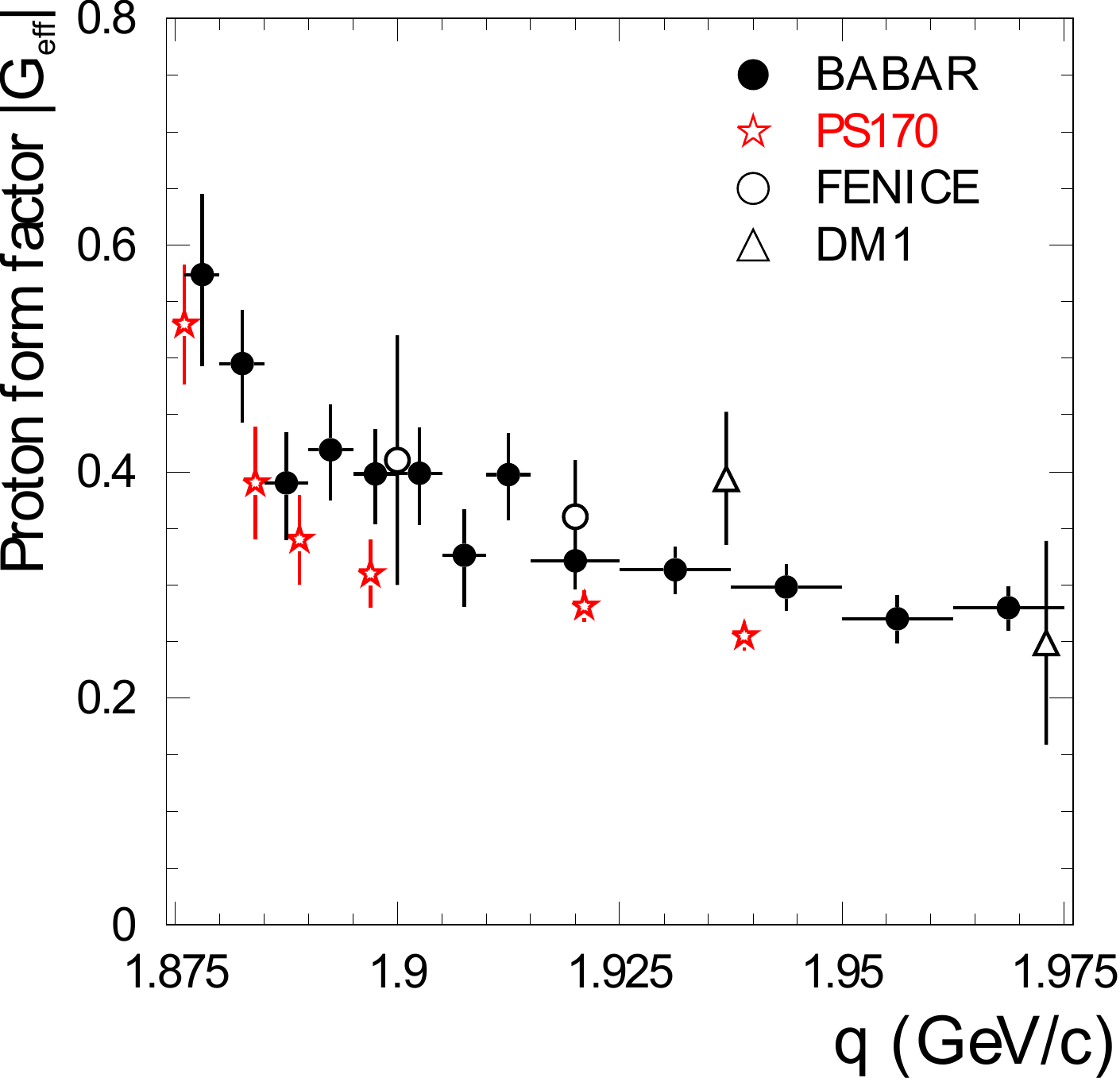}
\caption{Proton {\it effective} form factor measurements obtained in the energy range 
close to the $p\bar{p}$ threshold region: BaBar~\cite{BABAR}, PS170~\cite{PS170_94}, FENICE~\cite{Fenice}, DM1~\cite{DM179}. }
\label{fig:ff_p_threshold}
\end{center}
\end{figure}

\begin{itemize}

\item
A very valuable feature is the fact that in one single experiment 
the entire energy range of interest can be covered. 
This avoids the notorious problems associated with combinations of different data sets, which are usually 
having different normalization uncertainties. At BaBar, the 
proton FF could be measured over a very wide mass range from production threshold 
up to $\sqrt{s^\prime} = 4.5$ { GeV/c}, which has never been possible before and which exceeds the energy range covered by BES-II and the FNAL experiments.  

\item
In the baryon FF measurements performed at BaBar, the ISR photon has been tagged in the
em calorimeter (EMC) at relatively large polar angles. Events with small angle ISR photons, which a priori
have a much higher cross section, lead to an event signature, in which the 
hadronic system is { emitted opposite to the ISR photon. The hadronic particles are hence flying outside} the
detector acceptance. The lower the invariant of the hadronic system is, the higher is the probability to
lose at least one of the hadronic particles in the final state. However, for the tagged
photon approach this back-to-back signature has its own merits, namely it leads to
the fact that the geometrical acceptance for $p\bar{p}$ events is very flat compared to usual
energy scan experiments. This can be seen in Fig.~(\ref{fig:babar_analysis1}), where the angular acceptance for 
$e+e^- \rightarrow p \bar{p} \gamma$ events is shown. The almost uniform
angular distribution reduces the model dependence of the cross section measurement due to the
unknown $\left| G_E^p / G_M^p \right|$ ratio, which represents a major limitation in the data analysis 
of energy scan experiments.

\item
Furthermore, for ISR events -- differently from standard $e^+e^-$ experiments -- the momenta of the final state particles at threshold are non-zero. This does not only reduce the
systematic effects, which are usually unavoidable when dealing with low-momenta particles, but
it leads also to a flat reconstruction efficiency as a function of $M_{\rm hadr}=M_{pp}$, which is especially important if
the FF { at threshold is investigated}. In Fig.~(\ref{fig:babar_analysis2}) the mass resolution as a function of the hadronic mass
is shown. At low masses, {the mass resolution is as}
small as $1$ MeV/c$^2$. In order to improve the mass resolution and to reject background channels, a 
kinematic fit is usually performed.

\end{itemize}

\begin{figure}[t]
\begin{center}
\includegraphics[width=14.cm]{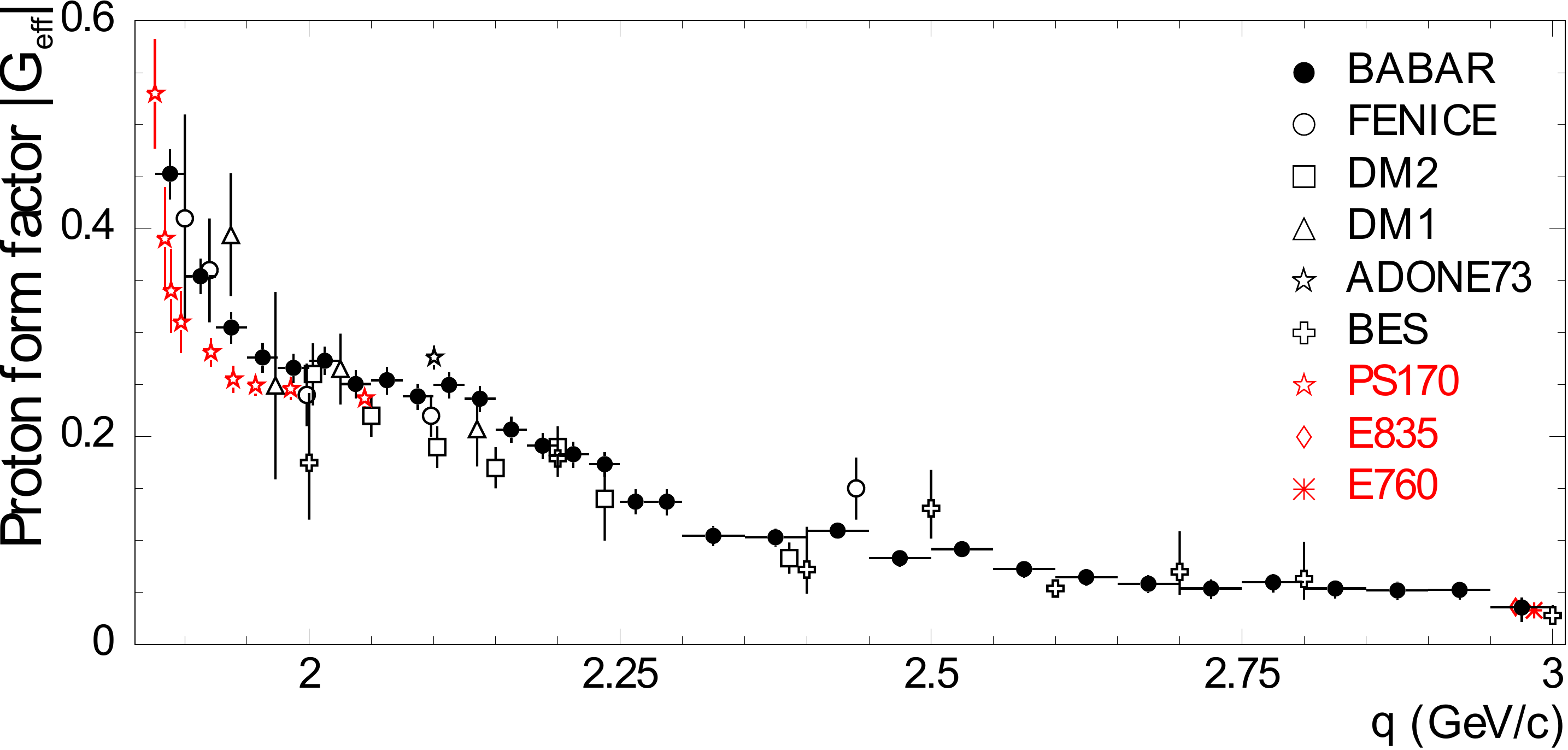}
\includegraphics[width=14.cm]{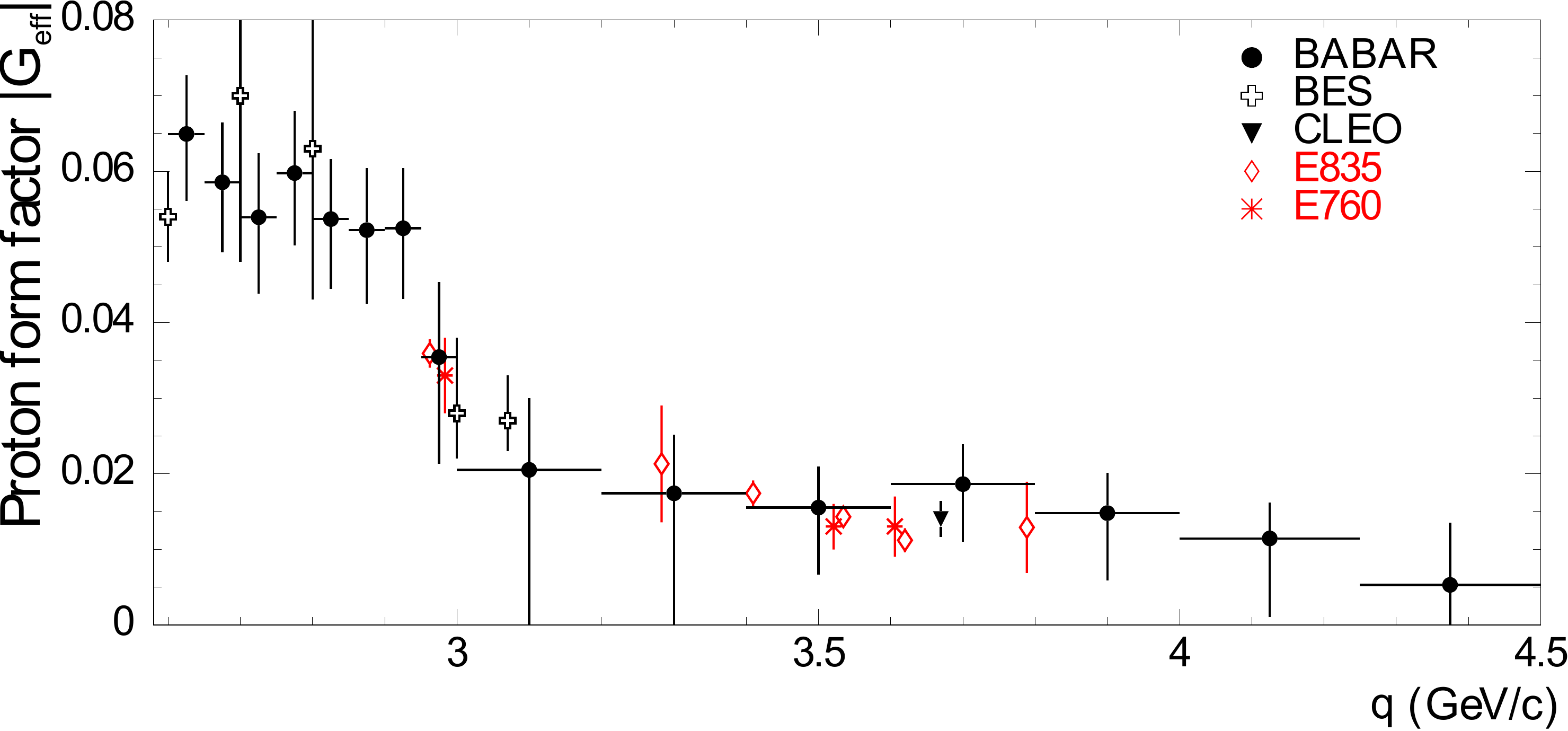}
\caption{Proton {\em effective} FF measurements 
plotted in linear scale for the energy range $1.85 - 3.01$ GeV (upper plot)  and
$2.58 - 4.5$ GeV (lower plot): 
BaBar~\cite{BABAR}, FENICE~\cite{Fenice}, DM2~\cite{DM283,DM290}, DM1~\cite{DM179}, ADONE73~\cite{Adone73}, BES~\cite{BES05}, CLEO~\cite{CLEO05}, PS170~\cite{PS170_94}, E835~\cite{E835_99,E835_03}, E760~\cite{E760_93}.}
\label{fig:ff_p_lin_2plots}
\end{center}
\end{figure}

We conclude that the ISR technique is a very competitive method for the measurement
of baryon TL FF's. Moreover, the
high statistics and accuracy of the BaBar measurement of the proton FF~\cite{BABAR} 
allowed a measurement of the 
$\left| G_E^p / G_M^p \right|$ ratio in five energy bins. Like in the case of 
the PS170 experiment, the angular distributions are investigated to disentangle
 $\left| G_E^p \right|$ 
and $\left| G_M^p \right|$. However, when using the ISR technique, the existence of an additional
high-energetic photon in the final state needs to be taken into account. In the BaBar analysis,
the differential cross section with respect to the cosine of the angle $\theta_{\rm P}$ has been fitted
(cf.\ Fig.~(\ref{as})), where $\theta_{\rm P}$ is the angle between the 
proton in the
$p\bar{p}$ rest frame and the momentum of the $p\bar{p}$ system in the 
$e^+e^-$ rest frame (remember that PEP-II is an asymmetric collider built for CP studies in 
the B-meson sector).
In Ref.~\cite{phokhara_LLbar} it has been 
suggested to use a different approach, namely to consider the proton angular distribution 
in the hadronic rest frame, with the z-axis aligned with the direction of the ISR 
photon and the y-axis in the plane determined by the beam and the photon directions.

Fig.~(\ref{fig:babar_det}) shows a sketch of the BaBar detector. Charged particles are detected in the BaBar tracking system,
which comprises a five-layer silicon vertex tracker (SVT) and a 40-layers cylindrical drift chamber (DCH) 
operating in a $1.5$ T solenoidal magnetic field. The momentum resolution for a $1$ GeV/c  track
is $\sigma_{p_t}/p_{t}=0.5$\%. Particle identification is provided by a an internally reflected 
Cerenkov detector (DIRC), as well as by $dE/dx$ information from the tracking system. The energy and
position of photons is measured by a CsI(Tl) em calorimeter (EMC); muons are identified
by a dedicated detection system inside the iron coil (instrumented flux return, IFR). No attempts have been
made at BaBar to measure the detection efficiency of neutrons inside the EMC; only
the proton-antiproton final state as well as hyperon channels, which decay into protons and pions, have been 
investigated. 

\begin{table}[b]
\begin{center}
\begin{tabular}{|l|c|c|c|c|c|c|c|} \hline
\textbf{Exp.} &	\multicolumn{1}{c|}{\textbf{Reaction}} & 	\textbf{\specialcell{Year \\ Publ.}} & \textbf{\specialcell{Mass \\ Binning}} & \textbf{Range [GeV]} & \textbf{\specialcell{$\mathcal{L}_\mathrm{int}$ \\ ~[pb${}^{-1}$]~}} & \multicolumn{1}{c|}{\textbf{Events}} & \textbf{Ref.} \\ \hline
\rule{0pt}{11pt}BaBar &  		$e^+e^- \rightarrow p\bar{p}$ & 			2005 & 	47 & $\text{threshold} - 4.5$ & 232$\cdot 10^3$ & 4025 &~\cite{BABAR} \\
\rule{0pt}{11pt}BaBar & $e^+e^- \rightarrow \Lambda\bar{\Lambda}$ & 	2007 & 	12 & $\text{threshold} - 3.0$ & 232$\cdot 10^3$ & 138 &~\cite{BABAR2} \\
\rule{0pt}{11pt}BaBar & $e^+e^- \rightarrow \Lambda\bar{\Sigma^0}$ & 	2007 & 	4 & $\text{threshold} - 2.9$ & 	232$\cdot 10^3$ & 24 &~\cite{BABAR2} \\
\rule{0pt}{11pt}BaBar & $e^+e^- \rightarrow \Sigma^0\bar{\Sigma^0}$ & 		2007 & 	5 & $\text{threshold} - 3.0$ & 	232$\cdot 10^3$ & 18 &~\cite{BABAR2} \\ \hline
\rule{0pt}{11pt}BELLE & $e^+e^- \rightarrow \Lambda_c^+\bar{\Lambda_c^-}$ & 	2008 & 50 & $\text{threshold} - 5.4$ & 	659$\cdot 10^3$ & not cited &~\cite{BELLE} \\ \hline
\end{tabular}
\end{center}
\caption{Summary of TL baryon FF results obtained via the ISR technique.}
\label{tab:isr}
\end{table}

Table~(\ref{tab:isr}) gives a summary of the ISR measurements of TL baryon FF's, which have been performed so far. Besides the BaBar results reported above,
the BELLE experiment at the Japanese B-factory project KEK-B has investigated
 the process $e^+e^- \to \Lambda_c^+ \Lambda_c^-$ in the energy range between
$4.5$ GeV and $5.4$ GeV using the ISR technique~\cite{BELLE}. The integrated luminosity collected at BELLE is
exceeding $1,000$ fb$^{-1}$ and is hence twice as large as the luminosity available at BaBar. 
Until now, the ISR programme at BELLE  has been entirely 
devoted to the production of charm and charmonium resonances in
the mass range above $4$ GeV/c$^2$. In the case of the $\Lambda_c^+ \Lambda_c^-$ analysis, a resonance at
$M=(4634^{+8}_{-7}({\rm stat.})^{+5}_{-8}({\rm syst.}))$ MeV/c$^2$ and a width of 
$\Gamma=(92^{+40}_{-24}({\rm stat.})^{+19}_{-21}({\rm syst.}))$ MeV has indeed been found.
\begin{figure}[h]
\begin{center}
\includegraphics[width=9.5cm]{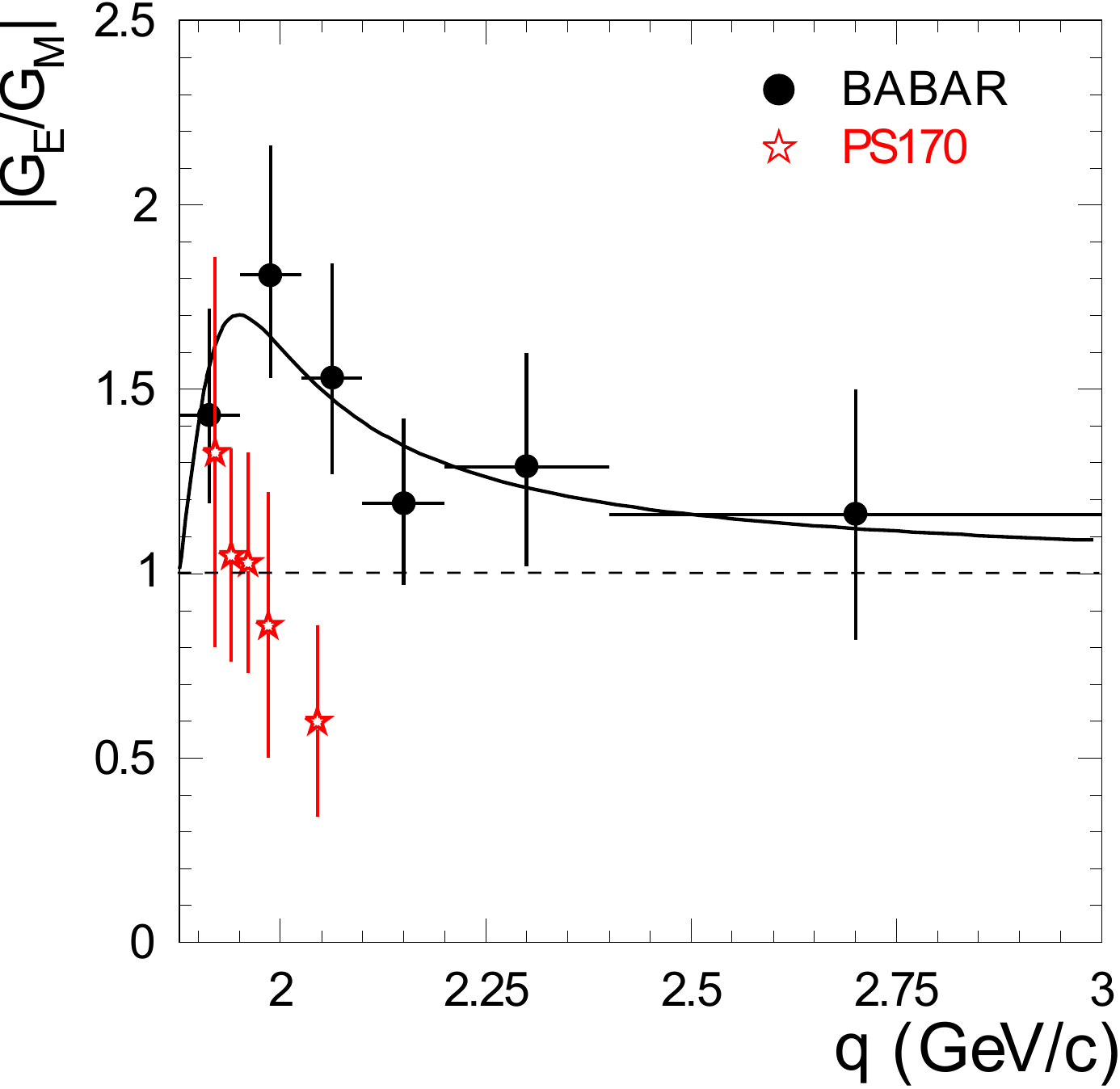}
\caption{Only the PS170~\cite{PS170_94} and the BaBar\cite{BABAR} experiments have been able to measure the ratio of the proton's electric to the magnetic 
form factor. The results are incompatible with each
other. A fit to the BaBar data is shown as well, cf.\ Ref.~\cite{BABAR}.}
\label{fig:GEGM_p}
\end{center}
\end{figure}

\subsection{Experimental results}
\label{expris}
In this Subsect. the experimental results obtained in in the extraction of 
 TL baryon FF's 
 are summarized. 
{ While the main focus of this review is the physics of the TL FF of the 
proton and of the neutron,
we also present the results} obtained by BaBar on hyperon FF's. 
Furthermore, we discuss some phenomenological
aspects related to the FF  measurements. The in-depth discussion of the theoretical
background follows in Sect.~\ref{theo}.

\subsubsection{TL proton electromagnetic form factor}

In Fig.~(\ref{fig:xsect_p_log}) the various measurements  of the  
cross section, performed at $e^+e^-$ colliders,
are summarized, see Tables~(\ref{tab:scan}) and~(\ref{tab:isr}). 
The energy range between the production threshold and $4.5$ GeV has been covered. 
Fig.~(\ref{fig:ff_p_log}) shows the {\it effective} FF for those $e^+e^-$
data sets and includes in addition the results obtained via proton-antiproton annihilation 
at CERN and FNAL, cf.\ Table~(\ref{tab:ppbar}).
We have omitted the older CERN results of
Refs.~\cite{PS170_91,PS170_91b,PS7677}
which are superseded by the high-statistics measurements of experiment 
PS170 (Ref.\cite{PS170_94}).
In general we can identify a good agreement among the different data sets. Nevertheless, the
complex shape of the TL proton FF  is largely not understood and 
has lead to many speculations. We briefly summarize the most relevant observations.

\begin{itemize}

\item
The FF shows a very steep rise toward threshold, which can be clearly identified
in the two data sets of BaBar and PS170. A zoom to the threshold region is shown in 
Fig.~\ref{fig:ff_p_threshold} (linear vertical scale), in which the BaBar spectrum is
plotted in a finer binning to take account of the rapid FF change. Although the
BaBar and PS170 spectra seem to indicate some normalization issues,
the threshold 
enhancement is clearly visible also in this plot. It is interesting to note that a similar behavior 
can be seen also in the $p\bar{p}$ invariant mass spectra of very different physics processes,
e.g.\ B-decays ($B^0 \rightarrow D^{(*)}p\bar{p}$, $B^+ \rightarrow K^+p\bar{p}$) measured
at BELLE~\cite{BELLE_B} and BaBar~\cite{BABAR_B} as well as 
the decay $J/\psi \rightarrow p\bar{p}\gamma$ measured at BES~\cite{BES_Psi}. 
It has been speculated 
whether the threshold enhancement might be due to the existence of a hypothetical, narrow
resonance with a mass just below threshold. Indeed, such a narrow $N\bar{N}$ bound state
could give rise to a dip in the energy dependence of the 
cross section $e^+e^- \rightarrow {\rm hadrons}$ as a result of an interference effect with a 
broad resonance~\cite{franzini}. Several experiments had observed such a dip in the energy spectrum
of the $6\pi$ state~\cite{6pi} and more recently also BaBar has confirmed this observation 
in the $3(\pi^+\pi^-)$  and $2(\pi^+\pi^-\pi^0)$ exclusive states~\cite{babar_6pi}. We will 
discuss the physics of the threshold region in more detail in Subsect.~\ref{thresh}.
 
\item
Fig.~(\ref{fig:ff_p_lin_2plots}) shows the energy dependence of $\left| G_{\rm eff}^N \right|$  
in the energy range from threshold to $3.01$ GeV (upper plot), and 
from $2.58$ GeV to $4.5$ GeV (lower plot). The events from $J/\psi$ and $\psi(2S)$ decays 
to $p\bar{p}$ are subtracted from the contents of the corresponding bins. 
Two rapid
decreases of the FF and cross section near $2.25$ GeV and $3$ GeV
are seen by BaBar. Rosner has pointed out~\cite{rosner} that these
steps are just below the respective thresholds for $p\bar{\Delta}(1232)$ and 
$N(1520)\bar{N}(1520)$ systems, respectively, and suggests an s-wave threshold effect to be responsible
for these structures.

\item
Finally, a comparison of the asymptotic behavior of the TL proton FF at 
high $q^2$ with the corresponding SL FF, measured in $e^-p$ scattering, represents another puzzle.
While perturbative QCD calculations and the application of the 
Phragm\'en-Lindel\"off 
theorem (cf.\ Subsect.~\ref{DR}) predict the asymptotic values for SL and TL FF's 
 to be identical at high energies, 
 the experimental investigation of this issue
 is confronted by the difficulties of disentangling the electric and magnetic
 FF in the TL region. If one assumes that the {\em effective} form factor could
 be an approximation of the TL magnetic form factor (cf. Eq.~(\ref{geffa})), 
 one finds that it is larger
 than the corresponding SL quantities by about a factor of two. 
 In order to perform this comparison, one has to assume that
   the TL magnetic FF is positive in the asymptotic region
 (cf. Subsect.~\ref{tlextre} for a more detailed discussion).

\end{itemize}

In Figs.~(\ref{as})and  ~(\ref{fig:ps170dsdO}) the results of 
BaBar~\cite{BABAR} and PS170~\cite{PS170_94} on the differential cross section
$d\sigma/d\Omega$ had already been presented. 
We remind that according to Eq.~\ref{eq:dsdo} the ratio $\left| G_E^p / G_M^p \right|$ becomes
accessible from the measurement of $d\sigma/d\Omega$.
Fig.~(\ref{fig:GEGM_p}) shows the results obtained by both collaborations on
this ratio. The experiment
PS170 has measured $\left| G_E^p / G_M^p \right|$ in five 
energy bins below $2.1$ GeV; BaBar has measured the ratio in six energy bins below $3$ GeV.
While the spectrum of the PS170 experiment seems to be compatible with the assumption 
$\left| G_E^p / G_M^p \right| =1$, the BaBar spectrum shows a relatively
large deviation from $1$ for intermediate energies.

\begin{figure}[h]
\begin{center}
\includegraphics[width=13.5cm]{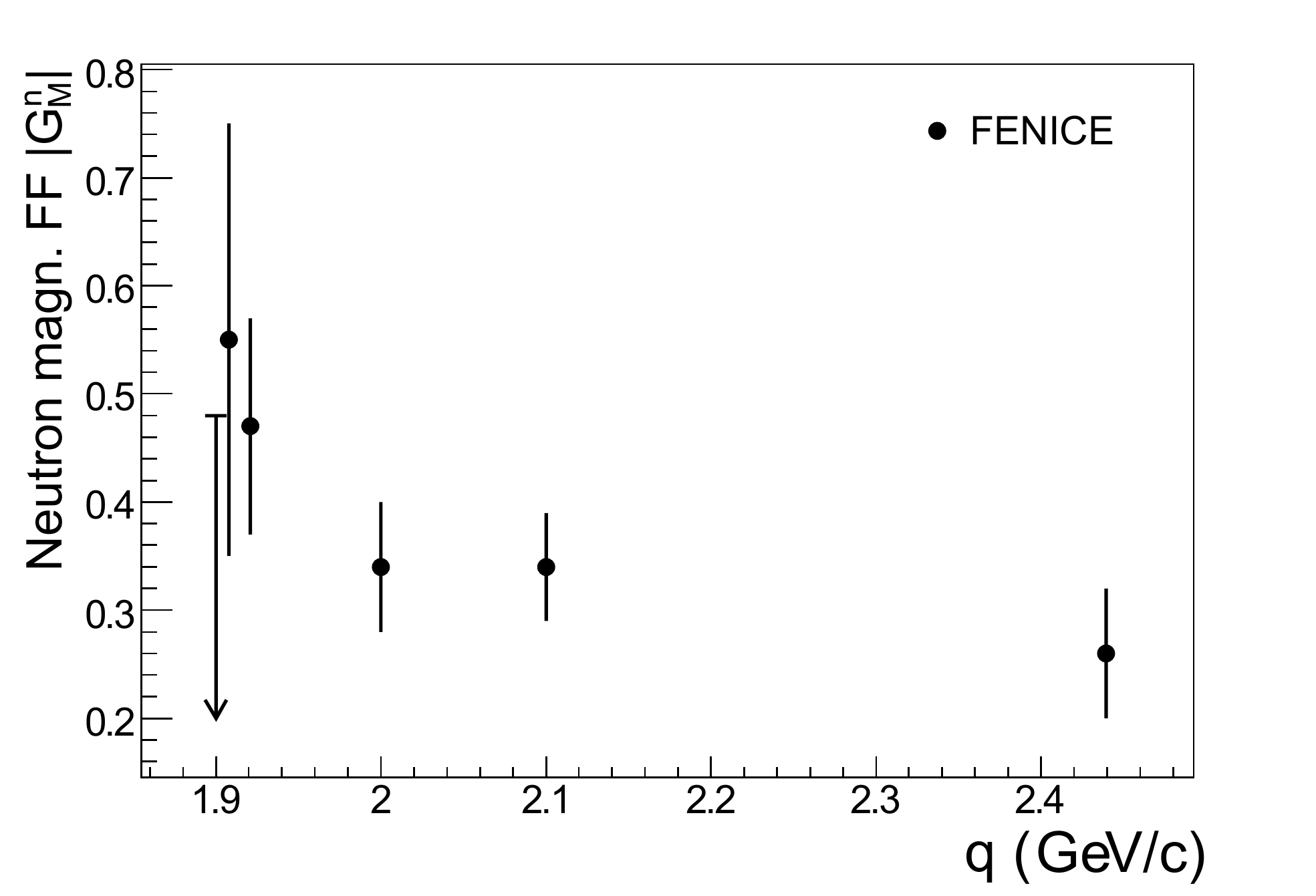}
\caption{Measurement by the FENICE collaboration~\cite{Fenice} of the energy dependence
of the magnetic FF of the neutron $\left|G_M^n\right|$. In this analysis it has been
assumed $\left|G_E^n\right|=0$, which was motivated by the angular distributions of the
neutron-antineutron events.
}
\label{fig:fenice_n}
\end{center}
\end{figure}

\subsubsection{TL neutron electromagnetic form factor}

As mentioned above, only FENICE has measured so far the FF for the reaction 
$e^+e^- \rightarrow n\bar{n}$. In addition to the usual choice of using an 
{\it effective} FF (cf.\ Eq.~(\ref{geffa})), the FENICE collaboration has 
also analyzed the neutron FF by assuming $\left| G_E^n \right|=0$ in the
one-photon cross-section, namely identifying the ratio
$$|G_{\rm eff}^{n} (s)|=\sqrt{\sigma_{e^+e^-\rightarrow n \bar{n}}(s)\over 
 {4\pi\alpha^2 
\beta \over 3s} }$$
with the magnetic FF of the neutron (see also Subsect.~\ref{TLFF}).
This was motivated by the angular distributions measured in the experiment, 
which were 
compatible with
a $(1+{\rm cos}^2\theta)$ shape, only. The data, hence, did not allow for an additional term proportional
to $\rm{sin}^2\theta$, which would have indicated a significant contribution from the electric FF,
see Eq.~(\ref{eq:dsdo}). 
In Fig.~(\ref{fig:fenice_n}), as in Ref.~\cite{Fenice}, the 
data are shown in terms of the previous ratio.
%

Although the process $e^+e^- \rightarrow n\bar{n}$ is measured with very low statistics, the  
FF is found to be 
systematically higher than in the case of the proton. This effect is not understood 
and { needs an experimental confirmation}, especially since naive quark models predict
the TL proton FF to be higher than the neutron one.

\subsubsection{TL hyperon electromagnetic form factors}

\begin{figure}[t]
\begin{center}
\includegraphics[width=9.5cm]{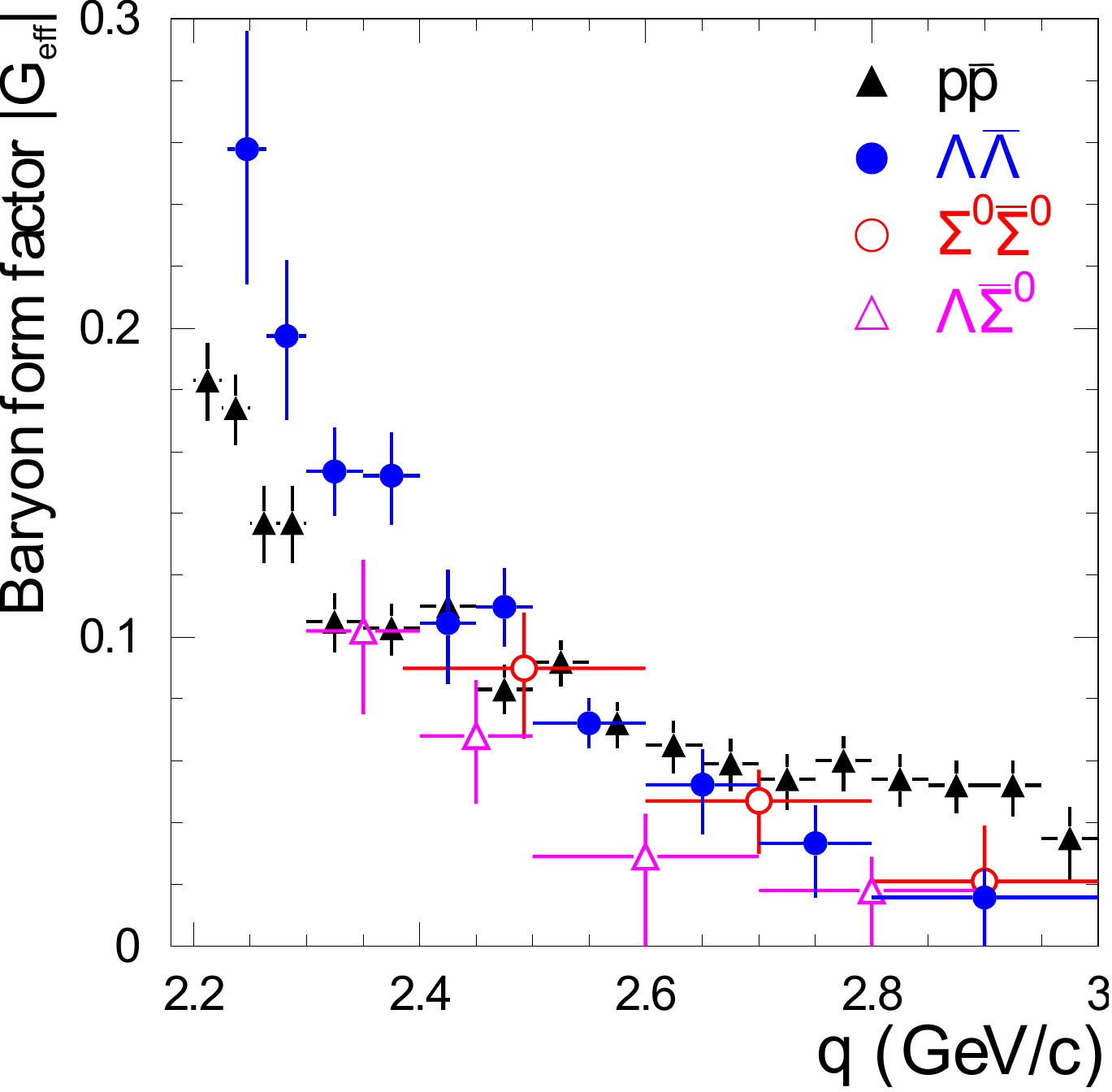}
\caption{The energy dependence of the em FF of several 
hyperon states ($\Lambda \bar{\Lambda}$, $\Lambda \bar{\Sigma}^0$ and $\Sigma^0 \bar{\Sigma}^0$)
measured by BaBar~\cite{BABAR2} is compared to BaBar's own measurement
of the proton form factor~\cite{BABAR}.}
\label{fig:hyperons}
\end{center}
\end{figure}

The BaBar measurements of hyperon {\it effective} FF's are presented in 
Fig.~\ref{fig:hyperons}. The results for the hyperon final states $\Lambda \bar{\Lambda}$,
$\Lambda \bar{\Sigma}^0$ and $\Sigma^0 \bar{\Sigma}^0$ are shown together with the
proton FF. Although, the hyperon FF's are having significantly larger
uncertainties confronted to the proton case, there is clear evidence that all measured
FF's are featuring a strong threshold enhancement. Again, this behavior
is not understood.

A nonzero relative phase $\phi$ between the electric and magnetic $\Lambda$ 
FF's 
$G_E^\Lambda$ and $G_M^\Lambda$ leads to a
polarization of the outgoing baryons with the polarization being proportional to ${\rm sin}\phi$. 
The $\Lambda$ polarization can be measured in the decay $\Lambda \rightarrow p \pi^-$ 
using the correlation between the direction of the $\Lambda$ polarization vector and 
the direction of the proton from the $\Lambda$ decay.
Due to the low statistics collected at BaBar, only a weak limit could be 
extracted for this phase: $-0.76 < {\rm sin}\phi < 0.98$. 
In Ref.~\cite{BABAR2} also the $\left| G_E^\Lambda / G_M^\Lambda\right|$ ratio was investigated
in two mass intervals. Again, the low statistics did only allow for measurements with large
uncertainties. $\left| G_E^\Lambda / G_M^\Lambda\right|$ is found to be $1.73_{-0.57}^{+0.99}$ 
for the mass range $2.23$ GeV $< M_{\Lambda\Lambda}\cdot c^2 < 2.40$ GeV and 
$0.71_{-0.71}^{+0.66}$ for $2.4$ GeV $< M_{\Lambda\Lambda} \cdot c^2 < 2.8$ GeV, where $M_{\Lambda\Lambda}$
is the invariant mass of the $\Lambda \bar{\Lambda}$ system.

\subsection{Experimental outlook}
\label{expoutlook}
At present, the knowledge of the em FF's of the nucleon in the TL regime remains largely 
not understood. This statement appears obvious, given the
various phenomenological puzzles mentioned above. 
The relevance of a better experimental data base for an improved theory prediction
will be discussed in detail in Sect.~\ref{theo}.
While experiments have been able to
measure the shape of a so-called {\it effective} FF of the proton, $\left| G_{\rm eff}^p \right|$, 
with a precision of a few percent at threshold, the uncertainties are much larger at higher momentum
transfer. The neutron FF remains almost {\it terra incognita}.
To receive a significant progress in our understanding of TL nucleon FF's, 
the future experimental programme, in our opinion, should concentrate 
on the following aspects:

\begin {itemize}

\item
Consolidate the existing FF measurements of the proton, and obtain first precision data
for the FF of the neutron (as well as for hyperons);

\item
Obtain statistically significant results for the ratio $\left| G_E^N / G_M^N \right|$;

\item
Measure the relative phase between the electric and magnetic Sachs FF's.

\end{itemize}
Fortunately, in a world-wide effort such an experimental programme is
 currently pursued. 
All experimental techniques presented in Subsect.~\ref{enscan} will be used: 
the $e^+e^-$ energy scan, the ISR technique, and the $p\bar{p}$ annihilation method.

Very recently the VEPP-2000 electron-positron
accelerator in Novosibirsk came into operation with c.m.\ energies of up to $2.0$ GeV
and with a design luminosity of $10^{32}$ cm$^{-2}$s$^{-1}$ at the highest beam energies.
First preliminary results by the 
CMD-3 and SND collaborations concerning TL nucleon FF's have already been presented 
at conferences~\cite{vepp2000}, both for the $p\bar{p}$ as well as
for the $n\bar{n}$ final states. 
These first results indicate that -- once the design 
luminosity will be achieved -- we can expect approximately $10^5$ $p\bar{p}$ and $n\bar{n}$ 
events produced close to threshold and the ratio $\left| G_E^N / G_M^N \right|$ to be 
measured with $10$\% accuracy.  

Furthermore, the second generation B-factory projects in Japan and Italy~\cite{sff}, which will
start operation in several years from now, 
foresee an increase of luminosity of up to two orders of magnitude compared to
{ the first generation. The ISR method will be applied at those facilities and the 
statistical errors for future FF measurements will be reduced by up to an order of magnitude.}

In the following, we will present in more detail two facilities for which a significant progress in the
field can be expected in the near-term future, especially at intermediate and large energies, where an
extraction of the $\left| G_E^N / G_M^N \right|$ ratio is of major importance. 
These are the BES-III experiment at Beijing and the PANDA experiment at the future FAIR facility
in Darmstadt.

\begin{figure}[h]
\begin{center}
\includegraphics[width=9.cm]{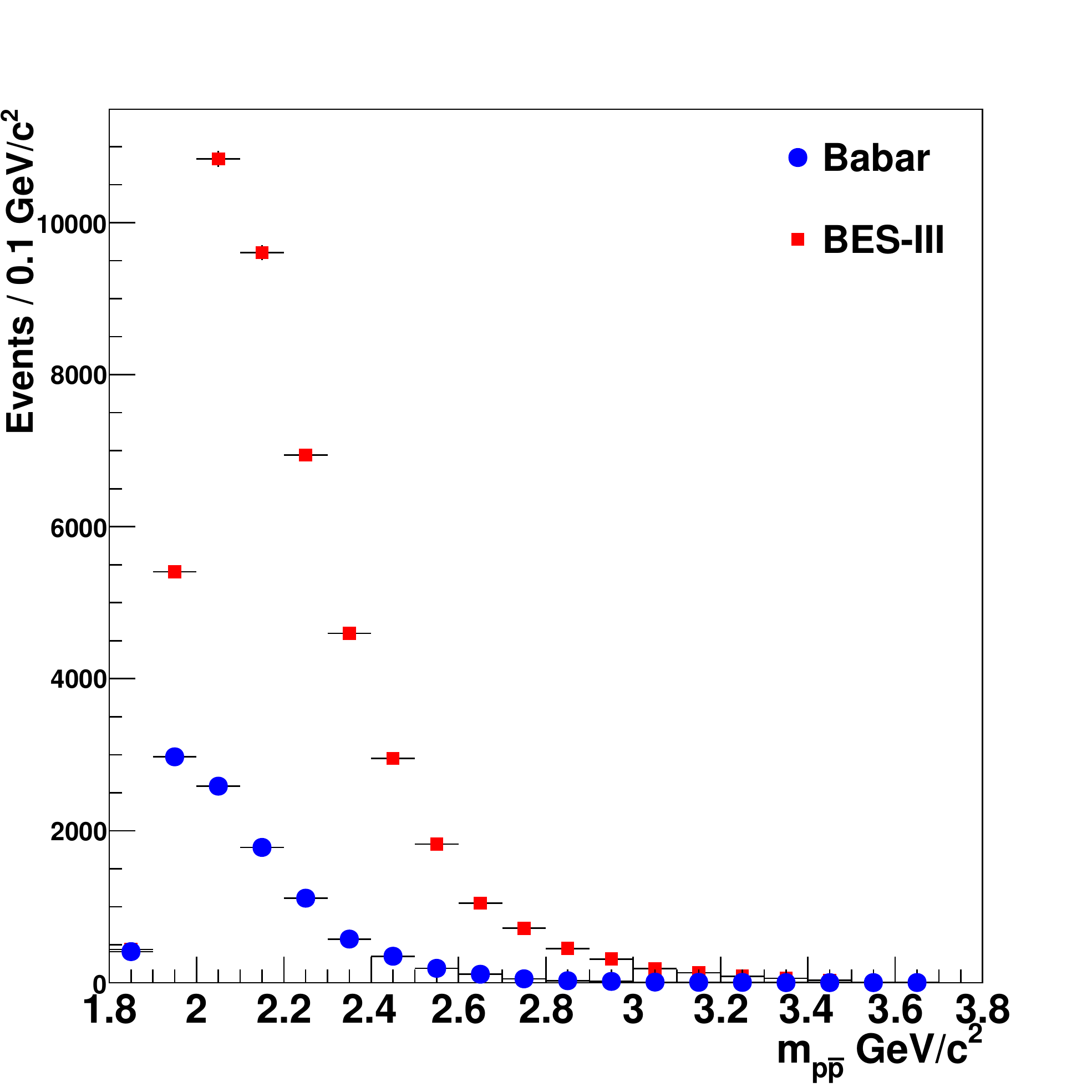}
\includegraphics[width=9.cm]{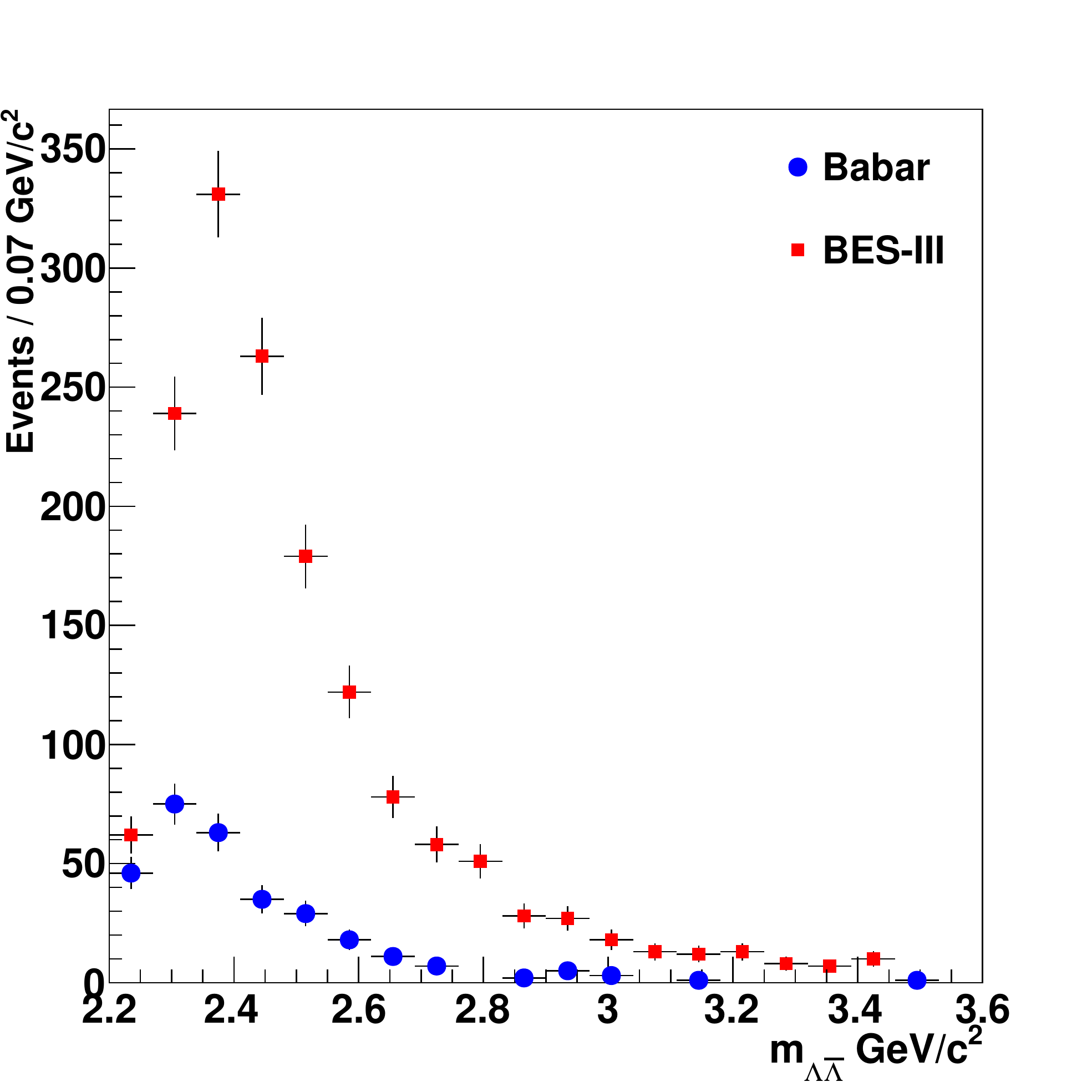}
\caption{The expected yield of $p\bar{p}$ events (left plot) and $\Lambda\bar{\Lambda}$ 
(right plot) using the ISR technique at $\sqrt{s}=3.77$ GeV at BES-III and assuming an integrated
luminosity of $10$ fb$^{-1}$. The number of produced events for the
BaBar kinematics, assuming an integrated luminosity of $232$ fb$^{-1}$, are also
shown, for the sake of comparison.
The PHOKHARA 
generator~\cite{phokhara_ppbar,phokhara_LLbar} has been used for these simulations.}
\label{fig:besvsbabar}
\end{center}
\end{figure}

\subsubsection{BES-III at the $e^+e^-$ collider BEPC-II} 

The BEPC-II $e^+e^-$ collider in Beijing is a next-generation $\tau$-charm factory
with c.m.\ energies corresponding to the mass of the charmonium resonances.
Data taking started
in 2009 with more than $4$~fb$^{-1}$ of integrated luminosity having been collected since then.
The world's largest statistics of $J/\psi$, $\psi^\prime$, $\psi(3770)$ and $\psi(4040)$ decays
is already available at BES-III, see Ref.~\cite{bes3pb} for a detailed summary of the physics programme.
The accelerator is designed to have the highest instantaneous luminosity on the $\psi(3770)$ 
resonance. Peak luminosities of $6 \cdot 10^{32}$cm$^{-2}$s$^{-1}$ have been achieved.
This corresponds to $\sim 60$\% of the design value. For lower and higher energies, the instantaneous
luminosity drops by up to an order of magnitude. Since the BEPC-II accelerator (differently 
from the B-factories) is designed
to operate not only at one single beam energy, an
excellent chance for a new precision measurement of the TL nucleon FF's
over a wide energy range is given.

Such an energy scan between $2.0$ to $4.5$ GeV is indeed scheduled for the coming years. 
In a first campaign, the step sizes will be approximately $100$ MeV wide with a finer
binning eventually at $2.25$ GeV and $3.0$ GeV, where steps in the spectrum of the proton
FF have been found (cf.\ Fig.~(\ref{fig:ff_p_lin_2plots})).
Based on the current
performance of the accelerator, the integrated luminosities per scan point can be estimated to be 
between $\sim 10$ pb$^{-1}$ (at $\sqrt{s}=2.0$ GeV) and up to $100$ pb$^{-1}$ (below and above
the $\psi(3770)$ resonance). 
Compared to the scan performed by the predecessor experiment BES-II (Ref.~\cite{BES05}), 
this implies not only a significant increase of the number of scan points 
as well as of a broadening of the energy range, 
but also an increase in statistics of about two orders of magnitude. According to the
current plannings, $ \sim 2,300$ detected $p\bar{p}$ events at $\sqrt{s}=2.0$ GeV and 
$\sim 500$ events at $\sqrt{s}=3.0$ GeV will be detected. Also concerning
the systematic uncertainties, which used to be below $10$\% in the case of BES-II, a major progress
can be expected, given the improved detector performance. 
The statistical error will however still be dominating the overall precision in most of the energy range.
 
A determination of the em FF of the neutron seems feasible as well
at BES-III. A recent measurement of the branching fraction
$\mathcal{B}(J/\psi \rightarrow n\bar{n})$~\cite{jpsinnbar} with $<10$\% uncertainty 
has demonstrated the sensitivity of the BES-III detector for neutron and antineutron 
detection, which is the major issue in this analysis.

The high integrated luminosities available at BES-III, will also allow for
a programme of ISR measurements of TL nucleon FF's. Notice that the threshold region
for $N\bar{N}$ production will only be accessible via ISR, given the 
fact that for technical reasons an operation of
the collider below $2$ GeV is most likely impossible. Background conditions appear to be
optimal for runs taken on the $\psi(3770)$ resonance, on which $2.9$ fb$^{-1}$ of 
integrated luminosity have already been taken and for which $10$ fb$^{-1}$ are expected
for the coming years. A feasibility study~\cite{cristina} shows, that the expected statistics at BES-III
will be competitive with the existing BaBar measurement. The fact that the c.m.\ energy of the
BEPC-II collider is very close to the hadronic mass range of interest, leads to a radiator
function $H_s(s^\prime)$ (see Eq.~(\ref{eq:hfunct})) which is advantageous with respect to the measurements performed 
at BaBar. This effect cannot be overcompensated by the higher integrated luminosities 
available at the B-factories. Moreover, the fact that the ISR photon at BES-III 
is much less energetic compared to BaBar, opens the possibility for
so-called {\it untagged} measurements, in which the detection
of the ISR photon is not explicitly required. Since the differential cross section for
ISR events increases significantly for low-angle photons, the {\it untagged} approach offers
a very high statistics at a good signal-to-background ratio~\footnote{The feasibility of untagged
measurements was already proven in ISR-analyses of the TL pion FF at the KLOE experiment
at DA$\Phi$NE~\cite{KLOE}.}.

The expected BES-III statistics in terms of {\it produced} events using the ISR technique can be seen in 
Fig.~(\ref{fig:besvsbabar}) for the $p\bar{p}$ channel (left plot) and for the  
$\Lambda \bar{\Lambda}$ channel (right plot). 
An integrated luminosity of $10$ fb$^{-1}$ 
has been assumed in these simulations. For the sake of comparison, the event yield 
for the BaBar case for an integrated luminosity of $232$ fb$^{-1}$ is shown as well. 
Since the geometrical
acceptances and detection efficiencies might be different between BES-III and BaBar, we only compare the
number of {\em produced} events\footnote{
We remind that the fully available BaBar statistics is about a 
factor two larger than the one used in the published results of Refs.~\cite{BABAR,BABAR2} and the one assumed in Fig.~(\ref{fig:besvsbabar}).}. 
In order to facilitate the detection of ISR-produced $n\bar{n}$ events, a dedicated tagging 
detector~\cite{bes_tagger}
has been installed in the BEPC-II beam line at very small polar angles. This tagging device is
intended to detect ISR-photons at very small polar angles. The precision of 
the future $n\bar{n}$ measurement remains however to be investigated.

\begin{figure}[t]
\begin{center}
\includegraphics[width=14.cm]{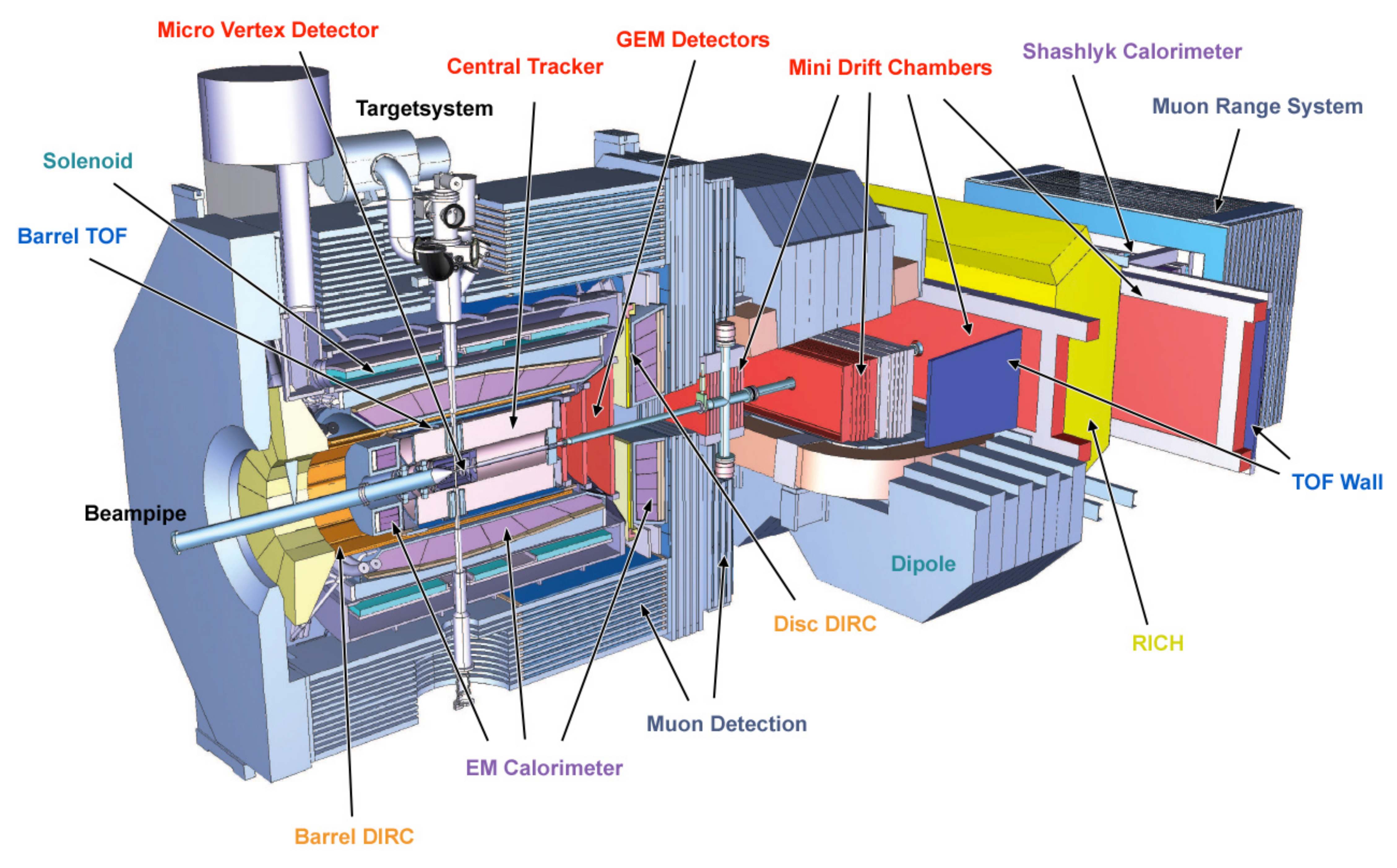}
\caption{A sketch of the PANDA detector.}
\label{fig:panda_det}
\end{center}
\end{figure}

\subsubsection{PANDA at the HESR antiproton beam at FAIR}

The HESR storage ring at FAIR~\cite{fair}, which will start operation in 2018, will provide a high-intensity and high-resolution 
antiproton beam in the multi-GeV energy range. The beam will be scattered on a proton
target (pellet or gas jet) within the PANDA detector, yielding luminosities of $10^{32}$cm$^{-2}$s$^{-1}$
and higher. This is an improvement of an order of magnitude compared to the luminosity, which used to be 
available at FNAL. 
 
The PANDA detector, which is shown in Fig.~(\ref{fig:panda_det}), consists of a barrel as well as a 
forward detector. The barrel detector resembles the geometry of typical collider  
experiments (see e.g.\ Fig.~(\ref{fig:babar_det})) with a solenoidal magnetic field. The forward 
spectrometer reflects the fixed target concept of the experiment and
contains a dipole magnet to allow for a momentum measurement of particles scattered
at small polar angles. 
The ${\rm PbWO}_{\rm 4}$ electromagnetic calorimeter of PANDA, the straw tube
tracking system together with a dedicated DIRC Cerenkov detector setup
{ will be of utmost relevance for a precision measurement} of the $p\bar{p} \to e^+e^-$ process.
With antiproton beam momenta between $1.5$ GeV/c and $15$ GeV/c available at the HESR, 
the em FF of the proton in 
the $q^2$ range between $5$ (GeV/c)$^2$ and $28$ (GeV/c)$^2$ can be accessed~\cite{panda_pb}. 

Detailed feasibility studies have been performed to prepare the measurement of the proton FF,
cf.\ Ref.~\cite{PANDA}. A serious background remains from
two-body processes like $p\bar{p} \rightarrow \pi^+\pi^-$ and $p\bar{p} \rightarrow \pi^0\pi^0$,
which happen to have up to six orders of magnitude higher cross sections compared to signal events. 
A special effort has indeed been made to provide realistic event generators for those
processes (cf.\ Ref.~\cite{PANDA}). 
In the $p\bar{p} \rightarrow \pi^0\pi^0$ background process, Dalitz decays
of neutral pions can be rejected by appropriate angular cuts. The $\pi^+\pi^-$ background channel
requires very good PID capabilities and a good understanding of their experimental efficiencies. 
Kinematic fits help to reject the hadronic two-body background channels as well as other background 
processes. Taking into account all these 
analysis items, and by performing realistic Monte-Carlo based detector simulations, 
it has been found that the process $p\bar{p} \to e^+e^-$ 
can be measured over the full kinematically accessible $q^2$ range with a 
reconstruction efficiency of
$>20$\%. The remaining background-to-signal ratios are expected to be as low as $0.1$\%. 
The measurement of the $e^+e^-$ yield will be
normalized to the integrated luminosity, for which a dedicated monitor system will be used at PANDA.

\begin{figure}[t]
\begin{minipage}[t]{0.485\linewidth}
\centering
\includegraphics[width=\textwidth]{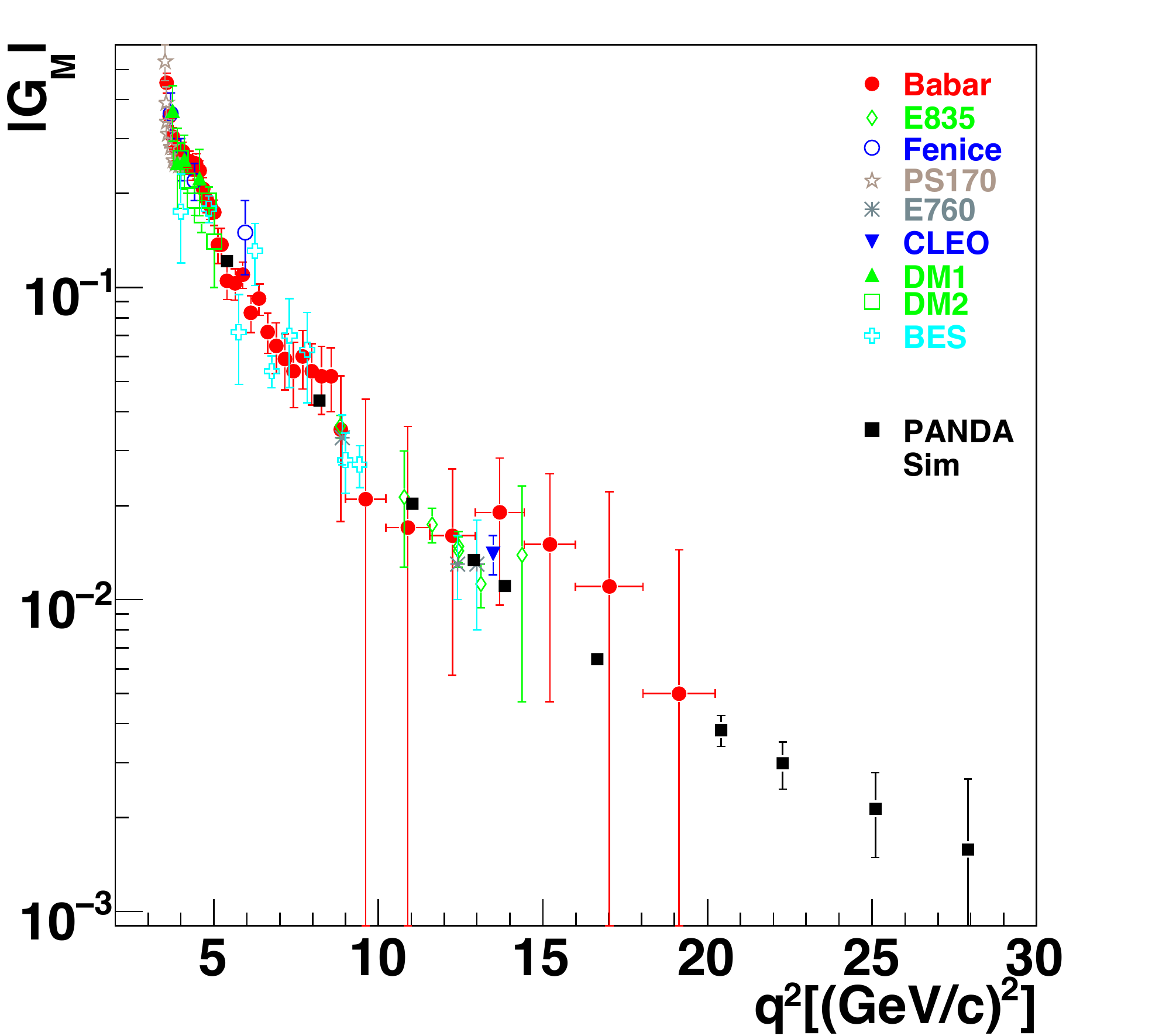}
\caption{Expected PANDA results on the TL magnetic FF of the proton and
a comparison with {\em effective} FF's measured by various experiments 
(Refs.~\cite{Fenice,DM179,DM283,BES05,CLEO05,PS170_94,E760_93,E835_99,E835_03})
In the PANDA simulation it has been assumed: 
$\left| G_E^p \right| = \left| G_M^p \right| = \left| G_{\rm eff}^p \right|$. 
Figure reprinted from Ref.~\cite{PANDA} with kind permission of The European 
Physical Journal (EPJ).}
\label{fig:panda_sim1}
\end{minipage}
\hspace{0.3cm}
\begin{minipage}[t]{0.485\linewidth}
\centering
\includegraphics[width=\textwidth]{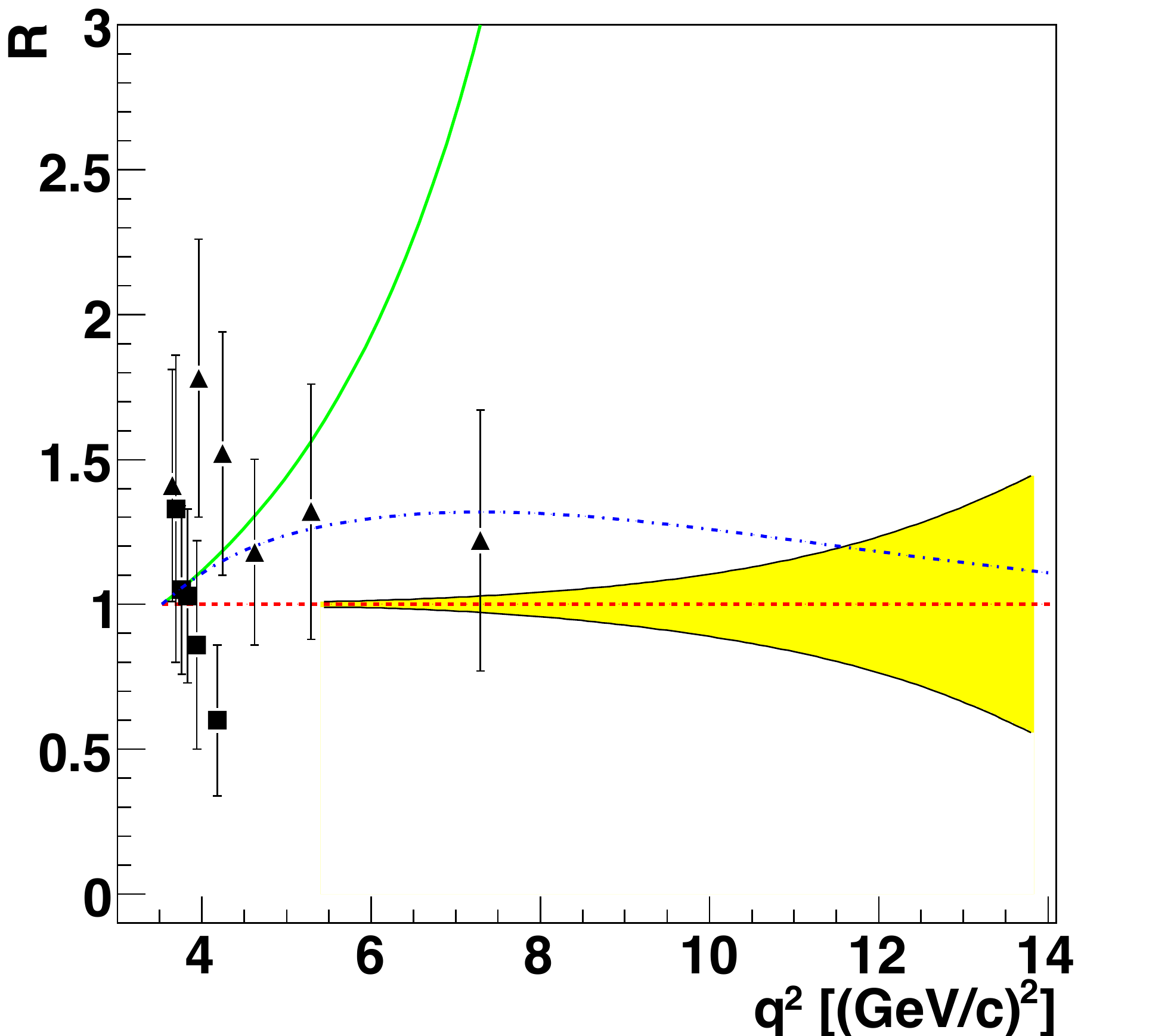}
\caption{Expected accuracy (yellow uncertainty band) at PANDA for the
ratio of the magnetic to the electric FF. The results of the PS170
~\cite{PS170_94} (squares)
and BaBar~\cite{BABAR} (triangles) collaborations are shown, as well as theoretical predictions, see text.
Figure reprinted from Ref.~\cite{PANDA} with kind permission of The European 
Physical Journal (EPJ).}
\label{fig:panda_sim2}
\end{minipage}
\end{figure}

In Figs.~(\ref{fig:panda_sim1}) and~(\ref{fig:panda_sim2}) the results of the 
above-mentioned feasibility studies are
summarized. Fig.~(\ref{fig:panda_sim1}) shows the expected accuracy for the {\it effective} FF
in comparison to already existing data. In the case of PANDA, 
an integrated luminosity of $2$ fb$^{-1}$ collected at
$10$ different $q^2$ values (i.e.\ scan points) was assumed; 
only the statistical error is shown. 
According to Fig.~(\ref{fig:panda_sim1}) we can expect a remarkable improvement with respect to the
so-far most precise measurement from BaBar.
Fig.~(\ref{fig:panda_sim2}) shows the accuracy (yellow error band), which is expected at PANDA for the 
ratio $R=\left| G_E^p / G_M^p \right|$ (for this simulation it was assumed $\left| G_E^p \right| = \left| G_M^p \right|$). 
In the $q^2$ range
between $5$ (GeV/c)$^2$ and $7$ (GeV/c)$^2$ the statistical error of the extracted ratio will be on the level 
{ of a few percent only}.
Above $14$ (GeV/c)$^2$ the extraction of the ratio is not feasible anymore due to the kinematic suppression of
$G_E^p$ (cf.\ Eq.~(\ref{eq:dsdo})). It is obvious from Fig.~(\ref{fig:panda_sim2}) that a precision measurement of the $\left| G_E^p / G_M^p \right|$ ratio at PANDA will be very 
sensitive to distinguish between different hadronic models. Three of those theoretical predictions
are shown in Fig.~(\ref{fig:panda_sim2}). The work corresponding to the 
red dashed line, cf.\ Ref.~\cite{BF} (Brodsky, and Farrar), is predicting $R=1$; the
green solid line and the  blue dot-dashed line are taken from Refs.~\cite{Iac73} (Iachello, Jackson, and Lande) and~\cite{Lomon02} (Lomon),
respectively. Those models will be further discussed in Sect.~\ref{theo}.

The PANDA collaboration has also suggested to measure the TL em FF of the proton in the unphysical region 
($q^2 < 4M_p^2$) by using the reaction $p\bar{p} \rightarrow e^+e^-\pi^0$, cf.\ Refs.~\cite{panda_unphysical}.
Furthermore, it has been demonstrated that TL measurements provide valuable
information to search for two-photon effects, cf.\ Ref.~\cite{panda_tp}. 

As mentioned above and as worked out in Ref.~\cite{Brod04}, future directions in experiment 
should concentrate on an experimental 
programme, which is aiming for a measurement of the relative phase between the electric
and the magnetic FF's. In order to achieve this goal, a polarized target, or -- as an alternative -- a polarized
antiproton beam is required\footnote{In principle this phase can be also measured in the case of 
$e^+e^- \rightarrow p\bar{p}$ annihilation using an unpolarized beam, but measuring the polarization 
of the outgoing proton. No experiment at an $e^+e^-$ collider has however taken this into account.
A polarimeter (e.g.\ made of carbon) would be required for this task.}.  At PANDA, feasibility studies for a polarized hydrogen 
gas target are underway~\cite{feher}, while within the PAX collaboration~\cite{PAX} at FAIR
solutions for antiproton polarization are being investigated. Spin filtering experiments
are carried out in this context { -- so far with protons --} indicating that this method
seems to be a promising 
method (cf.\ Ref.~\cite{spinfilt}) for nucleon polarization. Hopefully this approach can be tested 
soon also with antiprotons.
Double polarization experiments using both a polarized target as well as a polarized antiproton beam
will be giving further insights concerning the relative phases between the magnetic and the
electric FF's, cf.\ Refs.~\cite{Bilenki93,Tomasi05} for further details. 
In Sect.~\ref{TLFF} we will be discussing in detail the opportunities which are given by
the polarization degrees of freedom. \\
The various plannings and future 
directions outlined in this Subsect.\ make us
confident, that the experimental progress expected for the coming years will bring us new and enlightening
insights towards a quantum field theoretical understanding of the physics of 
TL nucleon FFs. { We stress once more that future polarization experiments are a must.}

\newpage
\section{Theoretical background}
\label{theo}
The theoretical investigation of the nucleon em
FF's in the TL
region
cannot be kept separate from the corresponding one in the SL region, given 
the complementary nature of the information on the nucleon internal structure,
that can be accessed through 
 the electron-nucleon elastic scattering and the annihilation processes 
 $N\bar N \leftrightarrow e^+ e^-$ or $e^+ e^-\leftrightarrow N \bar N $, all related via the crossing symmetry.
 Therefore, 
 it appears  useful to   briefly illustrate some generalities on
   the  nucleon 
FF's in
the SL  region. We remind that in a SL scattering process a virtual photon with a four-momentum
$q^\mu\equiv \{ \omega, {\bf q}\}$ hits a nucleon with four-momentum 
$p^\mu\equiv\{E=\sqrt{M^2_N+|{\bf p}|^2}, {\bf p}\}$. The following relation holds for SL processes: 
$q^2=\omega^2-|{\bf q}|^2\leq 0$.

\subsection{ Nucleon Form Factors in the SL region}
\label{SLFF}
\begin{figure}[htbp]

\centering{\includegraphics[width=5.0cm]{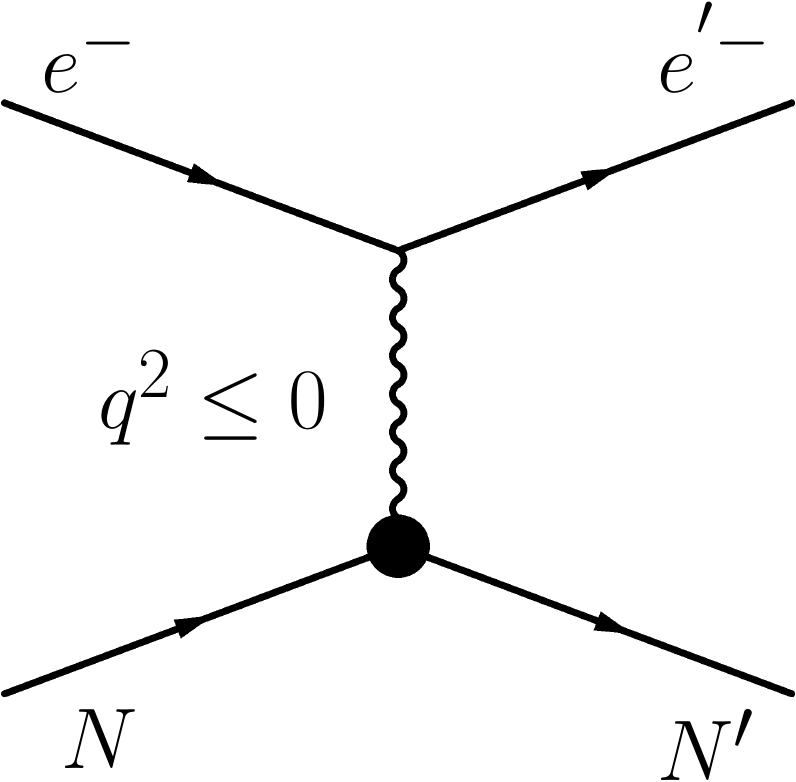}}

\caption{ The one-photon-exchange diagram for 
the electron-nucleon elastic scattering.  The black bubble indicates the
non-elementary nature of the photon-nucleon vertex.}
\label{fig:sldia}
\end{figure}

As it is well-known, in the SL region  the nucleon em FF's can be investigated through
the elastic reaction
$$e^- + N \rightarrow e^{\prime -} + N'$$
pictorially illustrated  by the  diagram in Fig.~(\ref{fig:sldia}), where the 
one-photon-exchange approximation is shown. Within such an approximation, a virtual photon 
($q^2\leq 0$) is exchanged between the incoming electron and the nucleon target,
in a $t$-channel process.  
It is compelling to  
immediately mention the relevant role, which the
polarization degrees of freedom played during  the last decade in the experimental studies of the proton 
FF's
 (see Ref.
\cite{SLrev} for recent reviews). In particular, 
by applying  a novel experimental technique, the so-called polarization-transfer
technique, it has been discovered 
  a totally unexpected 
 feature of the electric FF of the proton: a fall-off faster 
 than the dipole one, for $-q^2 > 2~
(GeV/c)^2$ (see  below).

For $q^2\leq 0$, the nucleon em  current operator is  
a Hermitian  operator, that behaves like a four-vector under
   Lorentz transformations,  and, moreover, fulfills the constraints imposed by 
   parity, time-reversal and  
current conservation. These general properties straightforwardly lead
 to the following covariant
expression  for the matrix elements in the SL region 
(see e.g.~\cite{Yennie},\cite{Hand} and~\cite{BD}).
\be
\sqrt {EE'\over M^2_N}~(2\pi)^3~\langle {\cal N} (p',s')|J_{em}^{\mu}(0)| {\cal N} (p,s)\rangle=\bar{u}_{s'}(p')\left [\gamma^{\mu}F_{1}^N(q^2)+
 i
{F_{2}^N(q^2)\over  2M_N}
\sigma^{\mu\nu}q_{\nu}\right ]u_{s}(p)=\nonu=
\bar{u}_{s'}(p')\left \{-\left({p'^{\mu}+p^{\mu}\over 2M_N}\right) F_{2}^N(q^2)+
\gamma^{\mu}\left[F_{1}^N(q^2)+F_{2}^N(q^2)\right]\right \}u_{s}(p)
\label{mesl}\ee
where 
i) $u_{s}(p)$ and $\bar{u}_{s'}(p')$ are the initial and final nucleon
 spinors (normalized as $\bar u_{s'}(p) u_s(p)=\delta_{s',s}$), and ii)
 $F_{1}^N$ and  $F_{2}^N$ are the Dirac 
and  Pauli form factors, respectively. They are scalar functions that depend upon  one scalar quantity,
 $q^2=2(M^2_N-p\cdot p')$, 
 namely the
  only, non trivial kinematical scalar constructed  through $p$ and $p'$.
The second line in  Eq.~(\ref{mesl}) has been obtained by applying 
the Gordon decomposition. It is
worth noting that, in the first line of  Eq.~(\ref{mesl}), the Dirac structure
 allows a physically
grounded classification of the two terms. As a matter of fact, one has a  helicity conserving 
contribution, that fulfills $[\gamma^0\gamma^\mu,\gamma_5]=0$  and  a helicity
 non-conserving term,that fulfills
$[\gamma^0\sigma^{\mu\nu},\gamma_5]\ne0$, (let us remind that  $(1\pm \gamma_5)/2$ 
is the helicity projector for an ultra-relativistic case).
 
The Dirac  and the Pauli form factors take into account the non-elementary nature 
of the nucleon, and they are normalized according to 
 the proton and  neutron charges for  $F_{1}^N$ and to the
anomalous magnetic moments for  $F_{2}^N$, viz 
\be
\begin{array}{ll}
F_{1}^p(0)= 1 &   F_{1}^n(0)= 0 \\
F_{2}^p(0)=1.792847351 \pm 0.000000028 & F_{2}^n(0)=-1.9130427 \pm 0.0000005
\end{array}\ee  
Notably, in the SL
 region, the  Dirac and Pauli FF's are real functions. This feature can be
  deduced by using  the  Hermiticity of both
 $J_{em}^{\mu}$  and  the Dirac structure 
 ($\gamma_0\gamma^\mu$ and $\gamma_0\sigma^{\mu\nu}$).
 One could  infer the   Hermiticity of  $J_{em}^{\mu}$, by observing that 
  the first-order approximation of the 
 T-matrix is a 
 Hermitian operator
 in the SL region,  since in the $t$-channel no real particles  can be  present in
 the intermediate states. In
 particular, 
   the  
 unitarity condition for the first-order T-matrix, $T^{(1)}$, reads
 \be
 T^{(1)}_{fi}(s,q^2)-\left[T^{(1)}_{if}(s,q^2)\right]^*=~-i~
 \sum_{a} (2 \pi)^4~\delta^4(p_a-q)~
 \left[T^{(1)}_{af}(s,q^2)\right]^*~T^{(1)}_{ai}(s,q^2)=0
 \label{unicon}\ee
 As mentioned above, the vanishing value of the rhs  is due to the unphysical
 nature of  the 
 intermediate states,
  $|a\rangle$. Then, the
 Hermiticity of $ J^\mu_{em}$  follows from the
  Hermiticity of $T^{(1)} \propto J_{em}\cdot j_e/q^2$.
 It is important to anticipate that, in the TL region, 
 the unitarity constraint has to be implemented by 
 taking into account the opening of   real-particle production channels, since 
 a
 $s$-channel reaction occurs. Therefore,  
   the TL nucleon FF's  are expected to be  complex functions.

Instead of using  Dirac and Pauli FF's, it is possible to introduce proper 
combinations of them 
\cite{esw,Hand} as follows  
\be
\label{sachs}
G_{E}^N(Q^2) = F_{1}^N(Q^2) - \tau  F_{2}^N(Q^2) ~,
\quad \quad
G_{M}^N(Q^2) = F_{1}^N(Q^2) +  F_{2}^N(Q^2) ~,
\ee
where $Q^2=-q^2~\geq ~0$, $\tau= Q^2/4M_N^2$. In Eq.~(\ref{sachs}),  $G_{E}^N(Q^2)$ and 
$G_{M}^N(Q^2)$ are the so-called Sachs electric and magnetic  form 
factors, respectively. The physical motivation of considering 
these combinations
 is given by  the analysis of the current operator performed 
 in the Breit frame, 
 where $Q^2
 =|{\bf q}|^2$ and ${\bf p}=-{\bf q}/2=-{\bf p}'$. In this frame, the 
  nucleon FF's can be expressed in terms of  
 three-dimensional Fourier transforms of functions to be interpreted
 as electric and magnetic densities, like  in the non relativistic framework, viz
 \be
\rho_{E (M)}^N(|{\bf r}|) = \int {d {\bf q}\over (2\pi)^3} 
e^{-i {\bf q}\cdot {\bf r}}
{ M_N\over E(|{\bf q}|)} G_{E (M)}^N(|{\bf q}|^2).
\label{density}
\ee
where $E(|{\bf q}|) = \sqrt{M_N^2 + |{\bf q}|^2/4}$.
If $E(|{\bf q}| ) \rightarrow M_N$, the non-relativistic limit is immediately recovered.
  Therefore, one could naively argue that the SL nucleon FF's
 yield information on  intrinsic properties of the nucleon. Unfortunately, such
 a correlation should be carefully reconsidered, taking properly
 into account the
 boosts to be applied to the  states of a composite system, like the nucleon
  (see, e.g.,~\cite{Kelly02,Krutov09}). Notice that, in the Breit frame, both initial 
  and final states have to be boosted when the matrix elements of the current
  are calculated.

Recently,  a new physical interpretation of 
 the nucleon FF's has emerged (see, e.g.~\cite{Miller07}). It is  based on
 the   field theoretical
 description of the constituents, inside the nucleon, given  in terms of
 the so-called Generalized Parton Distributions (GPD's)
 (see, e.g.,~\cite{Diehl03} for an extended review of the topic). 
 Such a formalism, developed within the Deep Inelastic Virtual Compton 
 framework, leads to a description 
 that fully satisfies the general principles,   allowing to correctly treat
   the relativistic-boost effects, as well. It should be 
  pointed out  that GPD's depend upon  
   three scalar invariants: $x= k\cdot n/([p+p')\cdot n]$, $\xi= (p'-p)\cdot n/[(p+p')\cdot n]$ and $\Delta^2=(p'-p)^2$,
   with  i) $k$ the average four momentum of 
   the active constituent, before and after interacting with an external probe, and 
   ii) $n$ a light-like four-vector. Furthermore, it turns out that, in the 
   Deep Inelastic Scattering (DIS) limit 
   where the variable $\Delta$ plays an analogous role of $q$, one can reobtain   the nucleon FF's from the unpolarized 
   quark GPD's, putting
  $\xi=0$ and integrating on $x$. To complete the analysis,  
  two-dimensional Fourier transforms of the relevant GPD's in the Drell-Yan frame (a particular Breit
  frame where the $q^0+q_z=0$) are introduced. In this frame, the  variable $\Delta^\mu$ has only transverse components,
  i.e. ${\mbf \Delta}_\perp={\bf q}_\perp$,
  (notice that in the Drell-Yan frame $Q^2=|{\bf q}_\perp|^2$) and it is conjugated to the impact parameter ${\bf b}_\perp$. 
  In conclusion, the main outcome   is
 given by the determination of   well-defined density distributions of the
 constituents in the transverse plane $\{b_x,b_y\}$. For instance,   the Dirac FF yields 
   \be
   \rho_{1}^N(|{\bf b}_\perp|)= \int {d{\mbf \Delta}_\perp \over (2 \pi)^2}~ 
  e^{{i\bf b}_\perp \cdot {\mbf \Delta}_\perp}~ \int_{-1}^1 dx~H^N(x,0,{\mbf \Delta}_\perp)= 
  \nonu =\int {d{\bf q}_\perp \over (2 \pi)^2}~  e^{{i\bf b}_\perp \cdot {\bf
  q}_\perp}~F_{1}^N(Q^2=|{\bf q}_\perp|^2)
   \label{rhogpd}\ee
 where $H^N(x,0,{\mbf \Delta}_\perp)$ is one of the unpolarized GPD's.
 Summarizing, $F_{1}^N$ can provide information on the  density 
 distributions in a transverse plane, once the effect of the boosts on the initial and final states of
 a composite system are correctly taken into account. This can be implemented
 by using the light-cone  description of both 
 the operator  current and the nucleon states 
 (see, e.g.,~\cite{Diehl03}).

The last issue to be mentioned,  that will be relevant also in the TL region when a model builder 
has to
constrain his own approach, is the asymptotic behavior of the nucleon FF's, 
as dictated by
QCD.
By considering both dimensional scaling 
laws and helicity conservation, perturbative QCD (pQCD) yields the  following asymptotic
behavior for 
the nucleon FF's (see, e.g.,~\cite{BF,BL})
\be
F_{i}^N(Q^2)\to \left[{1 \over Q^2}\right]^{i+1}~\left[ln\left({ Q^2 \over 
\Lambda^2_{QCD}}\right)\right]^{-\gamma}\quad \quad \quad i=1,2
\label{pqcd}\ee  
 where $\gamma$ is the anomalous dimension ($\gamma = 2.148, ~2.160,~ 2.173$, depending upon
 the active number of flavors $N_f=3,4,5$). It should be pointed that 
the asymptotic  behavior 
 of the FF's  is dictated by the perturbative gluon exchange, that provides the mechanism 
for sharing the momentum transfer among the constituents. In order to accomplish
this, at least two massless gluons must be
 exchanged, and each gluon propagator contributes with a factor $1/Q^2$. As to
 $F_{2}^N$, the spin-flip effect adds an extra factor $1/Q^2$.

To complete this snapshot of the SL region, it is impossible 
to avoid few words  on the 
  experimental technique that has given a new, strong 
  impetus to the investigation 
of the proton  FF's in this region. In particular, a technique 
 based on the measurement of the polarization of the final state
proton has experimentally proven that  the proton charge FF strikingly
 decreases faster   than the dipole  FF
 ~\cite{SLrev}. It is also worth
 noting that in the TL region, since the nucleon  FF's become complex functions,
   a $N \bar N$ pair, produced in the final state,  naturally acquires 
    polarization degrees of freedom,
even in absence of  polarized leptons in
 the initial state (see Subsect.~\ref{TLFF} and~\cite{Brod04}).
 
In the SL region, when a polarized  electron hits  an unpolarized nucleon 
target,  
 some polarization is transferred to the recoiling nucleon. Such a polarization can be 
 measured
with a suitable secondary analyzing reaction.
 The
 polarization-transfer method (see~\cite{SLrev} for more details) is  based on the measurement of the transverse and
 longitudinal components of  the  recoil proton polarization, in the elastic
 reaction
 \be
 \vec {e^-} +p \to e^{\prime -} +\vec {p^\prime} ~~,\label{polre}\ee
 where the incident electron has a longitudinal polarization $\pm h$ along the
 direction of the momentum transfer, $\hat q\equiv \hat z$ (indeed one could
 consider reactions with  more complicate polarization set-up~\cite{Akhiez58})).
 To understand why the final state proton gets a polarization, one should first
 recall that the polarization of a fermion (with four-momentum 
 $p$ and
 spin $s$) with respect to a given direction,
 $\hat {\bf n}$, is obtained from the following expectation value:
 \be
 \langle \mbf \sigma \cdot \hat {\bf n}\rangle ={1\over u^\dagger(p,s) u(p,s)}
  u^\dagger(p,s) \left(\begin{array} { c
 c}\mbf \sigma \cdot \hat {\bf n} & 0 \\ 0 & \mbf \sigma \cdot \hat {\bf n}
 \end{array}\right)
 u(p,s) ={1 \over \bar u(p,s) \gamma^0 u(p,s)}~
 \bar u(p,s) \gamma_5~\psla {\bf n} u(p,s)= \nonu =
 {M_N \over E} ~Tr \left({\psla p +M_N
 \over 2 M_N}~ {1 +\gamma_5 \psla s \over 2}~\gamma_5~\psla {\bf n} \right)
 ={M_N \over E}~{\bf s} \cdot{\bf n},\label{polar1}\ee
 where $n^\mu\equiv \{ 0,\hat {\bf n} \}$.
 In order to determine the polarization of the final proton,   Eq.~(\ref{polar1})
 has to be properly weighted  by using the polarization density of the final
 state, i.e. by exploiting  the transition probability to scatter 
 the final proton
 along a given direction. To this end, one has to use the density matrix of 
 the final nucleon, 
 formally written as $\rho_f= {\cal M} \rho_i {\cal M}^\dagger$, where
 the matrix  ${\cal M}$ is the T-matrix
   relevant for the process under scrutiny\footnote{In a non relativistic 
   framework, one 
 parametrizes  
  ${\cal M} = g-i {\bf h} \cdot \mbf \sigma$, with  $g$ and 
  $ {\bf h}$  complex functions, and   $\rho_f=
   (d\sigma/d\Omega)~[1 +{\bf P}\cdot \mbf \sigma]/2 $.} (see, e.g.,~\cite{Akhiez58} for a 
   discussion in
  one-photon approximation within a relativistic framework). It is worth
  noting that the density matrix of the final nucleon can lead to 
  a polarization in different cases:  i) in case 
  the initial nucleon has a polarization,  ii) if the incoming beam has a
  polarization, or iii) a complex-valued FF's of the nucleon 
  are present in
  the current (see~\cite{GW04} for the  non
 relativistic case, and~\cite{Brod04} for the relativistic one). 
 In conclusion, the above mentioned weighting of Eq.~(\ref{polar1}) amounts to
 obtain the polarization  of the final proton  by  
 evaluating   the 
 polarized differential cross section, where an average 
   on the
 spin of the initial proton  is performed (this means  an initial density matrix 
 purely scalar), and  the polarization of the incoming
 beam is taken into account. 
 Then, the experimental polarization, ${\bf P}\equiv\{P_x,P_y,P_z\}$, of  the recoiling proton 
 is given by (see,
 e.g.~\cite{rekalo02}  and~\cite{GW04})
 \be
 {\bf P}\cdot\hat {\bf n} ~{d\sigma \over d\Omega}= 
 K_f~ L_{\mu \nu}  {\cal W}^{\mu \nu}_p
 \ee
 where ${d\sigma / d\Omega}$ is the unpolarized cross section (representing
 the normalization of the polarization density matrix in the final state), $K_f$  a proper
 factor, $L_{\mu \nu}$ the leptonic tensor with both symmetric and  antisymmetric
 terms (since the electron beam is polarized in~(\ref{polre})), and   the proton tensor is given
 by
 $${\cal  W}^{\mu \nu}_p=Tr \left({\psla p_f +M_p
 \over 2 M_p}~ {1 +\gamma_5 \psla s_f \over 2}~J^\mu_p 
 {\psla p_i +M_p
 \over 2 M_p}~J^\nu_p~\gamma_5 \psla {\bf n} \right)$$
  It turns out that the proton  in the final state of the reaction~(\ref{polre}) gains 
  a polarization in
 the scattering plane, $(x,z)$, taking $ \hat q\equiv \hat z$. 
 The measured components of the
 polarization vector allows one to extract the ratio of the proton FF's
  as follows  
 \be
 {G_{E}^p \over G_{M}^p }= -{P_x \over P_z} {E_i+E_f \over 2 M_p}~tan
 \left({\theta_e \over 2}\right)
 \ee
 \begin{figure}[tbh]
\begin{center}

\vspace{-4cm}
\includegraphics[width=7.5cm]{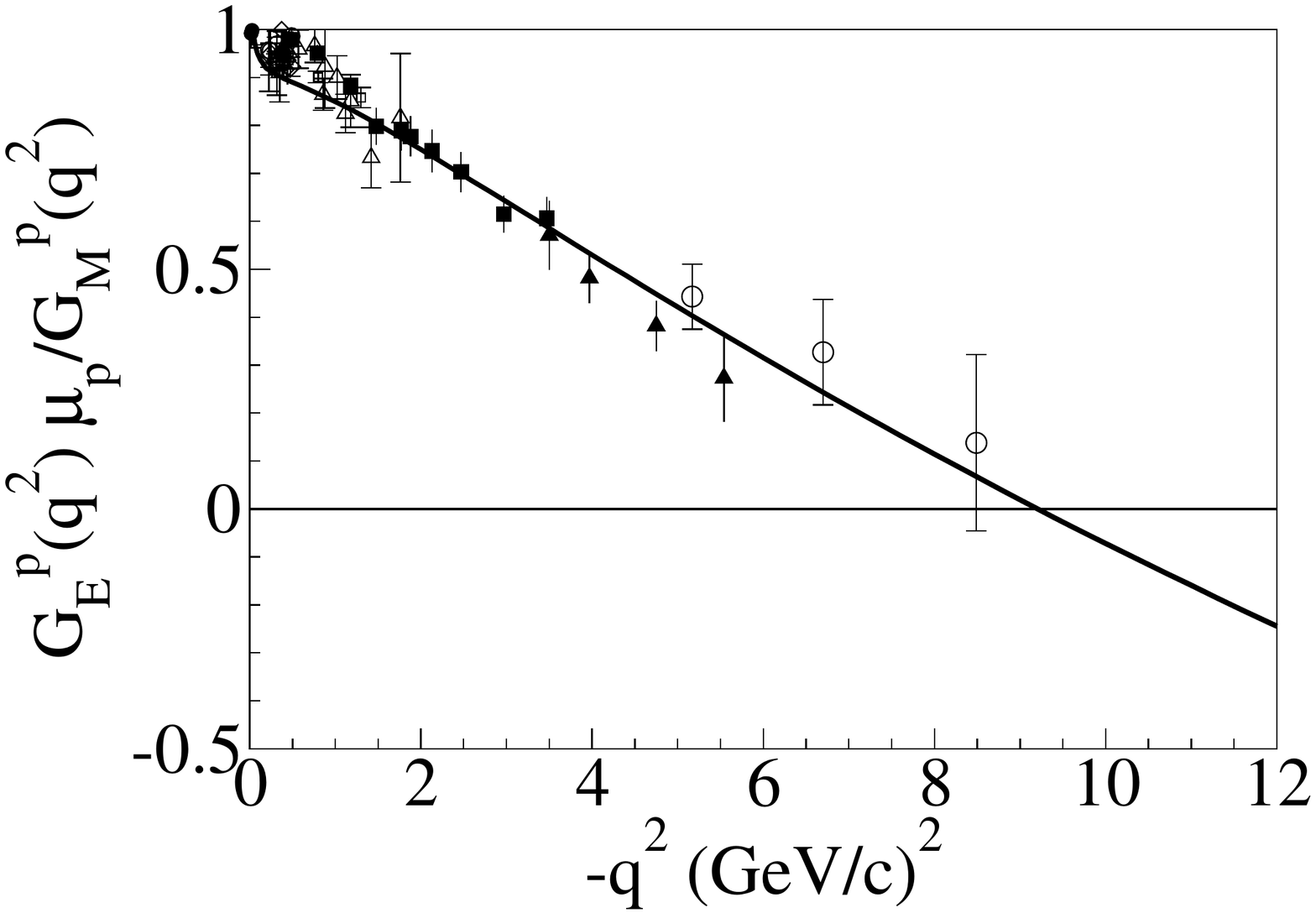} $~~~$
\includegraphics[width=7.5cm]{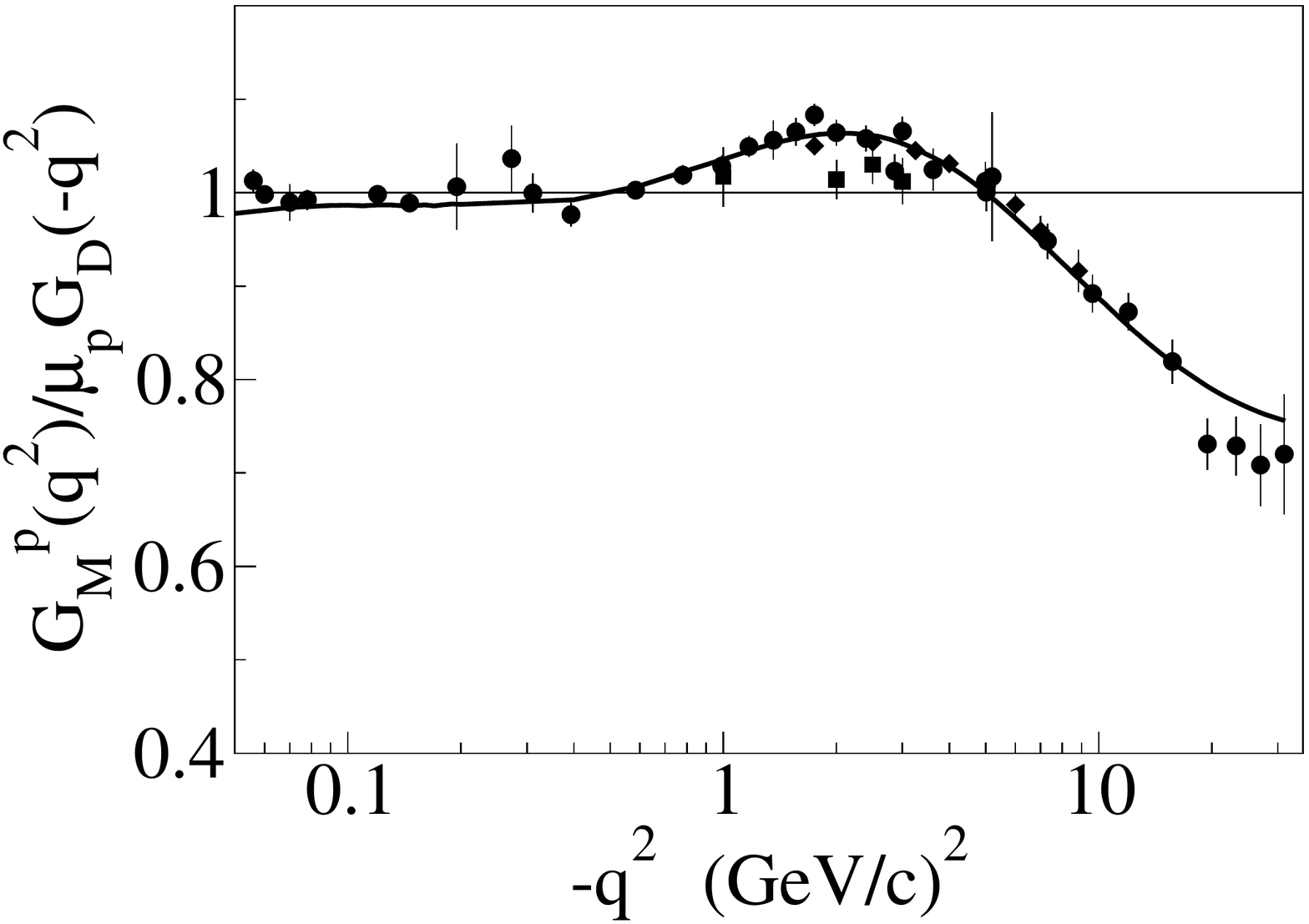}

\vspace{-3.3cm}
\includegraphics[width=7.5cm]{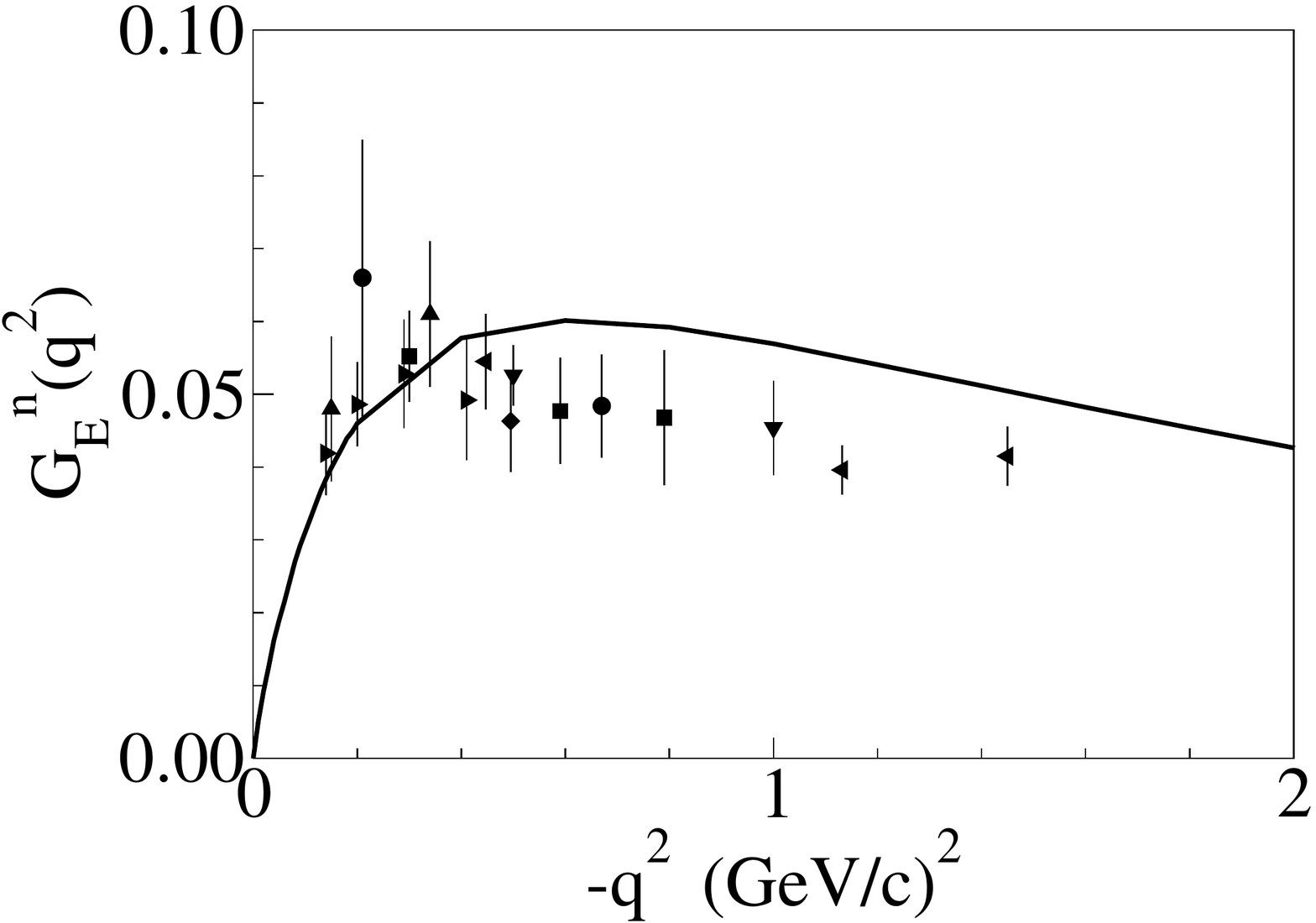}$~~~$
\includegraphics[width=7.5cm]{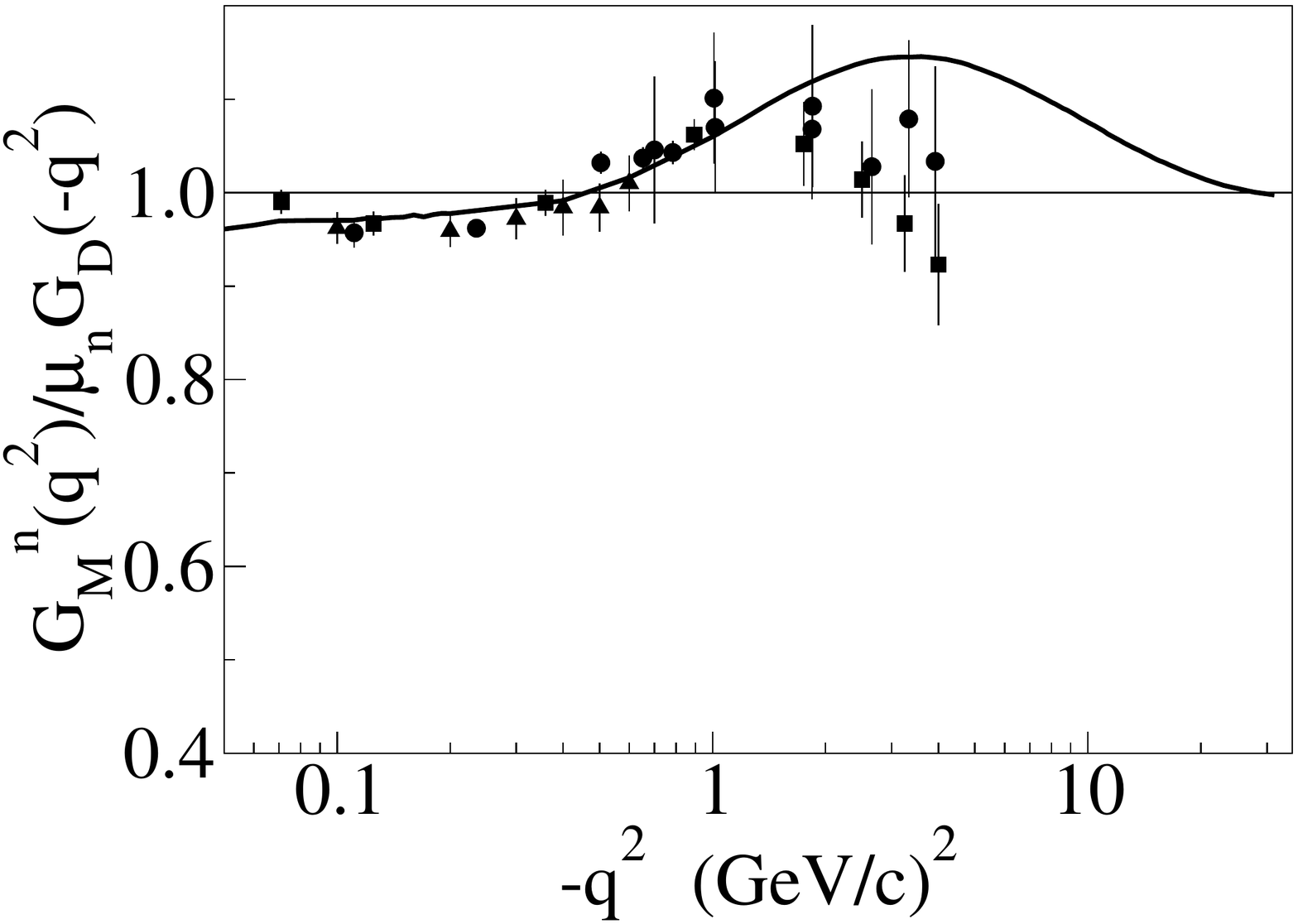}
\caption{ Space-like nucleon form factors vs. $Q^2=-q^2$. Left panel:   
the proton form
factors. Right panel: the neutron form factors. 
Data from the compilations in~\cite{SLrev}. The dipole form factor
$G_D(|q^2|)= 
[1 +| q^2|/(0.71 (GeV/c)^2)]^{-2}$ yields the reference line of the FF's, when
indicated. The solid lines are the calculations, with four free parameters, performed within the
relativistic Constituent Quark Model of Ref.~\cite{Demelo}.
(Adapted from Ref.~\cite{Demelo}).}
\label{fig:slffp}
\end{center}
\end{figure}
 Concluding this Subsection, for the sake of illustration, 
    an overview of 
 the  status of the analysis of the
 nucleon FF's in the SL region is presented in Fig.~(\ref{fig:slffp}). 
 It is also shown  a  recent theoretical 
 calculation
~\cite{Demelo}, based on a relativistic constituent quark (CQ) approach, 
 that has
 been extended to the TL region, as well (see Subsect.\ref{RCQM}). 
 The interested reader can be
 referred to, e.g., Ref.~\cite{SLrev} for a comprehensive and updated discussion 
 of the experimental measurements.

\subsection{Nucleon form factors in the 
TL region}
\label{TLFF}
\begin{figure}[ht]

\centering{\includegraphics[width=5.0cm]{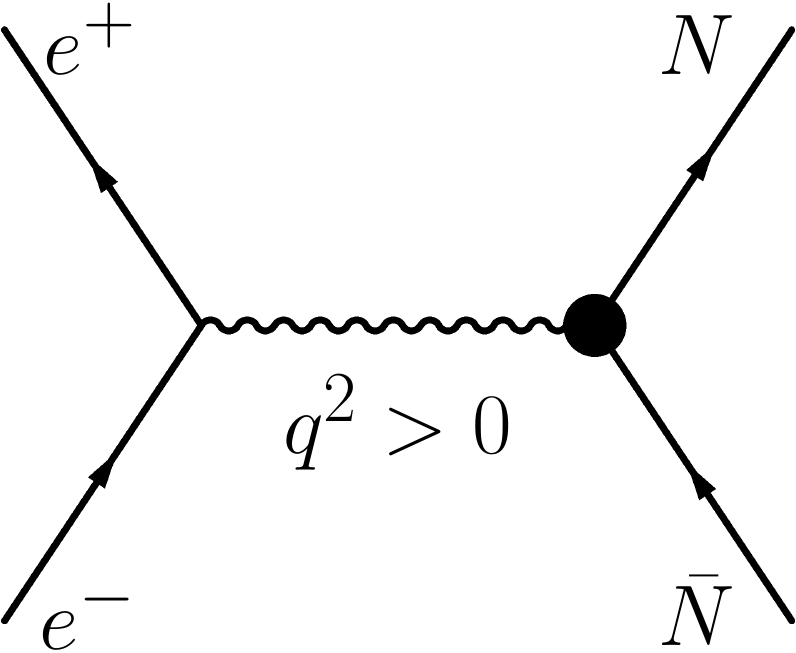}}

\caption{ The one-photon-exchange diagram for 
the electron-positron annihilation into a $N\bar N$ pair. Notice that a physical $ N\bar N$ pair is produced only for $q^2\geq 4
M^2_N$. As in the SL diagram, shown in Fig.~(\ref{fig:sldia}), the black bubble indicates the
non-elementary nature of the photon-nucleon vertex.}
\label{TL_diag}

\end{figure}
The  reaction where a $N \bar N$ pair annihilates into
 charged leptons, viz  \be
  N + \bar{N} \rightarrow \ell^+ +  \ell^-
\label{tlreac}
\ee
and the time-reversed process, viz
\be
 \ell^+ +  \ell^-\rightarrow N + \bar{N}
\label{tlreac1}
\ee
 represent a direct source of  information on the nucleon FF's in the TL region. For the
 sake of concreteness in what follows it will be considered the $e^+ e^-$ 
 annihilation.  
In Fig.~(\ref{TL_diag}), the one-photon approximation of the reaction 
(\ref{tlreac1}) is depicted. 

In the hadronic annihilation process,
a virtual photon with invariant square mass $q^2>0$  can materialize into 
 a quark-antiquark pair, since it can become active the hadronic
 part of the virtual-photon state, shortly given by 
 \be |\gamma^*\rangle=
 | point-like\rangle+\underbrace{|e^+ e^-\rangle +|\mu^- \mu^+ \rangle +....}  
+ ~\underbrace{|q \bar q 
\rangle +|q \bar q g\rangle+|q \bar q gg\rangle+....}\nonu
 \hspace{4.5cm} leptonic ~term 
 \hspace{1.5cm}hadronic~ term
 \label{phwf}\ee
An honored, and still very alive, approximation of the photon vertex is the {\em
vector dominance model} (VMD) (see, e.g., ~\cite{VMD} for a recent review), that describes the hadronic part 
of the photon wave-function through a
direct coupling between the virtual photon and a vector meson (with the proper mass, spin and parity).
 Such a model certainly allows one to construct a
very effective scheme of approximation, that has to be completed by a description of the
hadronic decay
of the vector meson, if one is interested in the hadronic FF's (i.e. $\gamma^*\to~ VM ~\to  h\bar h$). 
In view of the quark content of the hadronic term, one should also
consider a direct coupling to a $q\bar q$ pair, and therefore  more complicate production
channels (as it will discussed in what follows). With these  first physical insights in mind, we 
can start to deal with the analysis of the matrix elements of the nucleon current for  describing reaction
(\ref{tlreac1}).

The investigation of the process ~(\ref{tlreac1}) (as well as~(\ref{tlreac})) takes advantage of  the  crossing symmetry, and
therefore one  
 can exploit the previous analysis of the
elastic scattering in the SL region. 
In particular, the matrix elements of the  nucleon current operator
involved in the reactions   ~(\ref{tlreac1})  are written as
follows
\be
\sqrt {EE'\over M^2_N}~(2\pi)^3~\langle {\cal N} (p,s)\bar{\cal N} (p',s')|
J_{em}^{\mu}(0)| 0\rangle=
\bar{u}_{s}(p)\left [\gamma^{\mu}F_{1}^N(q^2)+
 i{F_{2}^N(q^2)\over  2M_N}
\sigma^{\mu\nu}(p'+p)_\nu\right ]v_{s'}(p')=\nonu=
\bar{u}_{s}(p)\left [-\left({p'^{\mu}-p^{\mu}\over 2M_N}\right) F_{2}^N(q^2)+
\gamma^{\mu}(F_{1}^N(q^2)+F_{2}^N(q^2))\right ]v_{s'}(p')
\label{metl}\ee
where  $q^2=(p'+p)^2=4E^2$,
$p^\mu\equiv \{E,{\bf p}\}$ and  $p^{\prime \mu}\equiv \{E,-{\bf p}\}$, in the center of mass. 
In the TL region, the current operator  is involved in the   transition from the 
vacuum to a state
with a nucleon-antinucleon pair, that becomes real beyond   the
proper threshold, given by  $q^0_{thr}= 2M_N$. This means that  production channels are  now 
 opened, and
 therefore the current operator is not more a Hermitian one. This can be
 verified from the unitarity condition, Eq.~(\ref{unicon}), applied to this 
 case, namely to an $s$-channel process.
 As a consequence, as it can be also recognized from a perturbative 
  analysis of the FF's themselves,  the nucleon FF's become complex functions, with a characteristic 
  multi-cut structure due to the free propagation of the
  relevant hadronic pair, beyond the proper production threshold.
Indeed, if we expand the analysis to include  the whole hadronic production,  
one can recognize that different channels can be opened, with different
energy thresholds.  In the interval $0< q^2 < 4m^2_\pi$, there is no purely hadron production 
 at all, while for increasing values of $q^2$, up to $q^2 < 4M^2_N$, one meets channels that contributes to the 
 production of a virtual $N \bar N$ pair. One  has in order : i) the isovector two-pion 
 channel, that  starts at 
 $q^2\geq  4m^2_\pi$, where $m_\pi$ is the pion mass, ii)
  the isoscalar three-pion  channel, that   begins at 
  $q^2\geq  9m^2_\pi$, iii) the $\phi$ meson production,  beyond $1~{\rm GeV^2}$, etc.
   The physical threshold for the $N\bar N$ is
 $q^2\geq 4 M^2_N$ and 
  the region $4m^2_\pi\leq q^2< 4 M^2_N$ is called the unphysical
 region, when we are dealing with the reaction~(\ref{tlreac1}): it contains many interesting
 information, particularly near the $N\bar N$ threshold (see Subsect.~\ref{thresh}).
  It should be pointed out that 
 the  opening of more production channels beyond the $N\bar N$
 threshold generates new overlapping  cuts in the FF's. As we will see below (Subsect.~\ref{DR}),
  one of the main issue in the
 general treatment of the TL FF's is precisely the
 evaluation of the discontinuity of the FF's across the cuts, namely  the imaginary
 part of the FF's. 
\begin{figure}[htbp]

\centering{\includegraphics[width=7.0cm]{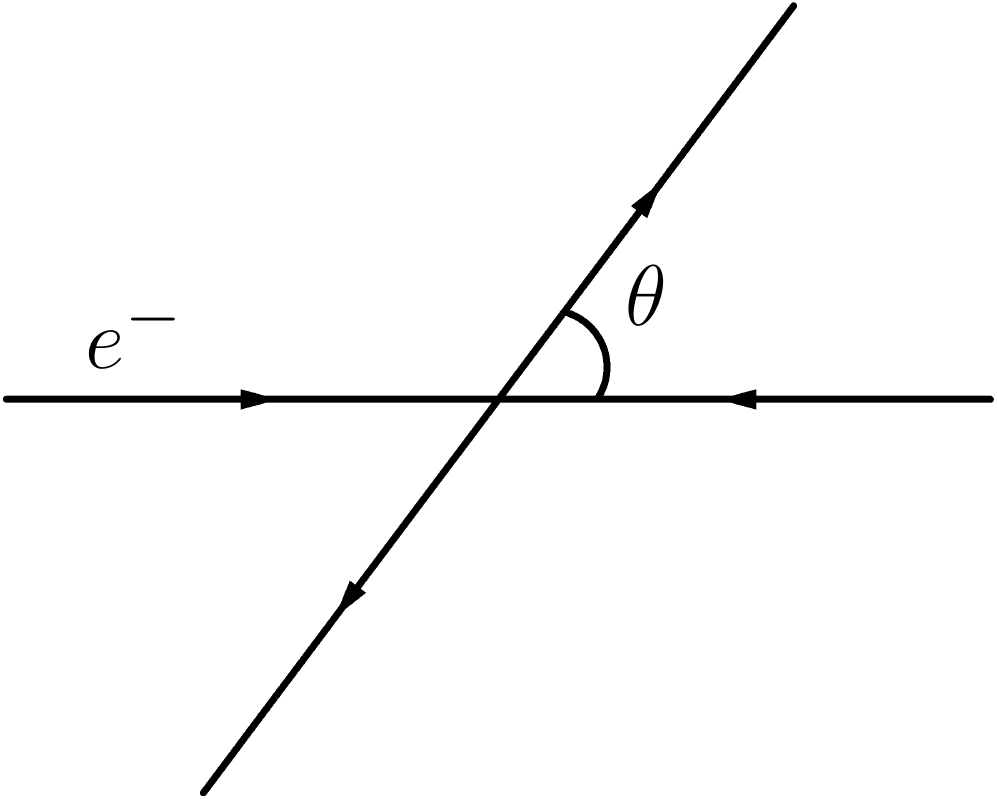}}

\caption{ Definition of the angle $\theta$ in the CM frame.}
\label{cmangle}

\end{figure}

By adopting the one-photon approximation and   performing the suitable traces,
 one gets (see also Eq.~(\ref{eq:dsdo})) the following differential cross
section  for the $N \bar N$-pair production, in the CM frame,
\cite{Cabibbo61} (for the cross section of the reaction $p\bar p \to e^+ e^-$
see Ref.~\cite{Zich62}, paying attention to the flux factor different from $1$,
in this case) \be
\left(\frac{d\sigma}{d\Omega}\right)_{e^+e^-\rightarrow N \bar{N}}^{th}=
\frac{\alpha^2 \beta }{4q^2} ~C_N(q^2)\left \{| G_{M}^N(q^2)|^2(1+\cos^2\theta)  
+| G_{E}^N(q^2)|^2~{1\over \tau}\sin^2\theta \right \}
\label{tlcross}\ee
where  
$\alpha=e^2/(4 \pi)\simeq 1/137$, $\theta$  is the angle between  the  
 direction of the incoming electron,
taken as $z$-axis,
and the produced nucleon (see Fig.~(\ref{cmangle})),   
$\beta=\sqrt{1 -4M^2_N/q^2}=|{\bf p}|/E$  the absolute value of the velocity of both $N$ and $\bar N$ 
(with  
$E=\sqrt{M^2_N +|{\bf p}|^2}$), $\tau= q^2/4M_N^2$ (notice the difference with
the definition in the SL region, below Eq. (\ref{sachs}))
and $C_N$ is a constant 
defined as follows
 \be C_N(y)=\left\{ \begin{array}{cc}  {y\over (1 -e^{-y})}& 
{\rm for~} p\bar p  \\ ~&~ \\
 1 &{\rm for~} n\bar n \end{array} \right.
\label{sommer} \ee 
with $$y= 2{M_p\over q} {\alpha\pi \over \beta }$$ 
(notice that the expression of the
 variable $y$ contains a factor $2$ at variance of what one can find 
  in Ref.~\cite{BABAR} where a mistyping is present~\cite{Druz11}).
   In the case of the $p\bar p$-pair, $C_p (y)$  is the s-wave 
Sommerfeld-Gamow factor that takes
into account the  QED leading-order correction   to the  wave-function of
the charged pair, and results to be  proportional
  to $|\Psi_{p\bar p}(0)|^2$, where $\Psi_{p\bar p}(0)$ 
  is  the relative wave-function in the continuum.
This factor specifically
affects the cross section near 
the threshold, where the relative 3-momentum of the charged hadronic pair is small (see, e.g., 
~\cite{Brod95,Hoan97}) and therefore $\beta \to 0$. In particular, it  makes the cross section different
  from 
 zero at threshold, since the factor $\beta$ in the denominator of $C_p (y) \sim   y $   
 balances a factor $\beta$ that comes from the phase-space calculation.
  As a final remark, it is worth noting that the  standard non relativistic expression of
  the variable $y$ is $$y={2\alpha\pi \over v_{r} }=
  2\alpha\pi ~{\mu_{rid}\over |{\bf p}| }~~~,$$
  where $v_r$ is the relative velocity, $|{\bf p}|$ the relative momentum and $\mu_{rid}$ the reduced mass ($M_p/2$).
   It can be heuristically generalized to  the
   relativistic framework~\cite{BABAR} as
  follows
   $$
   v_{r}=2 {p\over M_p}= {2 E \over M_p}~{|{\bf p}|\over E}~\Rightarrow~{q \over M_p}~ 
   \beta~~~,$$ 
obtaining the expression of the variable $y$ in Eq.~(\ref{sommer}) (notice that $c=1$ in the above
equations). 

 In Eq.~(\ref{tlcross}), as in the SL case,  one has  introduced 
  the Sachs FF's, 
 given by \be
\label{sachstl}
G_{E}^N(q^2) = F_{1}^N(q^2) + \tau  F_{2}^N(q^2) ~,
\quad \quad
G_{M}^N(Q^2) = F_{1}^N(q^2) +  F_{2}^N(q^2) ~.
\ee
Notably, at the threshold, i.e. $q^2=4 M^2_N$, 
 electric and magnetic Sachs FF's become equal by construction. 
This issue has been experimentally investigated for
 the proton case, by two different collaborations. As discussed in Subsect. \ref{expris},  the ratio 
 $|G_{E}^p(q^2)/G_{M}^p(q^2)|$ has been extracted from the  angular distributions (cf.\ Eq.
~(\ref{tlcross}))  measured in a wide  range  of momentum transfer, $ 2M_p\leq q \leq 3~ $(GeV/c),  
 by the BaBar collaboration~\cite{BABAR}. For $q$
up to $2.2~$(GeV/c), the ratio results to be significantly
greater than unity, in disagreement with  previous data, obtained by the 
PS170 collaboration ~\cite{PS170_94},
that are
compatible with the expected threshold value $\sim ~1$.
  
\begin{figure}[htb]
\begin{center}

\centering{\includegraphics[width=10.0cm]{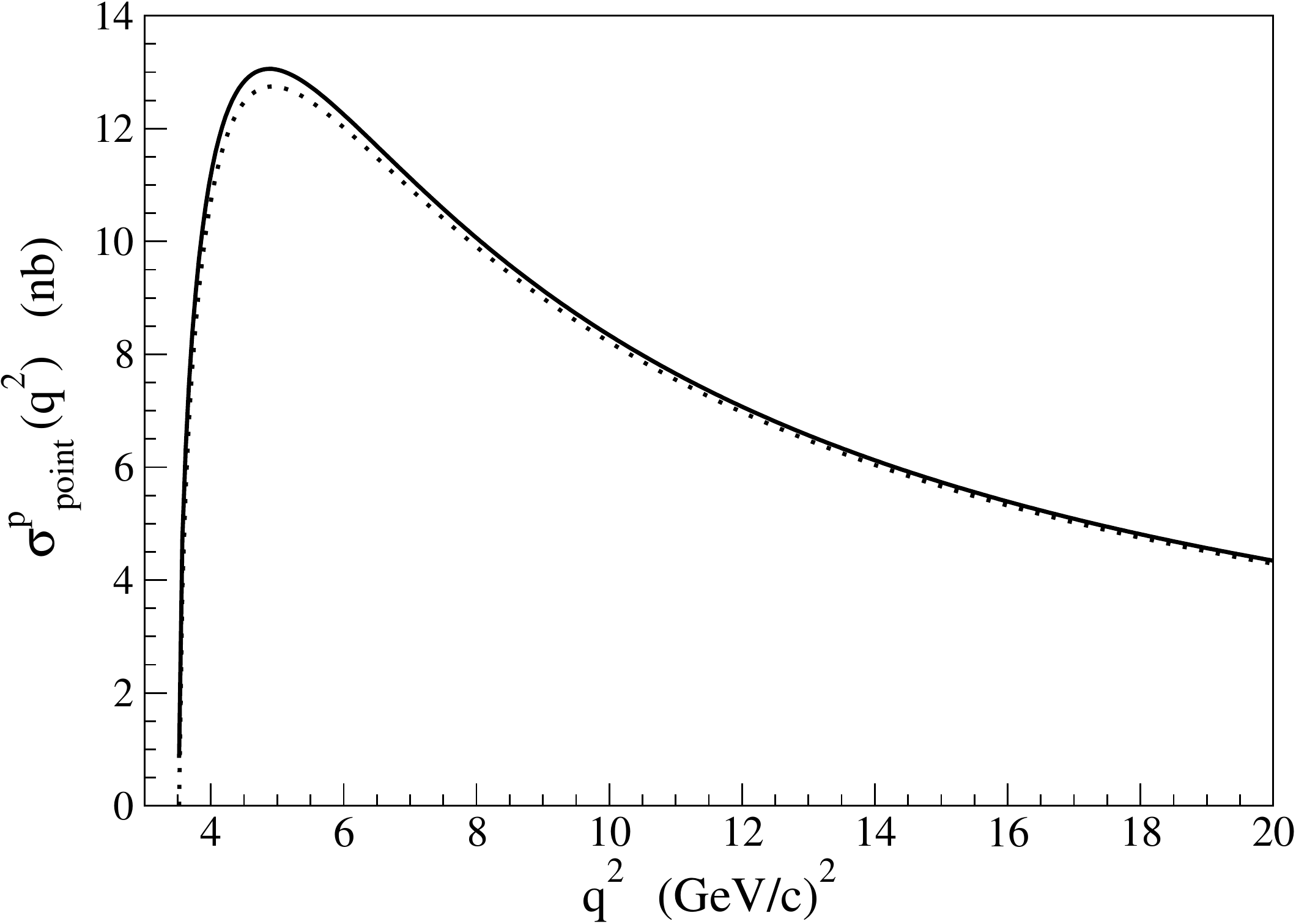}}
\caption{ Point-like nucleon  cross section, see Eq.~(\ref{sigmap}) and text, vs $q^2$.
Solid line: proton.
Dashed line: neutron. Notice the effect of the Sommerfeld-Gamow factor,
Eq.(\ref{sommer}), at the
threshold of the $p\bar p$ production.}
\label{sigma}
\end{center}
\end{figure}

The total cross section in one-photon approximation is obtained from Eq.~(\ref{tlcross})
by integrating over the solid angle $\Omega$ . For the sake of legibility, here we rewrite  Eq. 
(\ref{eq:ff_general}), but with some obvious changes, viz 
\be
\label{Eq:stot}
\sigma^{th}_{e^+e^-\rightarrow N \bar{N}}= {\alpha^2 \beta  \over 4q^2}~
C_N(q^2)\int d\Omega  \left [ |G_{M}^N(q^2)|^2(1+\cos^2\theta)  
+  |G_{E}^N(q^2)|^2{\sin^2\theta \over \tau}\right ]=\nonu
= 
{4\pi\alpha^2 \beta~ \over 3q^2} ~ C_N(q^2) \left [|G_{M}^N(q^2)  |^2
+\frac{|G_{E}^N(q^2)|^2}{2 \tau}\right ]~.
\ee
Such an expression suggests  
 to define an {\em effective } nucleon form
 factor in the TL region by dividing the actual total cross section
 shown in  Eq.~(\ref{Eq:stot}), by the
 point-like one, obtained from Eq.~(\ref{Eq:stot}) putting $|G_{E(M)}^2(q^2)|= 1$.  Then, one can write
 \be
 |G_{\rm eff}^N (q^2)|=\sqrt{\sigma^{th}_{e^+e^-\rightarrow N \bar{N}}(q^2)\over 
 \sigma^N_{point}(q^2)}=\sqrt{2 \tau~|G_{M}^N(q^2)  |^2+|G_{E}^N(q^2)|^2
\over 1 +2 \tau}
\label{geff}
 \ee
 where 
 \be
 \sigma^N_{point}(q^2)={4\pi\alpha^2 
\beta \over 3q^2}  C_N(q^2) \left [1
+\frac{1}{2 \tau}\right ]
 \label{sigmap}\ee
 The total  cross section becomes
 \be
 \sigma^{th}_{e^+e^-\rightarrow N \bar{N}}={\pi\alpha^2 \beta~C_N \over 3 M^2_p \tau}~
 \left[1 + {1 \over 2 \tau}\right]~|G_{\rm eff}^N (q^2)|^2=
 \sigma^N_{point}(q^2)~|G_{\rm eff}^N (q^2)|^2 
 \label{plsigma}\ee
The point-like cross sections for proton and neutron reduce at
threshold, i.e. 
$\tau=1$, to 
\be
\sigma^p_{point}(4M^2_p)={\pi^2\alpha^3  \over  2 M^2_p}=0.848~nb \quad \quad 
\sigma^n_{point}(4M^2_n)\to 0
\ee
where for the neutron $C_n \beta\to0$ for $q^2\to 4M^2_n$.
In Fig.~(\ref{sigma}), for illustrating the effect of the  Sommerfeld-Gamow factor and the 
behavior close to the threshold,  $\sigma^N_{point}(q^2)$ for both proton and
neutron, is shown in the  
range of
the present-day experiments. One should notice i) the very steep increasing between
the thresholds,  $q^2_{pth}\sim 3.52~{\rm (GeV/c)^2}$ and  $q^2_{nth}\sim
3.53~{\rm (GeV/c)^2}$ respectively,
and the maxima, around $q^2\sim 4.9~{\rm (GeV/c)^2}$, and ii) the  effect of
the Sommerfeld-Gamow factor that makes non vanishing the point-like cross 
section of the
proton. at the threshold.

Coming back to the definition of the {\em effective} FF, that in one-photon approximation  is a combination of electric
and magnetic Sachs FF's, it appears quite natural to  compare Eq.~(\ref{geff})
with the corresponding quantity defined through the experimental cross section 
$\sigma^{exp}_{e^+e^-\rightarrow N \bar{N}}$, namely Eq.~(\ref{geffa}). It should be pointed out 
that in both  numerator and  denominator in the middle
of  Eq.~(\ref{geffa})
  there are   quantities experimentally measured, and therefore one has an operative definition for 
$G_{\rm eff}^N (q^2)$, a
part the model dependence due to the Sommerfeld-Gamow factor. 
Moreover, in analogy with   the
 microscopic  interpretation of the SL  FF's 
 as Fourier transform   of electric and magnetic spatial distributions (cf Eq.
 (\ref{density}) with the caveat of Eq. (\ref{rhogpd})), one could develop a 
 simple physical picture 
 of the TL effective
  FF. Indeed, one could  consider 
 that it is microscopically 
 related, through a  Fourier transform in time, to the transition  amplitude
 from a virtual-photon state (cf Eq. (\ref{phwf}))  to  a hadronic state. 
 With this in mind, it becomes clear that in the TL
 region one can
 explore the time/energy structure of the hadronic Fock components of the
 photon state.
At this point,  a final short comment about $G_{\rm eff}^N (q^2)$ 
seems appropriate. One
should notice that,
 unfortunately, in the literature  one can find a shortcoming in the
 presentation of what the experiments provide. As a matter of fact, sometimes
 it is claimed that one measures the nucleon magnetic form
 factor {\em assuming}  $|G_{M}^2(q^2)|=|G_{E}^2(q^2)|$, but  this is not the
  case, unless $\tau=1$ (i.e. at the threshold).
 However,  it should be clarified
 that the proton database in the TL region, is entirely constructed 
 by introducing the {\em
 effective} proton form factors as defined in Eq.~(\ref{geffa}). For the neutron
 case, the Frascati data points~\cite{Fenice} are  shown  i) by using 
 Eq.~(\ref{geffa})  or ii)
 by {\em assuming}
 $|G_{E}^2(q^2)|=0$,  that amounts to  define for the neutron 
 \be
 |G^{\rm eff}_{Frn} (q^2)|=\sqrt{\sigma_{e^+e^-\rightarrow n \bar{n}}(q^2)\over 
 {4\pi\alpha^2 
\beta \over 3q^2} }
 \ee
In any case, in the experimental papers the procedure adopted for extracting the data
is always properly indicated.

 In the TL region, since the nucleon FF's
are complex functions, one has extra degrees of polarization for the produced
$N\bar N$ pair, even without a polarization of the incoming beams. This was already 
explained above, in
the brief introduction to the polarization-transfer method in the SL region.
For instance, one finds that a produced proton~\cite{Cabibbo61,Dubni96,Brod04}
gains a 
 polarization perpendicular to the scattering plane
 (recall that the  chosen $z$-axis 
is directed along    the incoming electron
beam) and, in one-photon approximation, it is given by 
\be
 P_y = ~-~{sin(2\theta)~\Imf\left\{G_{E}^p(q^2) G^*_{Mp}(q^2)\right\}\over D ~\sqrt{\tau}}
=\nonu=~-~{sin(2\theta)~|G_{E}^p(q^2)| ~~|G_{M}^p(q^2)|~sin(\varphi_E-\varphi_M)\over D ~\sqrt{\tau}}
\label{polartl}\ee
where $\varphi_{E(M)}$ is the phase of the complex-valued electric (magnetic) Sachs 
FF and $D$ is defined as:
 \be
D= |G_{M}^p(q^2)|^2(1 + cos^2\theta)  + |G_{E}^p(q^2)|^2
 {sin^2\theta\over \tau} 
 \label{denotl}\ee 
 The other two components of the polarization, $P_x$ and $P_z$,  lie
  on the scattering
 plane  and are different from zero only if 
 the electron beam has a non vanishing  longitudinal polarization, $P_e$, viz
 \be
 P_x = ~-~P_e~ {2sin(\theta)~\Ref\left\{G_{E}^p(q^2) G^*_{Mp}(q^2)\right\}\over D ~\sqrt{\tau}}
 \nonu
 P_z = ~P_e~ {2cos(\theta)~\left|G_{Mp}(q^2)\right|^2 \over D ~\sqrt{\tau}}
 \label{polartl1}\ee
 As one can see from Eqs.~(\ref{polartl}), ~(\ref{denotl}) and~(\ref{tlcross}),
   information on  absolute values and phases can be extracted 
   by  measuring both the
  angular distributions and  the normal (to the scattering plane) polarization.
  Indeed, one could consider also different cases, e.g.  
   a polarized  antiproton beam in the reaction~(\ref{tlreac})~\cite{PAX}, in order to extract
    double-spin polarization observables, that allows to access $\Ref\{
  G_{E}^NG_{M}^N\}$ (see  Refs.~\cite{Bilenki93} and~\cite{Tomasi05} 
  for details). 
   
  The previous expressions of both cross section and polarizations have 
  been obtained by considering only the
  one-photon-exchange approximation. The formalism for including the 
   two-photon-exchange effects can be found,
  e.g. in Ref. ~\cite{Adam07}  (for a recent review of the effects of the
  two-photon exchange in the SL region see Ref.~\cite{Arring11}), 
  while   Refs.~\cite{Tomasi08,Chen08} 
  give
  a first quantitative investigation of the effect,   showing a  reassuring smallness (few percents) 
 of the corrections.
 
 Finally the asymptotic behavior has to be the same as in the SL case (cf.\ Eq.
~(\ref{pqcd})) since it is governed by pQCD 
 (for an extended analysis of this issue 
 see also  Ref.~\cite{Pire}).

 \subsection{The $N\bar N$ production threshold}
 \label{thresh}
 The study of the reaction~(\ref{tlreac1}) near the threshold
 provides many relevant insights in the reaction mechanism that
 governs the transition from the unphysical to the physical regions.
 
 For the proton case,  
  the PS170 collaboration~\cite{PS170_94} observed for the
 first time   {\em  a steep
slope} of the {\em effective} FF when $q^0$  slightly increases from the  threshold
value, and more recently the BaBar collaboration 
~\cite{BABAR} has accurately 
 measured and confirmed the near-threshold peak. As to the neutron, the FENICE 
 collaboration~\cite{Fenice} (see also~\cite{Fenice1} for the proton
 measurements) made the first and 
 (till now) only measurement close to the
 threshold, showing also in this case, but with large error bars, an increasing
 value for $q^0\sim  q^0_{nth}$.

 The sizable and sharp rising of the cross section  close to the $p\bar p$  production, 
 given the 
 experimental efforts devoted to its investigation, has driven a
 lot of theoretical studies, pointing to elucidate the origin of such a
 surprising result. It is useful to remind that for the pion, one meets a more
 soft behavior, since a huge peak related to the $\rho$-meson production,
 around $770$ MeV, appears 
 far beyond the production threshold of a $\pi^+ \pi^-$ pair,
  $q_{\pi th}=2 m_\pi$.
 Even if the proposed mechanisms for explaining the proton case are different, all the approaches have 
 the common aim to clarify and constrain the $N\bar N$ interaction.
 In particular, the steep slope has been
 explained  i) by the $p\bar p$ final-state interaction  acting near the threshold
 (see, e.g.~\cite{Dalka10} and references quoted
therein), ii)    by the tail of 
a $p\bar p$ bound state (see, e.g.~\cite{Bruno09} for a recent
investigation of this issue by using the $N\bar N$ Paris potential and reference
quoted therein)
  or iii) by a narrow meson resonance
\cite{Fenice}, just
below the threshold. For instance, in Ref.~\cite{Haiden06} a
complex scattering length  in a   $^3S_1$ $p \bar p $ state 
is obtained from  the J\"ulich
$N\bar N$ potential~\cite{Juli}, allowing to achieve a good description of the
data. In Ref.~\cite{Dalka10}, the field of investigation is enlarged to cover
not only the proton case but also $n\bar n$ and other $B\bar B$ systems, 
by exploiting a phenomenological approach beyond
the scattering-length approximation, pointing to a resonant behavior near the
threshold. The  $p\bar p$ bound state appears a very intriguing possibility, but
such an explanation as mentioned above, is in competition with less exotic
effects.
However, if this is the case, one could address
the issue via reactions like $e^- + e^+\to \gamma +  p\bar p$ or even
$\bar p +d \to  e^- + e^+  + n$ as investigated in~\cite{Fonv09}, by taking
into account
 the off-mass-shellness of the nucleons inside the target nucleus.
 	
The neutron threshold deserves  some attention, in view of the existence
of only one data set~\cite{Fenice}. Those data posed a question about the
applicability of a naive pQCD analysis to the $q\bar q$ production involved in
the proton and neutron FF's. By using the FENICE data for both proton and
neutron, one has  
$|G_{\rm eff}^N(4M^2_n)|/|G^{\rm eff}_p(4M^2_n)|
\sim 1.4$, while 
  a
  naive perturbative description of the $e^+e^-$ annihilation~\cite{Ellis}
  leads to  $|G_{\rm eff}^N(4M^2_n)/G_{\rm eff}^p(4M^2_n)|\equiv~|e_d/e_u| =1/2$, namely
  almost a factor $3$ of difference. Fortunately
  the new data by the BaBar collaboration~\cite{BABAR} suggest 
  a path that reconciles experiment and naive expectation, since by considering
  the BaBar proton data near threshold, the difference between the experimental
  ratio and the naive pQCD prediction shrinks the difference to a factor of $2$, but with
  a large error band, due to the uncertainties of the FENICE neutron  data.

 \subsection{ Dispersion
 Relations: an overview}
 \label{DR}
 A truly microscopic calculations of the nucleon FF's have necessarily a
 prerequisite:  i) a knowledge  of the substructure of the nucleon and the interactions
 between the constituents, and ii)  a non-perturbative framework where elaborating the calculations.
 Such an observation makes clear how much challenging be the task.
 Historically, coming back to the fifties, the study of the nucleon FF's  had  great benefits
 from the extension of the
 investigation to the whole complex plane
 of
 the relevant variables. In this way, as briefly illustrated below,
  the non-perturbative regime of the strong interactions, based at that time on
 nucleonic and mesonic degrees of freedom, could be taken into account,
 by exploiting both i)  information gathered within a perturbative (diagrammatic)
 analysis of the needed amplitudes and  ii) more fundamental properties,
 like causality,  embedded and tightly related to the analytical behavior
 of the observables. As it is well-known, such an  analytic method has its origin in 
 the Kramers-Kr\"onig relation
 in optics~\cite{KK}.

  Local properties of the field 
  theories, as dictated by the microscopic causality, can be immediately
  translated into  the analytic behavior of the scattering amplitudes
   (see, e.g.~\cite{BD}), and in turn
    into  the analytic behavior   
   of the general Green functions of the interacting fields, if one  adopts    
   the Lehmann-Symanzik-Zimmermann 
   (LSZ)
   reduction formula~\cite{LSZ} for the S-matrix elements.
  
   The causality constraint  imposes a finite value for the speed of
   any  signal. This constraint can be  rephrased into  the request that
   space-time regions, separated by space-like distances, cannot communicate each
   other and therefore one has to require proper operatorial relations between
   fields, viz  \be
   \left [\phi(x'),\phi^\dagger(x)\right]=0\quad \quad 
   \left \{\psi(x'),\psi^\dagger(x)\right\}=0 \quad \quad {\rm for ~~}(x'-x)^2=c^2(t'-t)^2-
   ({\bf x}'-{\bf x})^2~<~0
   \label{com}
   \ee
   where $\phi(x)$ and $\psi(x)$ are the  local bosonic and 
   fermionic fields  of a given theory, respectively. In turn, 
  the commutation properties
    establish well-defined  analytic behavior of the 
   Fourier 
   transform of combinations of the field themselves, analogously to the basic investigation 
   performed in classical optics by Kramers and Kr\"onig, who analyzed the
   propagation of a wave front and found  
   the constraints induced by causality  on 
   the diffractive index (i.e. the em response of the composite system). A causal behavior means 
   that the response of the system will be vanishing for $t'<t_0$, where $t_0$ is 
   the
   instant when the system starts to be perturbed. If one takes into account
   such a constraint, one can rigorously deduce that 
   the Fourier transform of
   the system response must be analytic in the upper-half plane (see~\cite{BD}
   for details). Then,   
   an integral relation
    between the real and the imaginary
   part of such a Fourier transform  follows.
   For instance, if one has a complex function $F(z) =\Ref F(z) +i \Imf F(z)$,
   analytic in the upper-half plane,   one 
   can write 
   the  following  Cauchy integral
   \be
   F(z) ={1 \over 2 \pi i} \int_{\cal C} ~dz'~{F(z') \over z'-z}
   \label{cauchy}\ee
   with a contour, ${\cal C}$, composed by    the real axis and
   a semi-circle at infinity, in the upper half-plane. If $F(z)$ properly behaves  for $ |z|
   \to \infty$, namely 
    the
   contribution  from the arc at infinity is vanishing,
   one gets a simple expression 
   \be
   F(q)= \lim_{\epsilon \to 0_+} F(q+i\epsilon)={1 \over 2 \pi i} \int_{-\infty}^\infty dq
   {F(q')\over q'-q-i\epsilon}\ee
   otherwise one has to add the contribution of the semi-circle.
   Moreover, recalling the formal expression
   $$ {1\over q'-q\mp i\epsilon}= {{\cal P} \over  q'-q} \pm i\pi \delta (q'-q)~~.$$
    where ${\cal P}$ indicates the Cauchy principal value, 
    the following relations, called dispersion relations (DR's)\footnote{These
    relations are also
     known  as Hilbert transformations between $\Ref F(q)$ and $\Imf F(q)$.},
    can be straightforwardly deduced 
   \be
   \Ref F(q)= {1 \over \pi} {\cal P} \int_{-\infty}^\infty dq'~{{\Imf} F(q')\over
   q'-q} \label{realf}\\ &&
   \Imf F(q)= ~-~{1 \over \pi} {\cal P} \int_{-\infty}^\infty dq'~{{\Ref} F(q')\over
   q'-q}\label{imagf}
   \ee
      Therefore, the
   knowledge of the whole function $F(q)$ can be obtained from the 
   knowledge of
   only  one contribution: the real part or the imaginary one. The name dispersion
   relations  can be traced back to the work of Kramers and Kr\"onig~\cite{KK},
    where  
   the diffractive index was investigated, and relations between the real part
   (governing the dispersive behavior of the light) and
   the imaginary one (controlling its absorption)  were established, by using
    the  causality principle, as mentioned above.

   If the assumed quick fall-off  of 
   $F(z)$   does not hold, and therefore the integrals in Eqs.~(\ref{realf}) and ~(\ref{imagf}) are not well-defined, 
   then
     one can
   write down  new DR's, by introducing a proper 
   {\em subtraction} term. For instance, for  $F(0)\ne 0$, one can speed-up the fall-off by
   subtracting $F(0) / z$ and 
   repeating the same steps that lead to Eqs.~(\ref{realf}) and ~(\ref{imagf}),
   but  for $$ {F(z)-F(0) \over z}$$
   obtaining, e.g., for the real part 
   \be
   {\Ref F(q)\over q}= {\Ref F(0)\over q}+{1 \over \pi}
    {\cal P} \int_{-\infty}^\infty dq'~{{\Imf} F(q')\over q'( q'-q)}
  \label{subtr} \ee
   since ${\cal P} \int_{-\infty}^\infty dq'~1/ [q'( q'-q)]=0$.
   Such a subtraction procedure can be applied with the obvious
   changes, e.g. subtracting  $F'(0)/z$ from $[F(z)-F(0)] /z^2$, as many times   as  
  it needs in order to make   vanishing  the  contribution at infinity, in the integrals.
  
  In view of  the application to the nucleon FF's, it is interesting to
  briefly discuss the case when  $F(z)$ has a cut 
  for $q_0\leq \Ref z < \infty$, and it is real for $-\infty< \Ref z<q_0$. The 
  Schwarz reflection principle, 
   $F(z^*)=F^*(z)$, applied along the open interval  $-\infty< \Ref z<q_0$,
   allows one to analytically continue $F(z)$ in the lower half-plane. Then a
   different path ${\cal C}$ can be considered for writing down the Cauchy relation 
  ~(\ref{cauchy}). In particular, the integration path runs  along the lower rim   from 
   infinity to $q_0$ and along the
   upper one in the reversed direction, and it has to be completed by  a circle at infinity (to be
   integrated in
   the counter-clockwise direction). Then,  
   assuming for simplicity that the contribution   to the Cauchy integral for  $ |z|\to \infty$
  is vanishing (a subtraction procedure can be always  implemented),
  one gets
  \be
  F(z) ={1 \over 2 \pi i} \lim _{\eta \to 0_+} \int_{q_0}^\infty ~dq'~
  {F(q'+i\eta)-F(q'-i\eta)\over q'-z}
  = 
  {1 \over  \pi }  \int_{q_0}^\infty ~dq'~
  {\Imf F(q')\over q'-z}
  \label{drfin}\ee
  In conclusion, the function $F(z)$ is determined for any $z$ by the discontinuity across the
  cut. The relevance of this relation is given by  the possibility to link 
  the imaginary
  part of the nucleon FF's in the TL region to physical amplitudes, as
  discussed in the following Subsection.
  With simple steps, one can also write
  \be
  \Ref F(q) =\lim _{\epsilon \to 0_+} \Ref F(q+ i\epsilon) =
  {1 \over  \pi } {\cal P} \int_{q_0}^\infty ~dq'~
  {\Imf F(q')\over q'-q}
  \ee
  Such an equation is a generalization of Eq.~(\ref{realf}), where the
   analyticity
  of $F(z)$  was assumed only in the upper  half-plane. 
 In the case of a
  slow decreasing of $F(z)$ for $|z|\to \infty$, one can repeat the subtraction
  procedure as in the previous Eq.~(\ref{subtr}), obtaining
   the following one-time subtracted expression\be
  F(z)=  F(0)+{z \over  \pi }
   \lim _{\epsilon \to 0_+} \int_{q_0}^\infty ~dq'~
  {\Imf F(q')\over q'(q'-z -i\epsilon)}
  \label{drfinb}\ee

   In closing this Subsection, it is important for what follows to mention 
    one of
   the theorems that can  be deduced from the Phragm\'en-Lindel\"of principle
   that allows to extend the maximum-modulus principle  to unbounded domains
  ~\cite{Tit50}. The  theorem, relevant for putting in relation the asymptotic value of the
FF's in the TL region and in the SL one, states: 
   
  {\em If an analytic function $F(z)$, with $z=r e^{i\varphi}$, is such that i)
   $F(z) \to a$ as $z\to \infty$ along a 
   straight line starting from the origin and
$F(z) \to b$ as $z\to -\infty$ along another   straight line, 
again starting from the origin, ii)
 it is regular \footnote{A regular function is analytic and single-valued in a given region.}and bounded in one of the two sectors between these two straight
 lines.
Then $F(z=\infty) = F(z=-\infty)=a$ and $F(z)\to a$ uniformly in that sector as $|z| \to \infty$.} 
It is important to note the request that the analytic function must be bounded in one of the two sectors.

Summarizing the analysis in the complex plane, one should mention four items: 
i) causality determines the analytic behavior of an amplitude involved in a
given physical process, ii) the Schwarz reflection principle allows to analytically continue a
function from one side of the plane to the other side, if along a part of the
real axis it does not contain singularities, iii) in presence of a cut, the function
can be expressed in terms of the discontinuity across the cut, iv) a theorem from
the Phragm\'en-Lindel\"of principle leads to relate the asymptotic behavior of
the function for
$z\to \pm \infty$. One of the first  discussions 
of the application of
the Phragm\'en-Lindel\"off theorem (previously used for 
proving the Pomeranchuk theorem and
its generalizations) to the em nucleon FF's, putting in relation the asymptotic values in the SL
and TL regions,  can be found 
in Ref.~\cite{Bilenki93}.

\subsection{ Dispersion
 Relations and   the Nucleon Form Factors} 
 \label{DRFF}  
Refs.~\cite{Feder58,Chew58} (see also
 ~\cite{BD},~\cite{GW04} and~\cite{Drell60} for  more general  discussions)
 yielded a 
first description of how 
 the previous general properties, particularly causality, can be exploited for modeling the nucleon FF's
 in both the SL and the TL regions. Noteworthy, 
the same approach can be applied to any hadron. 

In  the seminal work  of Ref.
~\cite{Feder58}, dispersion relations for 
the SL and TL nucleon FF's were determined, after analyzing 
the matrix elements of the em current operator (needed for  the first order 
T-matrix) and  using  the LSZ
reduction 
formalism~\cite{LSZ} for the S-matrix elements (see, e.g.,~\cite{BD}).
 A direct and  fundamental consequence of such an 
investigation was the possibility of 
relating the  imaginary part of the FF's to matrix elements 
relevant for other physical processes, that, in turn, can offer experimental
inputs for an overall description of the FF's themselves.
   Let us briefly illustrate the issue in the TL case, by
   considering the 
   following matrix elements of the em current operator
   \be
\sqrt{{EE'\over M^2_N} } (2\pi)^3~\langle 0|
J_{em}^{\mu}(0)| {\cal N} (p,s)\bar{\cal N} (p',s')\rangle=
\bar{v}_{s'}(p')\left [\gamma^{\mu}F_{1}^N(q^2)+
 i
{F_{2}^N(q^2)\over  2M_N}
\sigma^{\mu\nu}(p'+p)_\nu\right ]u_{s}(p)=\nonu=
i\sqrt{{E\over M_N } } (2 \pi)^{3/2} \int d^4x~e^{-i p'\cdot x}~ \bar v_{s'}(p')
\langle 0|{\cal T}_+\{  S_N(x)J_{em}^{\mu}(0)\}|{\cal N} (p,s)\rangle
 \label{disp1}  \ee
where, due to the LSZ reduction, one has 
\be {\cal T}_+\{  S_N(x)J_{em}^{\mu}(0)\}=\theta(t)S_N(x)J_{em}^{\mu}(0) 
+\theta(-t)J_{em}^{\mu}(0) S_N(x)= \nonu=\theta(-t)\left [J_{em}^{\mu}(0),
S_N(x)\right]+ S_N(x)J_{em}^{\mu}(0)~~~, \label{timeor} \ee
 with $S_N(x)$  the source
of the nucleon interacting field, i.e.
\be
(i\Psla {\mbf \nabla} -m) \psi_N(x)=S_N(x)
\label{sour1}
\ee
It should be recalled that  $S_N(x)$, like $\psi_N(x)$, lowers the baryon number
by a unity. 

An important step for making more transparent the 
 investigation of the analytic behavior 
 of the FF's, is the substitution  
 of  the time-ordered sum ${\cal T}_+\{  S_N(x)J_{em}^{\mu}(0)\}$  
 by the commutator 
 $\left[J_{em}^{\mu}(0), S_N(x)\right]$, that allows one to implement the
 causality (cf.\ also Eq.~(\ref{com})). This can be accomplished by 
 noticing  that the matrix element
  $\langle 0| S_N(x)J_{em}^{\mu}(0)|{\cal N} (p,s)\rangle $, generated by the rightmost term in the
  second line of Eq.~(\ref{timeor}), does not contribute to the matrix element 
  of the current, Eq.~(\ref{disp1}).
   As a matter of fact,  since  
$$(i\Psla {\mbf \nabla} -m)\langle 0| \psi_N(x)|one~nucleon\rangle =0$$ the matrix
element
$\langle 0| S_N(x)|one~nucleon\rangle$ must vanish.
Therefore,   no contribution can be generated by a one-nucleon intermediate state, when a physical, complete
basis is used for spanning  the
 matrix element $\langle 0| S_N(x)J_{em}^{\mu}(0)|{\cal N} (p,s)\rangle$, viz 
 \be
 \langle 0| S_N(x)J_{em}^{\mu}(0)|{\cal N} (p,s)\rangle=
 {1 \over (2\pi)^4}\sum_{p_I;I\ne nucleon} \langle 0| S_N(x)|p_I; I,B=1\rangle  
 \langle p_I;I,B=1|J_{em}^{\mu}(0)|{\cal N} (p,s)\rangle=
 \nonu=
 {1 \over (2\pi)^4}\sum_{p_I;I\ne nucleon} e^{ip_I\cdot x}~\langle 0| S_N(0)| p_I;I,B=1\rangle  
 \langle p_I;I,B=1|J_{em}^{\mu}(0)|{\cal N} (p,s)\rangle
 \label{vanish}\ee
 where $\sum_{p_I;I}$ means an integration over the four-momentum and a sum over the discrete quantum numbers.
 It should be pointed out that the intermediate states spanning the matrix element in the lhs of Eq.~(\ref{vanish})
 must have baryon number 
 $B=1$, since both the em current conserves the
 baryonic number and $S_N(x)$ annihilates a nucleon in the initial state. Moreover,
  $p^2_I >M^2_N$ (the nucleon is the lowest $B=1$ state), and therefore the intermediate states must contain multi-hadron systems with $B=1$. Finally,
   since in Eq.~(\ref{disp1}) $(p')^2=M^2_N$, one immediately deduces that
  $\langle 0| S_N(x)J_{em}^{\mu}(0)|{\cal N} (p,s)\rangle$ does not contribute to $\langle 0|
J_{em}^{\mu}(0)| {\cal N} (p,s)\bar{\cal N} (p',s')\rangle$
  (see also~\cite{GW04} for details). As a matter of fact, one gets
   \be
 \int d^4 x~ e^{-ip'\cdot x}~
 \langle 0| S_N(x)J_{em}^{\mu}(0)|{\cal N} (p,s)\rangle=
   \sum_ {p_I;I\ne nucleon}~\delta^4(p'+p_I)~\nonu \times 
  \langle 0| S_N(0)| p_I;I,B=1\rangle  \langle p_I;I,B=1|J_{em}^{\mu}(0)|{\cal N} (p,s)\rangle
 =0
 \label{disp2} \ee
 Therefore, the matrix element in Eq.~(\ref{disp1}) can be rewritten as follows
 \be
 \sqrt{{EE'\over M^2_N} } (2\pi)^3~\langle 0|
J_{em}^{\mu}(0)| {\cal N} (p,s)\bar{\cal N} (p',s')\rangle=\nonu=
i\sqrt{{E\over M_N } } (2 \pi)^{3/2} \int d^4x~e^{-i p'\cdot x}~ \bar v_{s'}(p')
\theta(-t)\langle 0| \left [J_{em}^{\mu}(0),
S_N(x)\right]  |{\cal N} (p,s)\rangle
 \label{disp1b}  \ee 
  As mentioned above, the presence  of the commutator makes it possible to
   exploit rather directly the causality, since  
   $\left[J_{em}^{\mu}(0), S_N(x),\right]$,   is vanishing 
  for space-like separations, i.e.  $x^2<0$ (given the commutation rules of the local fields). Then, one has to deal with two theta
  functions in the four dimensional integral~(\ref{disp1b}): $\theta(-x^2)$ and
  $\theta(-t)$. A general discussion, that involves a careful analysis of the
  dependence in $exp(-ip'\cdot x)$, where
  $p'\cdot x=
  E' t-{\bf p}' \cdot {\bf x}$ and $|{\bf p}'|=\sqrt{E^{\prime 2} -M^2_N}$, 
  can be found in~\cite{GW04}. Here, following the simplified presentation of 
  Ref.~\cite{Feder58}, the case ${\bf p}'=0$ is assumed, and  the
  analytical dependence upon only $E'$, taken as a complex variable, is
  considered.   Given the function $\theta(-t)$, the analytic structure of the
  integrand in Eq.~(\ref{disp1b}) is governed by $exp(-|t|~\Imf E')$, and this 
  leads {\em to argue} that  the matrix elements of the current is analytic in
  the upper half-plane of $E'\equiv \Ref E' + i \Imf E'$.
  Straightforwardly, one can repeat the steps illustrated in the previous 
  Subsection and
   construct dispersion relations for the nucleon FF's. Indeed, if one would  
    use the dispersion relations
  (see  Eq.~(\ref{drfin}) or its subtracted version, 
  Eq.~(\ref{drfinb})), one has to fully determine   where the singularities are located,
  and in this task the help comes from the analysis in perturbation theory,
  namely by considering  
  the Feynman diagrams contributing to the FF's. Such a study (see, e.g.,
 ~\cite{BD}) shows that in the TL region one meets multiple cuts:  a first
  branch point located at 
   $s=4m^2_\pi$, i.e. in correspondence with the threshold of the
  two-pion production, that contributes to the isovector part of the nucleon
   FF's (in a VMD  framework~\cite{VMD}, one has the production of a 
   virtual $\rho$ meson,
  that decays in two pions), then another cut starting from the threshold of the 
  three-pion production, that contributes to the isoscalar part of the nucleon 
  FF's (in a VMD framework, one has the production of a virtual $\omega$ meson,
  that decays in three pions), 
   etc. (see below). In summary, within a DR approach one  obtains
   (cf.\  Eqs.~(\ref{drfin}) and ~(\ref{drfinb})) 
    the following  expressions for the nucleon FF's, valid for any $q^2$
   \be
   F_{1}^{IS}(q^2)= {1 \over 2} + {q^2 \over \pi} \int_{9 m^2_\pi}^\infty
   ds~{\Imf\left\{F_{1}^{IS}(s)\right\}\over s(s-q^2-i \epsilon)}
   \quad \quad \quad
   F_{1}^{IV}(q^2)= {1 \over 2} + {q^2 \over \pi} \int_{4 m^2_\pi}^\infty
   ds~{\Imf\left\{F_{1}^{IV}(s)\right\}\over s(s-q^2-i \epsilon)}
   \nonu
   F_{2}^{IS}(q^2)=  \mu^{IS} + {q^2 \over \pi} 
   \int_{9 m^2_\pi}^\infty
   ds~{\Imf\left\{F_{2}^{IS}(s)\right\}\over s(s-q^2-i \epsilon)}
   \quad \quad
   F_{2}^{IV}(q^2)= \mu^{IV} + {q^2 \over \pi} 
   \int_{4 m^2_\pi}^\infty
   ds~{\Imf\left\{F_{2}^{IV}(s)\right\}\over s(s-q^2-i \epsilon)}
   \label{drff1}\ee 
   where the normalizations of the FF's have been explicitly shown, with
   $$\mu^{IS}={\mu_p+\mu_n \over 2}\quad \quad \mu^{IV}={\mu_p-\mu_n \over 2}$$
   Moreover,
   the lowest thresholds for each isospin channel have been considered.  Three comments are in order: i) the
   main task should be the evaluation of the imaginary part of the form
   factors for $s\geq 4 m^2_\pi$, ii) one could expect more cuts starting at various thresholds,
   depending upon the particle productions one has to consider, iii) one should
   keep in mind that the dispersion integrals have contributions also from values of $s$
   below the $N\bar N$  production threshold ($4m^2_\pi<s<4
   M^2_N$), the so-called unphysical region. 
   
  Indeed, the analysis of the discontinuity across the real axis, 
  for $\Ref~q^2\geq 4 m^2_\pi$, of the current-operator matrix elements
    allows one to determine the imaginary part of the FF's, 
  establishing the announced link with physical  processes, like the hadronic
  decays of neutral vector mesons, the TL form
  factors of mesons (e.g. for describing the $\pi^+\pi^-$-continuum)
   or the physical $N \bar N$ annihilation into mesons. To show this, 
  still following Ref.\cite{Feder58}, one can write
  \be
  \sqrt{{EE'\over M^2_N} } (2\pi)^3~ 
 \langle 0|
J_{em}^{\mu}(0)| {\cal N} (p,s)\bar{\cal N} (p',s')\rangle =
\nonu=\sqrt{{E\over M_N } } (2 \pi)^{3/2} ~i~ \int d^4x~e^{-i p'\cdot
x}~\bar v_{s'}(p')\theta(-t)~
\langle 0|\left[ J_{em}^{\mu}(0), S_N(x)\right]| {\cal N}
(p,s)\rangle
=\nonu=
 \sqrt{{E\over M_N } } (2 \pi)^{3/2} { 1\over 2\pi } \int d\omega \int d^4x~e^{-i p'\cdot
x}~{ e^{-i\omega
 t} \over \omega -i \epsilon}~
\bar v_{s'}(p')~
\langle 0|\left[ J_{em}^{\mu}(0), S_N(x)\right]| {\cal N}
(p,s)\rangle
  =
\nonu=
 ~\sqrt{{E\over M_N } } (2 \pi)^{3/2} { 1\over  (2\pi)^5} \sum_{p_I;I} \int d\omega 
 \int d^4x~e^{-i (p'+p -p_I)\cdot x}~{ e^{-i\omega t} \over \omega -i \epsilon}
\nonu \times~
\bar v_{s'}(p')~
\langle 0| J_{em}^{\mu}(0)|p_I; I,B=0\rangle  \langle p_I;I,B=0| S_N(0)| {\cal N}
(p,s)\rangle
=\nonu=
 \sqrt{{E\over M_N } } (2 \pi)^{3/2} { 1\over  2\pi}\sum_{p_I;I}  
 \delta^3({\bf p}_I-{\bf p}'-{\bf p} )~{ 1 \over E_I-E'-E -i \epsilon}
\bar v_{s'}(p') \nonu \times ~
\langle 0| J_{em}^{\mu}(0)| p_I;I,B=0\rangle  \langle p_I;I,B=0| S_N(0)| {\cal N}
(p,s)\rangle
 =
\nonu=~
 \sqrt{{E\over M_N } } (2 \pi)^{3/2} { 1\over  2\pi}\sum_{p_I;I}  
 \delta^3({\bf p}_I-{\bf p}'-{\bf p} )~\left[{ {\cal P} \over E_I-E'-E} + i\pi 
 \delta(E_I-E'-E)\right]~\nonu
 \times~
\bar v_{s'}(p')~
\langle 0| J_{em}^{\mu}(0)| p_I;I,B=0\rangle  \langle p_I;I,B=0| S_N(0)| {\cal N}
(p,s)\rangle
\label{matrix0}\ee
 Notice that i) Eq.~(\ref{disp2}) has been used, and  ii) $S_N(0)$ lowers the baryon
 number by a unity and therefore the intermediate states must have $B=0$, 
 as also requested 
 by the matrix elements of the em current 
 $\langle 0| J_{em}^{\mu}(0)| p_I;I,B=0\rangle$.    If one repeats analogous steps (illustrated in
  Appendix \ref {matrix}) for the matrix elements
  of $\langle 0|
\left[J_{em}^{\mu}(0)\right]^\dagger| {\cal N} (p,s)\bar{\cal N} (p',s')\rangle=
 \langle  {\cal N} (p,s)\bar{\cal N} (p',s')|
J_{em}^{\mu}(0)| 0\rangle^*$, one can  single out the imaginary part of 
the nucleon FF's  by evaluating matrix elements of the skew-hermitian part of 
$J_{em}^{\mu}(0)$, as follows
\be
  \sqrt{{EE'\over M^2_N} } (2\pi)^3~{1 \over 2i}~
  \left[\langle  0|J_{em}^{\mu}(0)| {\cal N} (p,s)\bar{\cal N}
  (p',s')\rangle -\langle  0|\left[J_{em}^{\mu}(0)\right]^\dagger
  | {\cal N} (p,s)\bar{\cal N}
  (p',s')\rangle\right]=\nonu=
\sqrt{{E\over M_N } } (2 \pi)^{3/2} { 1\over  2}\sum_{p_I;I}  
 \delta^4({ p}_I-{ p}'-{ p} )
\bar v_{s'}(p')
\langle 0| J_{em}^{\mu}(0)| p_I;I,B=0\rangle  \langle p_I;I,B=0| S_N(0)| {\cal N}
(p,s)\rangle
=\nonu=
 \bar{v}_{s'}(p')\left [\gamma^{\mu}\Imf \left\{ F_{1}^N(q^2)\right \}+
 i
{\Imf \left\{F_{2}^N(q^2) \right \}\over  2M_N}
\sigma^{\mu\nu}(p'+p)_\nu\right ]u_{s}(p)
\label{disp3}\ee
From Eq.~(\ref{disp3}), by applying  the proper Dirac traces, one can obtain the needed relations 
for determining $\Imf \left\{ F_{1(2)}^N(q^2)\right \}$ (see the next
Subsection).
The attractive feature of Eq.~(\ref{disp3}) is the link with physical
processes, as already said. In particular, the matrix elements 
$\langle 0| J_{em}^{\mu}(0)|p_I; I,B=0\rangle$ is related to the em 
production of physical hadronic systems with $B=0$, while the matrix elements 
$\langle p_i; I,B=0| S_N(0)| {\cal N}(p,s)\rangle$ is proportional
to the   hadronic production of a $N\bar N$ pair (or the $N\bar N$ annihilation into hadronic states). 
It should be pointed out that, since the intermediate states have
$p^2_I=(p+p')^2>0$, given the presence of the four-momentum  conserving delta-function,  they 
represent physical states (differently from the SL case where $(p'-p)^2 \leq 0$, i.e. unphysical
intermediate states), and a spectral decomposition of the nucleon FF's
 can be devised. Strictly related to the value of   the masses of the 
intermediate states, there is   the question of the energy
thresholds of the multi-cut structure that must appear  in Eq.~(\ref{drff1}). 
One can immediately realize that the
em hadron production, $\langle 0| J_{em}^{\mu}(0)| p_I;I,B=0\rangle$,  has $2 m_\pi$ as the
lowest physical threshold (contributing to
the isovector FF's), 
 while the physical
threshold of the  hadronic production of the $N\bar N$ pair  
$\langle p_I; I,B=0| S_N(0)| {\cal N}(p,s)\rangle$, is $2M_N$. 
The region between these two
thresholds, as mentioned above, is  the unphysical region, that, however, 
contributes to the
dispersion integral, since the path of integration starts at  
$2m_\pi$, i.e.  where the free propagation  of the pions begins. 
Therefore, one of   the issues in the DR method is
given by the    evaluation of the 
matrix elements $\langle p_I;I,B=0| S_N(0)| {\cal N}(p,s)\rangle$ in the unphysical
 region. This can be overcome   through an analytic continuation of the 
  physical processes and  
   by considering the  crossing symmetry (see~\cite{Feder58} and
  ~\cite{Hammer07} for a more recent discussion). 
   Below the threshold $2 M_N$, there are other possible energy thresholds
 related to the em productions of   three pions, $K\bar K$-pair, four pions, 
 $K\bar K \pi$, etc. 
 In particular, one has to remind that  the intermediate states, $|p_i;I\rangle$, 
 must  have  vanishing 
 baryonic number and strangeness, and therefore they  can contain two or more 
  mesons,  with proper quantum numbers, or baryon-antibaryon pairs. 
 Moreover, given the presence of the matrix elements
  of the current operator $\langle 0| J_{em}^{\mu}(0)| p_I;I,B=0, S=0\rangle$, 
  the intermediate states must have the proper isospin 
  quantum
  numbers dictated by  $J_{em}^{\mu}(0)$, that contains an isoscalar term and an 
  isovector one.
  For instance, the first term  filters non-strange meson states with $I^G
  (J^{PC})= 0^- (1^{--})$, i.e  states with an  odd number
  of pions,  while the second term
  can single out non-strange
  meson states with $I^G(J^{PC})= 1^+ (1^{--})$, i.e  states with an even number
  of pions.  Since the  G-parity does not affect states  made by  strange mesons, 
  the lightest state with a kaon and an antikaon can contribute to both isospin channels.

\subsection{TL Nucleon  form factors and the dispersion relation approach}
\label{TLDR}
The DR approach has been widely adopted for describing the SL nucleon FF's
(see, e.g.,~\cite{Mergel96} and references quoted therein). Particular care
was devoted to modeling the spectral representation of the imaginary part 
of the FF's (cf.\ the sum in  Eq.~(\ref{disp3})). One of the main
ingredients for achieving such a description was represented  by
the phenomenological
input supplied by the VMD (see, e.g.,~\cite{VMD}), since one could use a physically meaningful scheme and exploit
experimental quantities for describing $\Imf\{F^N_{1(2)}(q^2) \}$. Moreover, the extension to the TL FF's was straightforward, since 
the DR's hold for any value of $q^2$.
In particular,   
 the DR approach  adopted in  Ref.~\cite{Mergel96} for a 
very accurate
description  of the SL FF's (as  known  at the date),
  was extended in Ref.~\cite{Hammer96} to
the TL region, without changing the number of the free parameters. For brevity we call this approach
the Mainz model.
In the spectral decomposition of $\Imf\{F^N_{1(2)}(q^2) \}$, it was considered: i) 
the same numbers of VM poles as in~\cite{Mergel96}, ii) the $2\pi$ continuum,
already used in the SL region for constraining the overall description through
unitarity,  and
iii) the asymptotic 
behavior suggested by pQCD (cf.\ Eq.~(\ref{pqcd})),  including  in this way 
 both hadron and quark degrees of freedom. For the sake of concreteness, 
 let us
 mention that
  the spectral 
  decomposition in terms of  
   the  VMD model   amounts to 
    a finite sum on the intermediate states with masses 
  equal to the relevant vector mesons, viz
  \be
\Imf F^I_{VMD;i}(t')=~\pi~\sum_n ~a_{i,n}^{I}~\delta(t'- (m^{I}_{n})^2)
\ee where $i=1,2$ and $I= IS,IV$, and $a_{i,n}^{I}$ contains all the information about $\gamma^*-VM$ coupling and
the transition $VM\to N\bar N$.
The above result follows  from the  VMD contribution to the whole FF's, viz
\be
F^I_{VMD; i} (t)=\sum_n ~{a_{i,n}^{I}\over(m^{I}_n)^2 -t} 
\ee
where $t=q^2$. Notice that the denominators, that come from the VM propagators, vanish
only for $t>0$, and produce the well-known pole pattern, experimentally seen  in the TL region(cf Subsect.
\ref{expris}).

\begin{figure}[ht]
\begin{center}

\includegraphics[width=10.0cm]{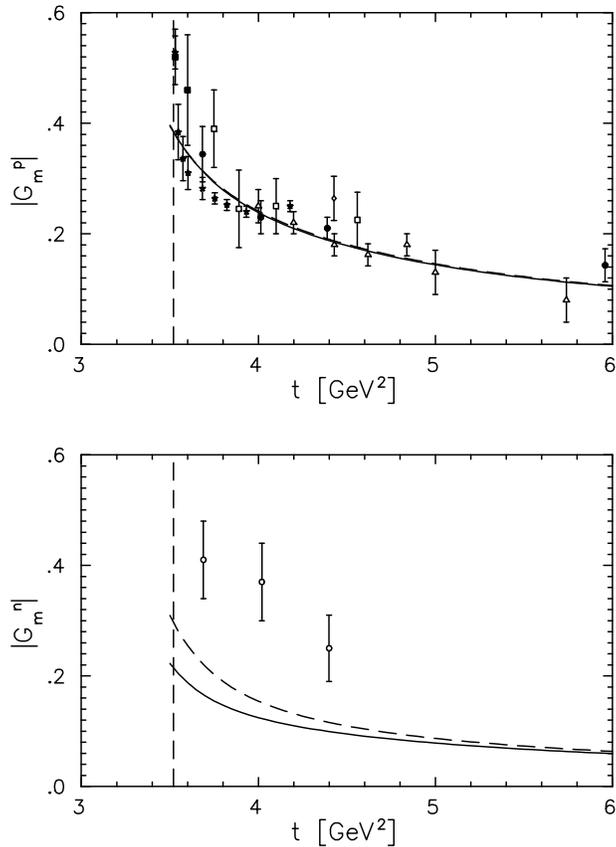}
\caption{ TL nucleon  form factors within the Mainz model~\cite{Hammer96}. 
 Solid lines:  calculations with three meson poles, for each isospin channel, 
 and  $\Lambda^2 = 12.0~GeV^2$ (see text). Dashed lines:
the same as the solid lines, but including the contribution from 
$\phi (1680)$ as well.  
Notice that the experimental data are presented as the absolute value of 
magnetic form factors, but the experimental data correspond to {\em effective} proton form factor
given by Eq. (\ref{geffa}). Figure taken from Ref.~\cite{Hammer96}.
}
\label{HamPLB96fig1b}
\end{center}
\end{figure}

In~\cite{Hammer96}, it was assumed an unsubtracted form for DR's, that allows
one to write 
 \be
F_i^{IS} (t)=  \left[ \sum_{n}
\frac{a_ {i,n}^{IS} ~ L^{-1}\left[(m^{IS}_{n})^2\right]}{(m^{IS}_{n})^2 - t }
\right] ~ \left[
\ln \left( \frac{\Lambda^2 - t}{Q_0^2} \right)\right]^{-\gamma}
\nonu
F_i^{IV} (t)=  \left[F_i^\rho(t)~L(t)
 + \sum_{n\ne \rho}
\frac{a_{i,n}^{IV} ~ L^{-1}\left[(m^{IV}_n)^2\right]}{(m^{IV}_n)^2 - t }\right] ~ \left[
\ln \left( \frac{\Lambda^2 - t}{Q_0^2} \right)\right]^{-\gamma}
\label{drech1} \ee
where the function $L(t)$,  given by 
\be
L(t) = \left[ \ln \left(\frac{\Lambda^2 - t}{Q_0^2} \right)
\right]^{-\gamma}~~~,
\label{drech2}
\ee
 takes into account the pQCD prescription for the asymptotic behavior. In the second line of Eq.~(\ref{drech1}), 
 $F_i^\rho(t) $  describes  the two-pion
contribution,  including both the term from the $\rho$ pole and the $2\pi$
continuum. This last term is parametrized  through the pion form
factor and the P-wave $\pi \pi  \to N\bar N  $ partial wave amplitude, 
without free parameters (see~\cite{Mergel96} for details).
It is worth recalling that by using Eq.~(\ref{drech2}) one can recover the standard
 pQCD
behavior for $t\to -\infty$, viz
\be
\Ref\{F_i (t)\} \to  {1 \over 
t^{(i+1)}} \, \left[ \ln\left(\frac{|t|}{Q_0^2}\right) \right]^{-\gamma}
\label{pqcdtl}
\ee
that, in turn, can be properly continued in the TL region.
In Eq~(\ref{drech2}), $\gamma$  is the anomalous dimension, 
 $\Lambda$ controls the transition to the pQCD behavior and  
  $Q_0 \sim \Lambda_{ QCD}$. In the actual calculations,  the
 constants were chosen as follows: 
 $\gamma=2.148$, corresponding to a number of flavors $N_f=3$,  
 $Q^2_0=0.35~{\rm GeV^2} $, while the adjusted parameters were: i) 
 the quantity  $\Lambda$ and ii) the residues $a^I_{i,n}$,
  corresponding to the
 contributions from three meson poles for each isospin channel. In particular
 the masses of the vector mesons were:
  $m(\rho')=1.40~{\rm GeV}$, $m(\rho'')=1.45~{\rm GeV}$, $m(\rho''')=1.69~{\rm GeV}$,
 $m(\omega)=0.782~{\rm GeV}$, $m(\phi)=1.019~{\rm GeV}$ and  $m(S')=1.60~{\rm GeV}$. Results for
 the TL nucleon FF's, obtained by applying the Mainz model, are shown in Fig.~(\ref{HamPLB96fig1b}).

\begin{figure}[htbp]
\begin{center}

\vspace{-2.5cm}
\includegraphics[width=17.0cm,angle=0]{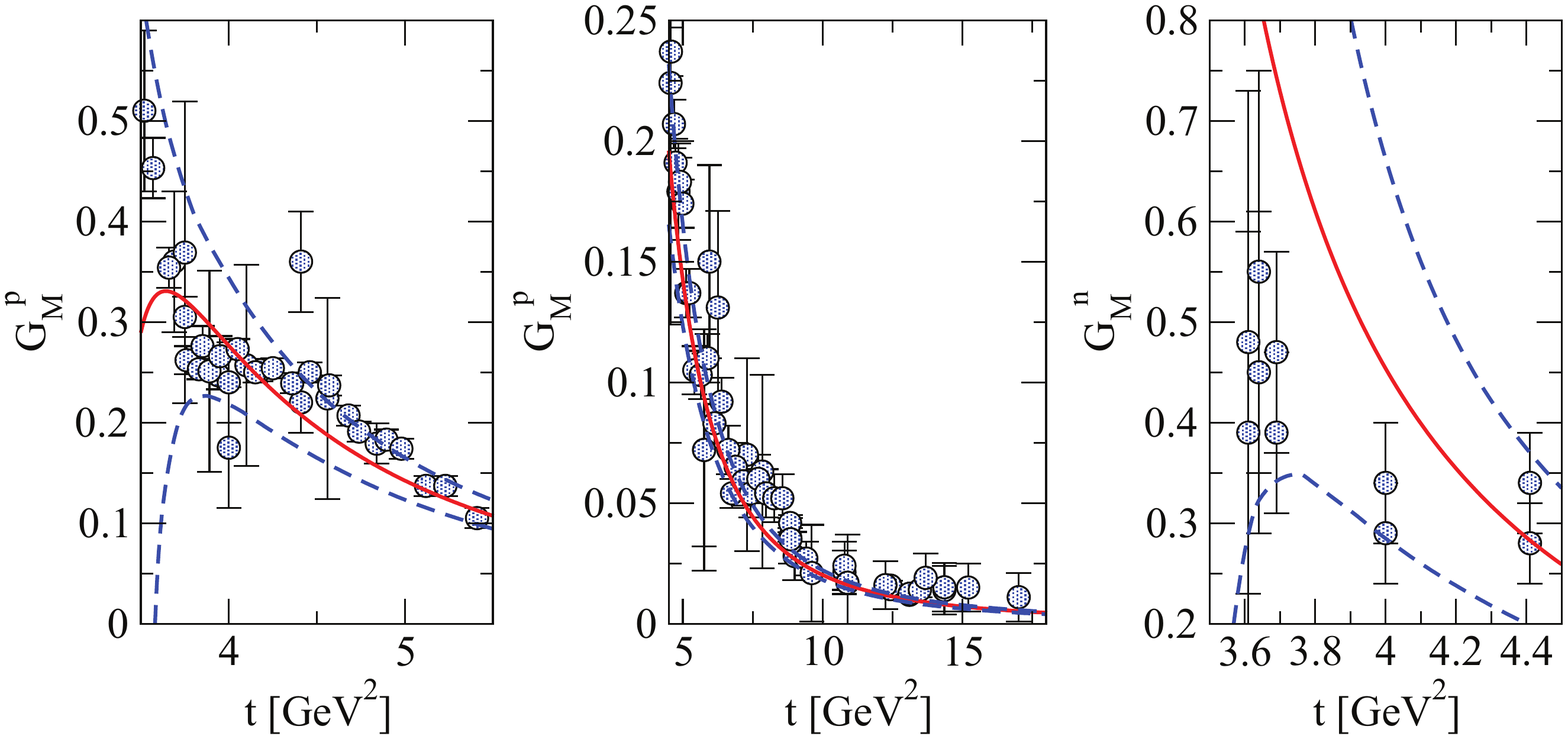}

\vspace{-2.cm}
\caption{TL nucleon  form factors evaluated within the Mainz-Bonn-J\"ulich 
Super Convergent dispersion relation approach~\cite{Hammer07}.
 Left panel: proton  form factor in the range $3.5<t\leq 5.5~ {\rm GeV^2}$. 
  Middle panel: the same as the left panel, but in the range $5.5<t<20~ {\rm GeV^2}$.
Right panel:  neutron  form factor.
Solid lines: best fit.
Dashed lines:  error bands
assigned by allowing some variation to the $\chi^2/DOF$ 
(see~\cite{Hammer07} for
details). Notice that the experimental data are presented as the absolute value of 
magnetic form factors, but they corresponds to the {\em effective} nucleon FF, 
introduced in Eq.~(\ref{geffa}). Figure reprinted with permission from 
 Ref.~\cite{Hammer07}, \copyright 2007 American Physical Society.}
\label{fig:hammer}
\end{center}
\end{figure}

The work described in Ref.~\cite{Hammer96} has been  steadily improved
 by the Mainz-Bonn-J\"ulich collaboration~\cite{Hammer06,Hammer07},  
 enriching the previous description both  i) by adding the $K\bar K$ as well as the 
  $\rho \pi$ continuum, and ii) by updating all the experimental
 inputs with latest experimental results. Moreover,  different
 ways for implementing the asymptotic behavior requested by   pQCD 
 (see Refs.~\cite{Hammer06,Hammer07} for more details) were  considered. 
 In particular, in the most recent
 SL and TL calculations presented in Ref.~\cite{Hammer07}, it was introduced the so-called
 superconvergent approach for controlling the asymptotic behavior of the TL nucleon FF's, 
 besides the standard one given by Eq.~(\ref{pqcdtl}).
 The  superconvergent approach amounts to impose  a proper fall-off not only to the real part 
 (cf.\ Eq.~(\ref{pqcdtl})), 
 but also to
 the imaginary part of 
  Dirac and Pauli FF's, i.e.
 \be
 \int_{t_0}^\infty dt~\Imf\{ F_1(t)\}=0\nonu
 \int_{t_0}^\infty dt~\Imf\{ F_2(t)\}=0 \quad \quad \int_{t_0}^\infty dt~t~\Imf\{ F_2(t)\}=0
\label{supc} \ee 
It should be pointed out that these constraints allow to take into account 
the application of the Phragm\'en-Lindel\"off theorem,
as illustrated in Subsect. \ref{DR}.
 The TL nucleon FF's, calculated by using   such a superconvergent approach are 
   shown in Fig.~(\ref{fig:hammer}). 
  Notice that  i) the SL and TL proton data and ii) the neutron SL FF's have been
   included  in a
 17-parameter fit, and therefore  the neutron TL calculation (solid curve in the rightmost panel of
 Fig.~(\ref{fig:hammer})) represents a genuine 
prediction. 
In Ref.~\cite{Hammer07}, it was also presented an explicit pQCD continuum
approach, based on a phenomenological Ansatz that fulfills the pQCD behavior
(\ref{pqcdtl}). In this case the set of  free parameters is reduced to 14, 
but the quality of the fit become worst, as
pointed out in the same work.

\begin{figure}[htbp]
\begin{center} 
 \includegraphics[width=6.5cm]{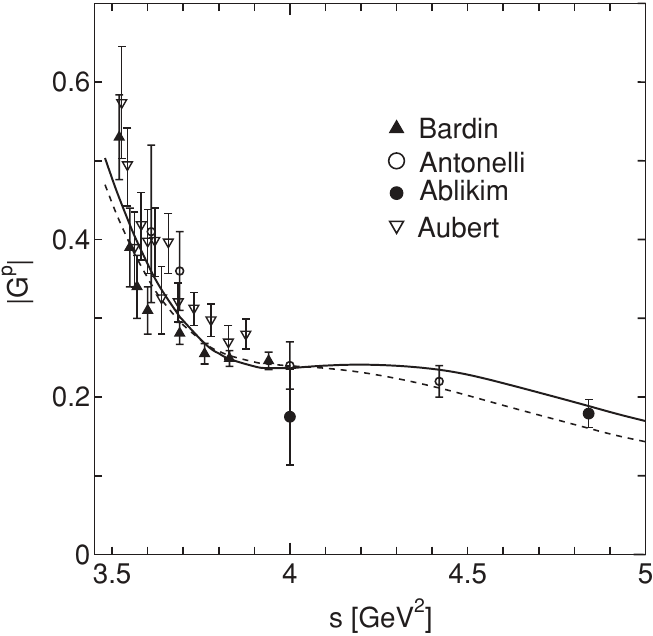} $~~~~~~~~~~~$ 
\includegraphics[width=6.5cm]{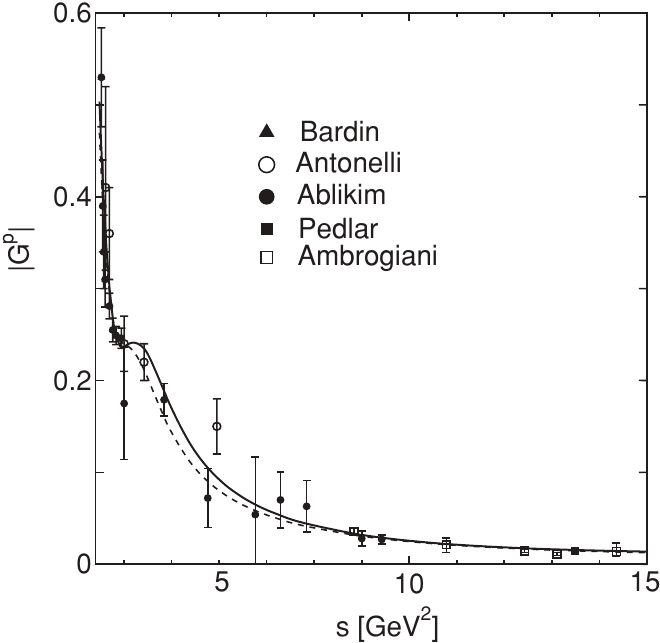}
\caption{ TL proton {\em effective} form factor, Eq.~(\ref{geff}), within the fitting
procedure of Ref.~\cite{Furuichi10}, based on the DR approach and a
super-convergence condition (cf.\ Eq.~(\ref{supcf})), vs the Mandelstam variable
$s=q^2$.
Left
panel:  $s^2<5 ~GeV^2$. Right panel: the whole range of $s$. Figure reprinted with permission 
from  Ref.
~\cite{Furuichi10}, \copyright 2010 American Physical Society.} 
  \label{Furup}

\vspace{0.5cm}
\includegraphics[width=6.5cm]{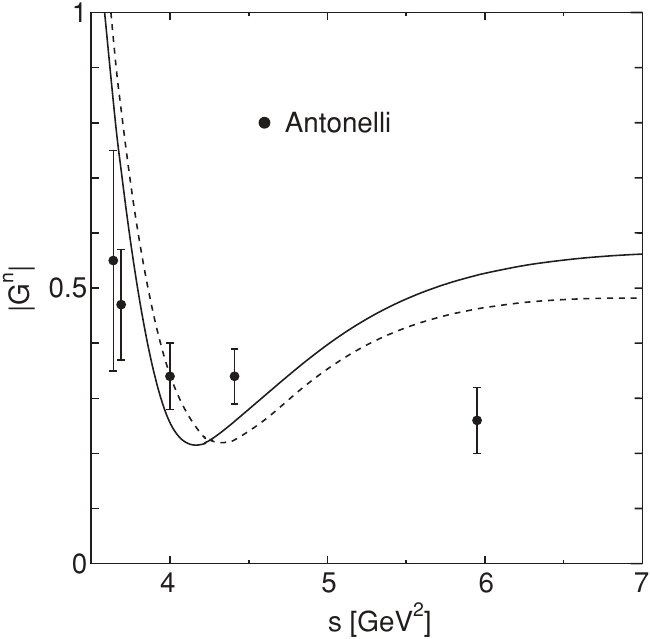}
\end{center}
\caption{ The same as in Fig.~(\ref{Furup}), but for the  neutron 
effective form factor, Eq.~(\ref{geff}). Figure reprinted with permission 
from  Ref.
~\cite{Furuichi10}, \copyright 2010 American Physical Society.}
 \label{Furun}
\end{figure}

More recently, the above illustrated approach has been improved  by
the Tokyo-Chiba group,
adding more vector-meson poles and more parameters for modeling the pQCD
contribution. 
In Ref.~\cite{Furuichi10}, an unsubtracted DR has been assumed for both Dirac and
 Pauli form factors, in the SL and TL regions. The 
 spectral decomposition of the $\Imf \{F^N_{1(2)}(q^2)$ is described in an analogous way as in
 Ref. ~\cite{Hammer07}. It was considered i) a $2\pi$ continuum with a helicity
 decomposition of the $\pi\pi\to N\bar N$ amplitude (notice that the
 tabulated data for the $\pi N$ scattering adopted in~\cite{Furuichi10}, i.e. 
 the ones by H\"oheler et al 
 ~\cite{Hohl},
 already contain   the $\rho$
 contribution), and ii) up to 5 meson resonances for each isospin channel, taking
 into account both mass (up to $m_{VM} \sim 2.05~GeV$) and width in
  the Breit-Wigner function that describes
 each 
 vector-meson pole contribution. A particular care was devoted for modeling 
 the asymptotic
 behavior suggested by pQCD, both considering contributions up 
 to $\alpha^4_s(t)$ and 
   implementing the  superconvergent Ansatz  in a slightly 
 different way  with  respect to  Ref.~\cite{Hammer07} (cf.\ Eq.
~(\ref{supc})), viz 
 \be
 {1\over \pi}\int_{4m^2_\pi}^\infty d t'~(t')^n~\Imf\left\{F^I_1(t')\right\}=0\quad \quad n=0,1\nonu
 {1\over \pi}\int_{4m^2_\pi}^\infty d t'~(t')^n~\Imf\left\{F^I_2(t')\right\}=0\quad \quad n=0,1,2
 \label{supcf}\ee
 where $\Imf\left\{F^I_i(t)\right\}$ has to satisfy the  asymptotic condition
 \be
 \Imf\left\{F^I_i(t)\right\} \to {c \over t^i} \left [ln {t\over Q^2_0}\right]^{-\gamma -1}
 \ee
 with $c$ and $\gamma$ constants. In the actual calculation it was used $\gamma=2$. With the 
 previous
 constraints, $F_i(t)$ has the standard pQCD behavior, a part the value of $\gamma$ different from the standard anomalous
 dimension.
 In the approach of Ref.
~\cite{Furuichi10}, the number of free parameters has substantially 
 increased, since one has 5 masses and widths for each isospin channel,
 and correspondingly 10 residues for the vector meson poles of both  the Dirac  
 and  Pauli FF's.
 Therefore the VMD part contains 40 parameters. Finally,  one has to add 12
 adjusted coefficients for the QCD part,   3 parameters that enter a
 convergence factor,  ensuring the super-convergence condition and another one
  that
 shifts slightly the masses of the resonances in the Breit-Wigner width. 
 The
  56 parameters are fixed by fitting:  i) the most recent SL and TL proton 
  data, even considering the ratio $ |G_{E}^p/G_{M}^p|$ measured by
  the BaBar collaboration~\cite{BABAR} in the TL region (notice a mistyping in 
 ~\cite{Furuichi10}, where the experimental ratio is indicated with 
  $\mu_p|G_{E}^p/G_{M}^p|$) and ii) the SL and TL neutron data.
  The reasonably good description of the nucleon FF's,
   achieved by the fitting procedure of Ref.~\cite{Furuichi10}, is illustrated in Figs.~(\ref{Furup}) and~(\ref{Furun}).  
  
  { To summarize the  investigations presented in this Subsection, one has to stress the richness of
information in the TL region, that can  be accessed through $\Imf \{ F_{1(2)}^N\}$. Noteworthy,  the experimental
determination of 
 the relevant phases 
 could strongly motivate future efforts in developing more and
  more refined theoretical analysis, possibly tightly related to microscopic
  calculations, in order to establish a close vision on the non-perturbative
  regime of QCD.}

\subsection{Phenomenological DR's and TL extrema}
\label{tlextre}
In Ref.~\cite{Bald99} the nucleon magnetic FF's 
below the $N\bar N$ threshold, i.e. in the 
region $0 < q^2 \leq 4M^2_N$, has been investigated in an almost model-independent way.
 The DR's are adopted for determining an integral equation
 that relates $log ~G_{N}(q^2)$  and $log |G_{N}(q^2)|$ where $G_{N}(q^2)=
 G_{M_N}(q^2)/\mu_N$ (therefore $G_{N}(0)=1$), following the modulus representation of the FF discussed
 in Ref.~\cite{Gourdin}.
 In the TL region one writes the following subtracted 
 (recalling that $log ~G_{N}(0)=0$) dispersion relation 
 \be
 log ~G_{N}(q^2) = log ~|G_{N}(q^2)| +i \Phi_{G_N}(q^2)={q^2 \sqrt{4m^2_\pi-q^2} \over \pi}~\int_{4m^2_\pi}^\infty
 dt ~ {log |G_{N}(t)| \over t~(t-q^2) \sqrt{t-4m^2_\pi}}
 \label{baldie}\ee
 where $\Phi_{G_N}(q^2)$ is the phase of $G_{N}(q^2)$.
The range of integration can be   split into two parts,  ${4m^2_\pi}\leq t\leq {4M^2_N}$ and
${4M^2_N}\leq t<\infty$. The integral in the second interval, i.e. the physical region, 
has been estimated by adopting a suitable function for $|G_{N}(q^2)|$, 
with the expected asymptotic 
behavior,
 and fitting the
available experimental data,  assumed to directly give 
$|G_{M}^N(q^2)|$ (cf.\ Eq.~(\ref{geff}) for the proper caveat).  
Then the value of $log~ G_{N}(q^2)$ in the unphysical
interval, 
$0 < q^2 \leq 4M^2_N$, is obtained by  solving the  following integral
equation, through a very detailed procedure in order
 to overcome many numerical subtleties,
\be
log ~G_{N}(q^2) -I_{ph}(q^2)={q^2 \sqrt{4m^2_\pi-q^2} \over \pi}~\int_{4m^2_\pi}^{4M^2_N}
 dt ~ {log |G_{N}(t)| \over t~(t-q^2) \sqrt{t-4m^2_\pi}}
\ee
where $I_{ph}(q^2)$ indicates the above mentioned integral that depends upon 
$|G_{N}(q^2)|$ in the physical region.
 Interesting structures, related to 
peaks for i)  $\rho(770)$, ii) 
$\rho^\prime(1600)$    and iii) near the $N\bar N$ threshold 
are automatically 
generated. Noteworthy,  no evidence of a bump at the
 $\phi(1020)$
 mass has been found, in spite of the expectation that such a peak should 
 exist if a sizable polarized $s\bar s$ content in
the nucleon were present.

Within the DR spirit, 
in Refs.~\cite{Bald06,Pacetti07}, the proton ratio $R_p(q^2)=\mu_p
G_{E_p}(q^2)/G_{M_p}(q^2)$ has been investigated by considering the following 
 dispersion relation
 \be
 R_p(q^2)= R_p(0)+{q^2 \over \pi}\int_{4m^2_\pi}^\infty
 dt ~ { \Imf \{ R_p(t)\}\over t~(t-q^2) }  
 \label{baldie1}\ee
 where $R_p(0)=1$. Moreover, the Phragm\'en-Lindel\"off theorem was implemented 
by imposing a vanishing asymptotic value for $\Imf \{ R_p(q^2)\}$, in order to recover a real function
when $q^2 \to \infty$ (as in the SL region). 
  As a final step, the  imaginary part  has been spanned 
onto the ortho-normal basis given
by 
the Chebyshev polynomials, with independent coefficients for the unphysical and
the physical regions. The two set of coefficients are fixed by means of  the
available  experimental data and the 
theoretical  constraints in 
both the SL and the TL~\cite{Bald06,Pacetti07}, obtaining a description 
in  the unphysical region
and  also in a wide range of $q^2$. Unfortunately, 
 the fitting
procedure  cannot be fully constrained, and this is reflected in the wide band
associated to the solution of the integral equation~(\ref{baldie1}). 
However, interesting results have been
obtained, both near the threshold and in the asymptotic region.
In particular,  a dominance
of  $G_{E}^p(q^2)$ near threshold has been found, as indicated by a growing ratio
up to $\sim 11$ for $q^2\sim 4~{\rm (GeV/c)^2}$. As to the asymptotic region, the moduli of the ratio are quite similar,
but the prediction based on the Phragm\'en-Lindel\"off theorem cannot be immediately tested. As a matter of fact,
the SL result
for $R_p(Q^2)$  is negative for $Q^2=-q^2\to \infty$, and in the TL region 
the extracted value is only the modulus $|R_p(q^2)|$. 
 Therefore  the phase
$\Phi(\infty)$  of $R_p(q^2)=|R_p(q^2)|~exp[i\Phi(\infty)]$ for 
$q^2\to \infty$  has to be
determined, for a full comparison.
 Indeed, in order to fulfill the equality between the SL and TL
asymptotic values, i.e.  $\lim_{ q^2 \to \infty}~\Ref \{R_p(q^2)\}=\Ref 
\{R_p(-q^2)\}$ (as
dictated by the application of  the Phragm\'en-Lindel\"off theorem; cf.\ Subsect.
\ref{DR}), one should expect that $ \Phi(\infty)=(2n+1) ~\pi$, since 
$\Ref \{R_p(-q^2) \}$
asymptotically is negative. From such an observation, one can deduce the
following interesting result. If one applies the argument  theorem ~\cite{Tit50}
to a closed path, composed by a circle at infinity
and the  upper and lower rims of the cut $[4m_\pi^2,\infty]$, and  reminds 
also the    
Schwarz reflection principle, one gets  that $\Phi(\infty)- 
\Phi(4m^2) =\pi (N-P)$, where $N(P)$ is the number of zeros (poles) inside the
closed  path.  Notice that $P=0$, given the
assumed analyticity inside the closed path and one can choose $\Phi(4m^2)=0$. Therefore,  $N$ must be 
 an odd number to fit a negative value for $\Ref
 \{R_p(q^2)\}=|R_p(q^2)|~cos[\Phi(\infty)]$. This seems to be in nice agreement with
 the SL
data suggesting the existence of at least one zero (cf.\ $R_p(Q^2)$ in Fig.~(\ref{fig:slffp})). 
Further analyses of the issue  have been carried out 
in  Ref.~\cite{Pacetti07},  where it has been  discussed 
 the large uncertainty that affects the
modulus of the previous ratio, due to the choice of the data set that 
one adopts for constraining the
high momentum
part of $\Imf\{R_p(q^2)\}$. In particular, one gets  
$$ \left|{G_{E}^p(q^2) \over G_{M}^p(q^2)}\right|=~\left\{\begin{array}{c r}
0.95 \pm 0.20
& {\rm from~}BaBar ~\cite{BABAR}\\
2.3 \pm 0.7 & {\rm from~}Lear~\cite{PS170_94}\end{array}\right.$$

Summarizing, all the previous results indicate that the phenomenological studies
performed through the DR's are a source of highly non trivial  information in both the extrema of
the TL region, given the powerful tool represented by the complex analysis.

\subsection{The analytic continuation of the SL Nucleon FF's} 
\label{tlcont}
In the DR approach, in its full glory, the main issue is given by the attempt of 
exploiting physical processes for 
 modeling  the imaginary part of the nucleon FF's, or better 
the discontinuity of the
current matrix elements across the multi-cut region for $\Ref ~q^2> 4m^2_\pi$. 
As we have seen
in the previous Subsections, the knowledge of the imaginary part of 
the nucleon FF's allows one to determine the
FF's themselves in the whole range of $q^2$.
 A different
program is presented in Refs.~\cite{Iac04,Bij04}, where it was proposed a fit of the TL data, 
starting from a well-known  model of  the SL FF's, like the one in
Ref.~\cite{Iac73}, devised in the 1970's for describing, through a VMD approach (see, e.g.,
Ref.~\cite{VMD}), the available
SL nucleon FF's, but able to predict the fall-off of the SL ratio $\mu_p
G_{E}^p(Q^2)/G_{M}^p(Q^2)$ recently observed (cf.\ Subsect.~\ref{SLFF} 
and~\cite{SLrev}). The simple strategy for 
generating the TL  nucleon FF's, that are complex functions, is based on the analytical
continuation of the SL FF's, that are real functions of $\Ref~ q^2 \leq 0$, 
   replacing  the dependence 
upon $Q^2=-q^2$, with $-q^2~exp(i\theta)$ and then  using the phase $\theta$ 
as a free parameter in 
the fitting procedure to obtain the TL nucleon FF's. VMD plays a fundamental role in the DR approach
for reconstructing $\Imf F_{1(2)}(q^2)$, since it provides the pole structure,
to which one has to add the multi-cuts generated by the free propagation of the
hadronic states. Therefore the VMD content of the SL FF's allows one a first
approximation, that can be enriched by using the adjusted parameter $\theta$.  
 Indeed, only an infinitely accurate knowledge of the SL FF's (i.e.\ only in a
  mathematical realm) could univocally determine the FF's in their analytical
   domain, but this is not the case
  within the realm of the phenomenology. 
  
  Refs.~\cite{Iac04,Bij04} have generalized  the  
SL model of Ref.~\cite{Iac73},
where the external photon couples to
both an intrinsic structure, described by a proper   form factor
 and a meson cloud, obtained within the VMD framework.
It should be pointed out that the presence of the intrinsic form factor has 
a clear physical
motivation: 
 the size of the constituent quarks (CQ's) inside the nucleon (see, e.g.,   Ref.~\cite{CPSS95} 
for a discussion  of the CQ FF in  a relativistic description of  the SL  
FF's with a realistic nucleon wave-function). 
 In particular, the SL  model is given by
\be
F_1^S(q^2)= {1 \over 2} g(q^2)~\left[1-\beta_\omega - \beta_\phi+ 
\beta_\omega {m^2_\omega \over m^2_\omega-q^2}+\beta_\phi {m^2_\phi \over
m^2_\phi-q^2}\right]
\nonu
F_1^V(q^2)= {1 \over 2} g(q^2)~\left[1-\beta_\rho  
\beta_\rho {m^2_\rho \over m^2_\rho-q^2}\right]
\nonu
F_2^S(q^2)= {1 \over 2} g(q^2)~\left[(\mu_p+\mu_n-1-\alpha_\phi )
{m^2_\omega \over m^2_\omega-q^2}+ 
\alpha_\phi {m^2_\phi \over
m^2_\phi-q^2}\right]
\nonu
F_2^V(q^2)= {1 \over 2} g(q^2)~\left[{(\mu_p+\mu_n-1-\alpha_\phi )
 \over 1-\gamma~q^2}+ 
\alpha_\rho {m^2_\rho \over
m^2_\rho-q^2}\right]
\label{iac1}\ee
where
$$ g(q^2)= { 1\over (1-\gamma~q^2)^2}$$ is 
the intrinsic CQ FF,  
$m_\rho=0.776~{\rm GeV}$, $m_\omega =0.783~{\rm GeV}$, $m_\phi
=1.019~{\rm GeV}$.  The six free parameters, $\beta_\rho$,  $\beta_\omega$, $\beta_\phi$, 
$\alpha_\rho$, $\alpha_\phi$ and $\gamma$ are fitted to the SL data. 
\begin{figure}[t]
\includegraphics[width=17.0cm,angle=0]{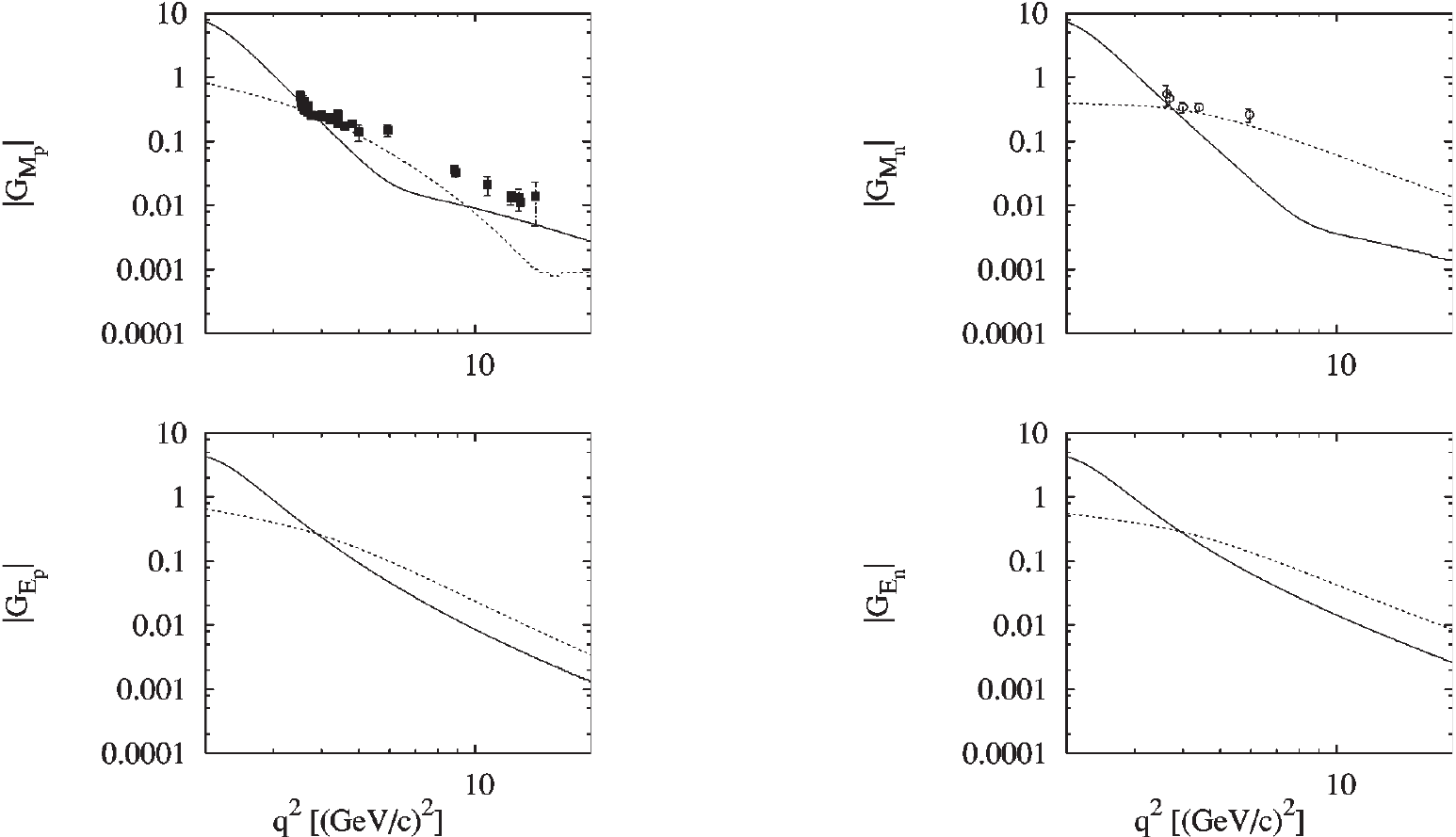}

\caption{ TL
nucleon FF's calculated within the  approach  of Iachello et al.~\cite{Iac04,Bij04}(see text). Upper panel: the  magnetic form
factors $|G_{M_p}(q^2)|$ and $|G_{M_n}(q^2)|$.  Lower panel: the  electric form
factors $|G_{E_p}(q^2)|$ and $|G_{E_n}(q^2)|$. The solid lines are from the  analysis of   Ref.~\cite{Bij04} and the
dashed ones are from  previous calculations presented in Ref.~\cite{Iac04}. The experimental data
correspond to the {\em effective} TL nucleon FF's, as given in Eq.~(\ref{geffa}). 
Figure reprinted with permission from Ref.~\cite{Bij04}, \copyright 2004
American Physical Society.}
\label{fig:iachello}
\end{figure}
It is worth noting that the asymptotic behavior of the previous parametrization
is the same predicted by pQCD, without  logarithmic corrections. Moreover, in order to take into 
account the non negligible  width of the $\rho$ meson,
the propagator is changed accordingly to
$${m^2_\rho \over m^2_\rho-q^2} \to {m^2_\rho + 8 \Gamma_\rho m_\pi /\pi
 \over m^2_\rho-q^2+(4 m^2_\pi -q^2) \Gamma_\rho[\alpha(q^2) -i\beta(q^2)]/m_\pi 
 }$$ with $  \alpha(q^2)$ and $\beta(q^2)$ being {\em adhoc} functions. The value of the fitted phase was
 found to be  $\theta \sim 53^\circ$
 in Ref.~\cite{Iac04}, and  $\theta \sim 22.7^\circ$ 
 in Ref.~\cite{Bij04}, where the proton data at the largest $q^2$ values were not included.
 In the upper  panel of Fig.~(\ref{fig:iachello}), the theoretical magnetic FF's,
 $|G_{M_p}(q^2)|$ and 
 $|G_{M_n}(q^2)|$, are presented
 and compared with  the experimental data, that correspond to the 
 {\em effective} nucleon FF's (see Eq.
(\ref{geff})
and the relative caveat). In  the lower panel, the  TL electric  
FF's are also shown.
 Unfortunately, as illustrated in the case of $|G_{M_p}(q^2)|$, the model does not 
contain the  
structure in resonances shown by the TL data for $q^2> 4 M^2_N$ (the adopted
VMD stops at the $\phi$-meson), and therefore the
comparison with the experimental data is not satisfactory.  As to the neutron, it should be
noted the large value reached at the threshold, in agreement with the FENICE
 measurements~\cite{Fenice}. 
\begin{figure}[t]
\begin{center}
\includegraphics[width=12.0cm,angle=-90]{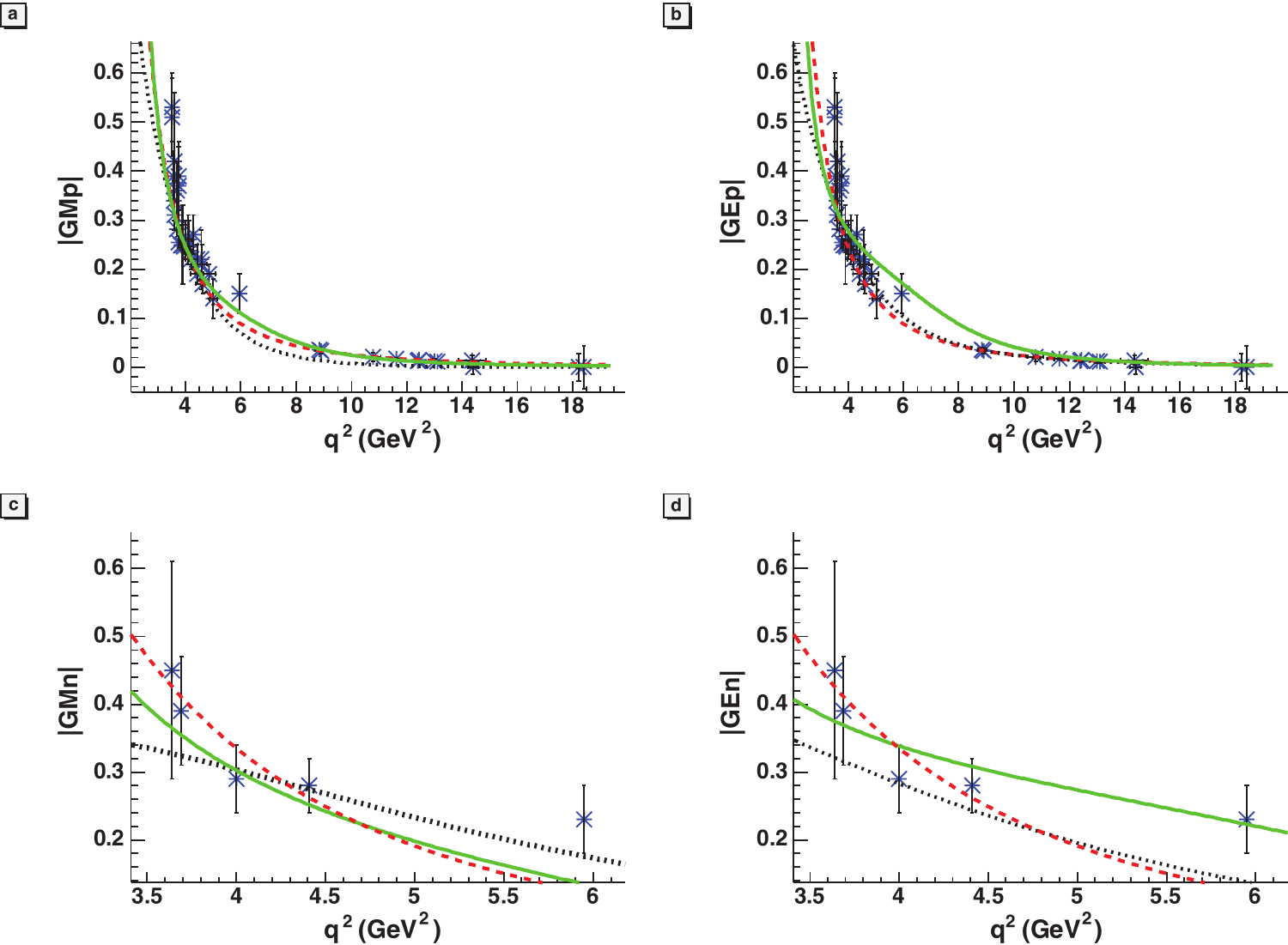}
\caption{ TL nucleon FF's evaluated  within the approach
 of Tomasi-Gustafsson
et al.~\cite{Tomasi05}. Dashed lines:  
pQCD-inspired model given in  Eq.~(\ref{tompqcd}). Dotted
lines: TL extension of the SL VMD model of Ref.~\cite{Iac73}.   
Solid  lines: as the dotted lines, but for the 
 SL model of Ref.~\cite{Lomon02}. Notice that the experimental data corresponding to the {\em effective} FF given in Eq.
~(\ref{geffa}), are separately compared to the magnetic and electric Sachs FF's.
Figure reprinted from Ref.~\cite{Tomasi05} with kind permission of The European 
Physical Journal (EPJ).
}\label{fig:tomasi}
\end{center}
\end{figure} 

In the same spirit as the previous approach,  in Ref.~\cite{Tomasi05}   
the TL generalizations of  both the SL model of Ref.~\cite{Iac73} and 
the SL model of Ref.~\cite{Lomon02} 
have been  elaborated. 
In the last case, the 14 parameters present in the SL nucleon FF's 
have been updated by  fitting the most recent  SL and TL nucleon FF's 
data-base (see Ref.~\cite{Tomasi05} for
the full list). In particular,
the generalization to the TL region was performed exploiting  the following relations
\be
Q^2 =-q^2=q^2~e ^{-i\pi} \to \left \{ \begin{array}{c} ln~Q^2=ln [q^2] -i\pi\\
\sqrt{Q^2}=~e ^{-i\pi/2}~\sqrt{q^2} \end{array} \right.
\label{tomcont}\ee
 Together with the above mentioned models, the analytical continuation of a pQCD-inspired model was also proposed,
based on counting rules and following the dipole behavior
  in the SL region. Such a model becomes in the    TL region 
\be
|G_{M_n}(q^2)| = {A(N)\over  q^4 ln^2(q^2/\Lambda^2)}
\label{tompqcd}\ee
 where $\Lambda=0.3~{\rm GeV}$ is the QCD scale parameter,  $A(p)=56.3~{\rm GeV^4}$ and  
$A(n)=77.15~{\rm GeV^4}$ are extracted from the fitting procedure to the TL data. The
results of the approach of Ref.~\cite{Tomasi05} are presented in Fig.~(\ref{fig:tomasi}). Notice that the SL model of Ref.~\cite{Lomon02} contains two
extra vector mesons with masses: $m_{\rho'}=1.45$ GeV and $m_{\omega'}=1.419$ GeV.
Unfortunately, also in this case, the TL data, corresponding to the {\it effective} nucleon FF's, given by Eq.
(\ref{geffa}), have been compared with both $|G_{M}^N|$ and  $|G_{E}^N|$, separately.
 In view of this, the comparison proposed in Fig.~(\ref{fig:tomasi}) 
 should be carefully considered at large
values of $q^2$.

Very recently, within the analytic continuation framework, 
an updated version of the FF's  of Ref.~\cite{Lomon02} 
have been applied in Ref.~\cite{LomonTL}
 for obtaining 
a unified fit of both SL and TL FF's.

\subsection{Nucleon FF's within a Relativistic Constituent Quark Model}
\label{RCQM}
A more microscopic description  of the nucleon FF's, in both SL and TL
regions, was  proposed in Ref.~\cite{Demelo}, where  it was adopted a   relativistic 
Constituent Quark Model (RCQM),  already successfully applied to the
 pion FF in
the whole kinematical range~\cite{CQMpion}.  The aim of the approach is 
 to calculate  
  as many
quantities as possible in terms of quark degrees of freedom and to  perform
a direct evaluation of the SL and TL FF's,
without taking care of the DR constraints. At the end, after fixing only 
four  adjusted parameters 
by using the SL experimental data, one can  predict the proton and
neutron  FF's in the TL region. In  Ref.~\cite{Demelo}, the following 
 ingredients  were chosen: i) the Mandelstam formula~\cite{Mandel55}
for calculating the matrix elements of the current operator  in
 terms of the nucleon 
Bethe-Salpeter amplitude, 
  ii) a microscopic relativistic
Hamiltonian that is able to describe reasonably well the isovector and isoscalar 
vector meson spectra~\cite{TobVM}, to
be included in the VMD part of the quark-photon vertex, iii) a quark-photon vertex decomposed in a bare
term and a VMD term, iv) a Light-front (LF)
framework~\cite{Brodsky98}, for embedding the previous ingredients and obtaining
a  relativistic (LF-boost invariant) CQM. 

In the RCQM of Ref.~\cite{Demelo}, the quark-nucleon vertex  is described through the Bethe-Salpeter amplitude, 
with a properly symmetrized 
Dirac structure,  approximated by
\be\Phi^s(k_1,k_2,k_3,p) 
=~\imath 
\left [~S(k_1)~\tau_y ~ \gamma^5 ~ S_C(k_2)C ~\otimes~ S(k_3)~+
~S(k_3)~ \tau_y ~ \gamma^5 ~S_C(k_1)C ~\otimes~ S(k_2)~+ \right. \nonu \left.
 +~S(k_3)~ \tau_y ~ \gamma^5 ~S_C(k_2)C~\otimes~S(k_1)
 \right ]  
 \Lambda(k_1,k_2,k_3) ~ \chi_{\tau_N} ~ u_s(p)
\label{qnvert}\ee
where 
  $\chi_{\tau_N}$ is the isospin spinor, $S(k_i)$ the Dirac propagator of the quark,
$S_C(k_i)=C~S^T(k_i)~C^{-1}=\gamma_5 S(k_i) \gamma_5$ with $C$ the charge 
conjugation and  $\Lambda(k_1,k_2,k_3)$ yields  the symmetrized
  dependence of the
vertex function upon the quark momenta,  $k_i$.

In the TL region (for the SL expression see Ref.~\cite{Demelo}),  
the matrix elements of the {\em macroscopic}  current can be evaluated
{\em microscopically} by means of the Mandelstam formula~\cite{Mandel55}, i.e.
\be
\langle {\cal N} (p,s)\bar{\cal N} (p',s')|
J_{em}^{\mu}(0)| 0\rangle=3N_c
\nonu
\times~ 
\int {d^4k_1 \over (2\pi)^4}\int {d^4k_2 \over (2\pi)^4} ~ 
Tr\left \{S^{-1}_C(k_2) \bar \Phi^{s'}_N(k_1,k_2,k'_3,p')
 ~ S^{-1}(k_1) ~{\cal I}^\mu(k_3,q)~
 \Phi^s(k_1,k_2,k_3,-p)\right \}
\ee
where in $\Phi^s(k_1,k_2,k_3,-p)$ appears $u_s(-p)\equiv v_s(p)$, ${\cal I}^\mu(k_3,q)$ 
is the quark-photon vertex, $N_c$ the number of colors and the 
trace  runs over isospin
and Dirac indexes. Notice that $k_1+k_2+k_3=p$ and $k_1+k_2+k'_3=-p'$.

The quark-photon vertex  is decomposed as follows
$$
  {\cal I}^\mu=~{\cal I}^\mu_{IS} +\tau_z
  {\cal I}^\mu_{IV} ~~~,$$
 where $\tau_z$ is the third component of the isospin operator and
  each term has  a purely valence contribution. acting 
  in the SL region only, and a contribution corresponding to the 
 em  $q\bar q$-pair production (non valence contribution). 
  In turn, the pair-production
    contribution, i.e. the relevant one in the TL region,
     can be decomposed in a bare term plus a 
  VMD term (according to    the photon-state expansion given in Eq.
 ~(\ref{phwf})). For instance, in the SL region, for a CQ  with momentum $k^\mu$ one has 
\be
 {\cal I}^\mu_{I}(k,q) = {\cal N}_{I} \underbrace{\theta(p^+-k^+)\theta(k^+)
   \gamma^\mu}+   
  \underbrace{\theta({q}^+ + k^+)
 \theta(-k^+)~\left \{{Z_B}~{\cal N}_{I}  \gamma^\mu+ 
   {Z^I_{VM}~}\Gamma^\mu_I(k,q)\right\} }\nonu
   \hspace{3.5 cm} valence \hspace {4.5 cm} non~valence
 \label{VMD2}  \ee
  where $k^+=k^0+k^3$ is the so-called plus component of the quark 
  four-momentum (see Ref.~\cite{Brodsky98} for a detailed review of the
  LF framework), 
   $I = IS, IV$, ${\cal N}_{IS}=1/6$ and ${\cal N}_{IV}=1/2$. 
  It has to be pointed out that the constraints
 $\theta(k^+)$ and $\theta(-k^+)$ single out a quark or an antiquark,
 respectively. The constants in the non valence term,
 ${Z_B}$ (bare term) and ${Z^I_{VM}}$ (VMD term), are unknown weights 
  determined from
 a fit to  the SL data-base (indeed, in the actual fitting procedure
 $Z_B=Z_{VM}^{IV}$ was
 assumed, see below). 
 
\begin{figure}[thb]

\centering\includegraphics[width=11cm,]{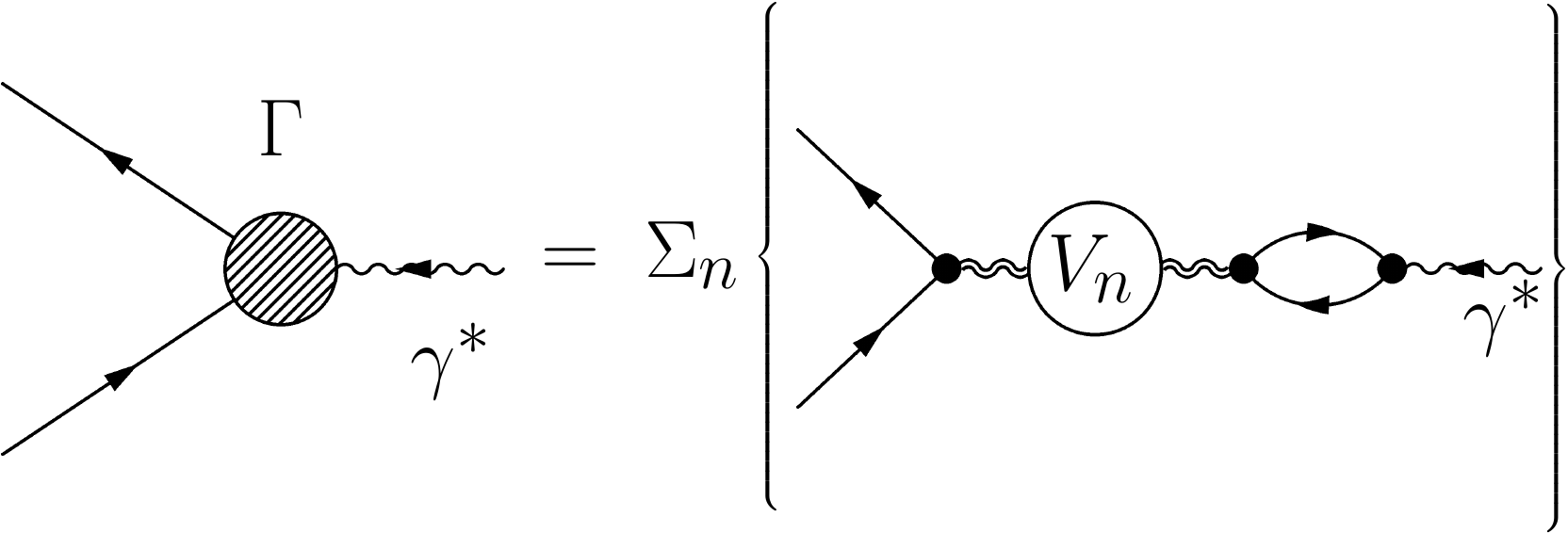}

\caption{VMD contribution to the quark-photon vertex. The double-wiggly
lines represent the Green function describing the
 propagation of the n-th  VM. The loop on the right indicates
 the decay constant of the n-th VM. The LF-time
 flows from the right toward the left. Figure reprinted with permission from 
 Ref.~\cite{CQMpion}, \copyright 2006 American Physical Society.} 
\label{figcqmvm}
\end{figure}

 In Eq.~(\ref{VMD2}),  $\Gamma^\mu_I(k,q)$ yields  the VMD contribution to the quark-photon vertex
 and it has the same structure of  the one 
 adopted in the pion case~\cite{CQMpion},
 but   including the isoscalar  vector mesons. For a given isospin channel, $\Gamma^\mu_I(k,q)$ 
 is approximated as
 pictorially illustrated
 in Fig.~(\ref{figcqmvm}) (cf.\ Ref.~\cite{CQMpion} for more details in the pion case), viz
 \be
\Gamma^{\mu}_I(k,q) = \sqrt{2} \sum_{n, \lambda}
\left [ \epsilon_{\lambda} (P_{I;n})\cdot \widehat{V}_{n}(k,k-P_{I;n})  \right ]
\Lambda_{I;n}(k,P_{I;n}) ~ { [\epsilon ^{\mu}_{\lambda}(P_{I;n})]^* f_{I;n} \over \left [ q^2 -
m^2_{I;n} + \imath ~m_{I;n} \tilde{\Gamma}_{I;n}(q^2)\right ]} \ ,
\label{cur7}
\ee
where $\epsilon_{\lambda}(P_{I;n})$  is
the VM polarization,
 $f_{I;n}$ is the em decay constant of the $n$-th vector
meson, $m_{I;n}$ the corresponding
 mass, $\Lambda_{I;n}(k, P_{I;n})$ gives the momentum dependence of the VM vertex function
and $\widehat{V}^\nu_{I;n }(k,k-P_{I;n})$ is the corresponding Dirac structure,
while $\tilde{\Gamma}_{I;n}(q^2)$ is the running  total decay width. It has to point out that in 
  Eq.~(\ref{cur7})  all the quantities appearing in the numerator have been 
  approximated through the corresponding values calculated
  at the respective pole (i.e. the residue at $q^2= P^2_{I;n}=m^2_{I;n}$), obtaining, therefore, 
  the correct result when the contribution is maximum. In turn, 
  such an approximation allows one to evaluate $\left [ \epsilon_{\lambda}(P_n) \cdot \widehat{V}_{n}(k,k-P_n)  \right ]  ~
\Lambda_{n}(k,P_n)$, (our Ansatz for the  VM vertex function, i.e. the
  amputated VM 
   Bethe-Salpeter amplitude),
   by using the eigenvalues and the eigenvectors of the mass operator  
for  vector mesons, proposed in
Ref.~\cite{TobVM}.

The running total decay width in the denominator of Eq.~(\ref{cur7}), $\tilde{\Gamma}_{I;n}(q^2)$, 
is vanishing in the SL
 region, while in the TL region it is assumed to be 
\be
\tilde{\Gamma}_{I;n}(q^2) = \Gamma_{I;n} ~ \left[ { q^2 - 4 m_\pi^2\over m_{I;n}^2 - 4 m_\pi^2} \right]^{3/2} 
\left[ { m_{I;n}^2 \over q^2} \right]^{1/2}
\label{Gamma} 
\ee 
where  $\Gamma_{I;n}$  is the total decay width (see, e.g., PDG\cite{PDG06}).

Once the on-mass-shell approximation has been introduced in Eq.~(\ref{cur7}),  
  $\Gamma^{\mu}_I(k,q)$  is not anymore a four vector,
  but since only the plus component $\Gamma^{+}_I(k,q)=\Gamma^{0}_I(k,q)+\Gamma^{3}_I(k,q)$ has to be  used in the evaluation of 
  the nucleon FF's, 
  it is possible to generate matrix elements of the current
  operator that correctly transform under the LF-boosts (these boosts act on the plus component 
  of a given four-vector without
  mixing other components, differently from the standard boosts).

 Up to 20 isoscalar and isovector VM's have been considered, since in the case of the pion
 that number of isoscalar mesons was necessary in order to get fully  convergent
 results for the TL pion FF  at the largest experimental values of $q^2$ 
 ($q^2 \sim 10~{\rm (GeV/c)^2}$). The first four (three) meson
 masses for the isovector (isoscalar) channel are taken from PDG~\cite{PDG06} together with the 
 corresponding 
 total decay widths $\Gamma_{I;n}$, while all the other masses come from the solutions of 
 the mass eigenequation~\cite{TobVM}, and  
  $\Gamma_{I;n}=0.150~{\rm GeV}$ is taken, as in the pion case. Noteworthy, all the em decay constants
   $f_{I;n}$  are calculated from
  the corresponding eigenfunctions, and they allow  the evaluation 
  of  the em decay widths $\Gamma^{I;n}_{e^+e^-}$, that result  in a very nice agreement with the
  experimental values, when available.
  Finally, as in the case of the pion, the quark mass $m_q = 200 ~ MeV$ is
  adopted. In conclusion, the VMD part of
  the photon-quark vertex is  fully constrained, a part the relative weights $Z^{IS}_{VM}$ and $Z^{IV}_{VM}=Z_B$.

The momentum dependence of the nucleon vertex function in Eq.~(\ref{qnvert}) is 
  differently modeled in
the valence region  (necessary for evaluating the SL nucleon FF's) and in the non valence one. 
In the valence sector,  
 the spectator quarks are on their-own
mass-shell, and the 
3-momentum dependence,  
is approximated through a
nucleon wave function \`{a} la Brodsky-Lepage (pQCD inspired)~\cite{Brodsky98}, given by
\be 
 {\Lambda_{val}(k_1,k_2,k_3) \over [M_N^2 - M^2_0(1,2,3)]} = 
   ~  {\cal {N}}~
{~(9~m^2)^{7/2}
 \over (\xi_1\xi_2\xi_3)^{p}~ \left[\beta^2 + M^2_0(1,2,3)\right]^{7/2}}
\label{val}
\ee
where $M_0(1,2,3)$ is the free mass of the three-quark system,  
$\xi_i = k^+_i/P^+_N$ 
and  ${\cal N}$  a normalization constant, fixed through the normalization of the proton charge FF.
The power  $ 7/2 $ and the parameter  $ p = 0.13 $   are chosen for obtaining
an asymptotic decrease of the valence  contribution to the SL nucleon FF's faster than the dipole.
 The value
$\beta = 0.65 ~GeV$ can be fixed through the evaluation of the anomalous
magnetic moments, $\mu_p$ and $\mu_n$, obtaining for those quantities,
$\mu_p= 2.87$  (Exp.= 2.793) and  
 $\mu_n=   -1.85$  (Exp.= -1.913).

The non valence vertex, that is fundamental for obtaining a zero in the SL ratio
$\mu_p G_{E}^p(Q^2)/ G_{M}^p(Q^2)$ (cf.\ Fig.~(\ref{fig:slffp})) and  essential for evaluating 
the TL FF's,
 can depend on the available invariants,
e.g.  in the SL region one has  i) the free mass of the spectator quarks, $M_0(1,2)$, 
and ii) the free mass of the N-$\bar {q}$  system 
$M_0(N,\bar {3})$.
Then, in the SL region, the non valence vertex  is approximated by
\be
\Lambda_{NV}^{SL}(k_1,k_2,k_3) = [g_{12}]^{2}~
[g_{N\bar {3}}]^{7/2-2}~
 \left [{k_{12}^+ \over P^{\prime +}_N }  \right ]
  \left [ P^+_N \over k_{\overline {3}}^+ \right ]^r
  \left [{P^{\prime +}_N  \over  k_{\overline {3}}^+ }  \right ]^{r} 
\label{nval1}\ee
where $
 k_{12}^+ = k_1^+ +  k_2^+$and $g_{AB} = (m_A ~ m_B) / \left
[\beta^2+M^2_0(A,B)\right]$. The power $r$ is a free parameter.
In the TL region,  
 the non valence vertex can depend on the mass  
of the nucleon-$\bar q \bar q$ system, $M_0(N, \overline {12})$, and it is given
by
  \be
    {\Lambda}^{TL}_{NV}= 2
  [g_{ \bar 1,\bar 2}]^{2}[g_{ N, \overline {1 2}}]^{7/2-2}    
 \left [{- k_{12}^+ \over P^+_{\bar N} }\right ] 
 \left [P^+_{\bar N} \over {k'}^+_3 \right ]^r 
 \left [P^+_{ N }\over {k'}^+_3 \right ]^r
\label{offtl}
   \ee
where the factor  of $2$ counts the possible patterns for the gluon emission by the
produced $q \bar q$ pair.  

\begin{figure}[tb]

\centering{
\includegraphics[width=8cm]{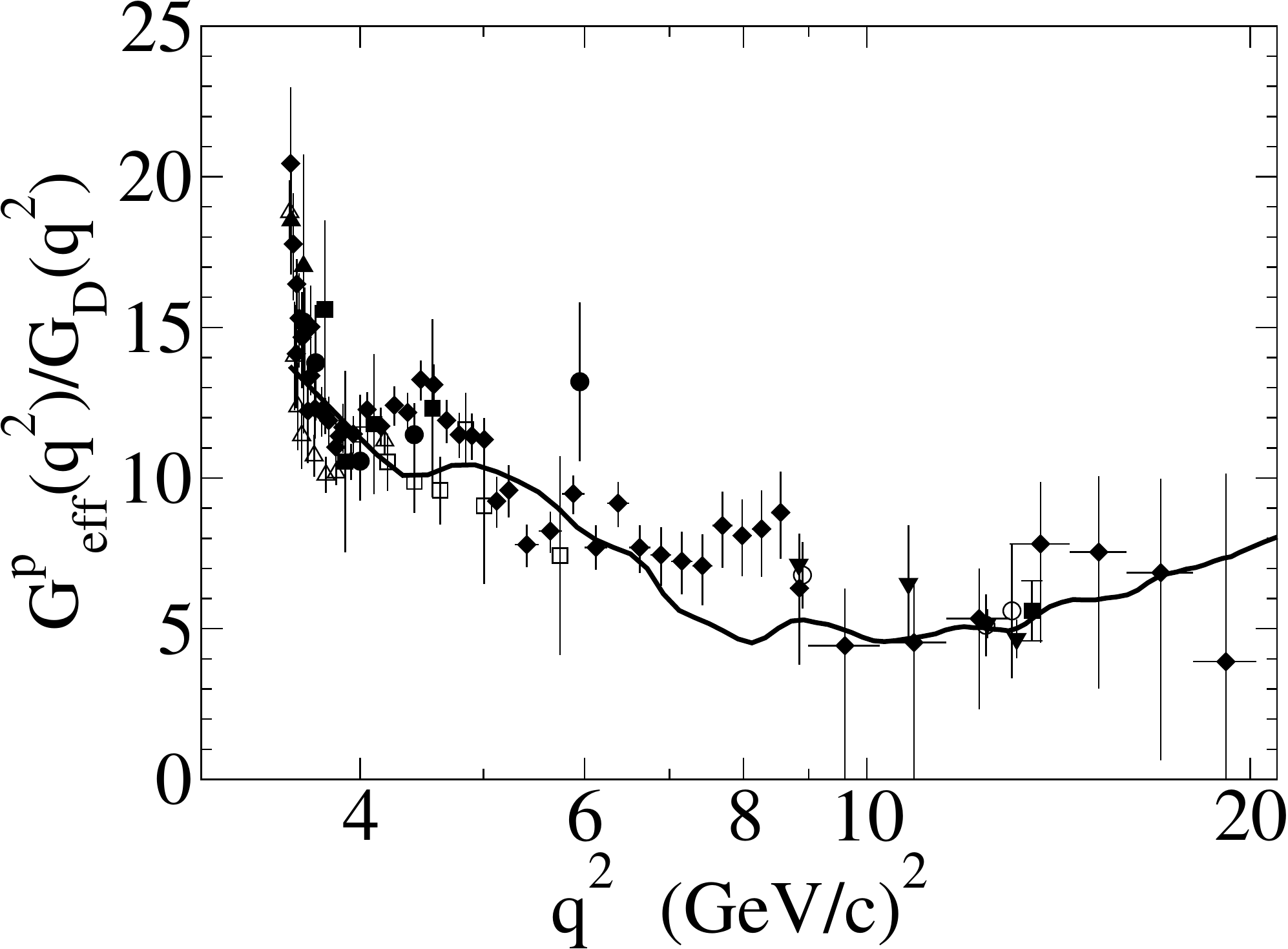} $~~~$
\includegraphics[width=8cm]{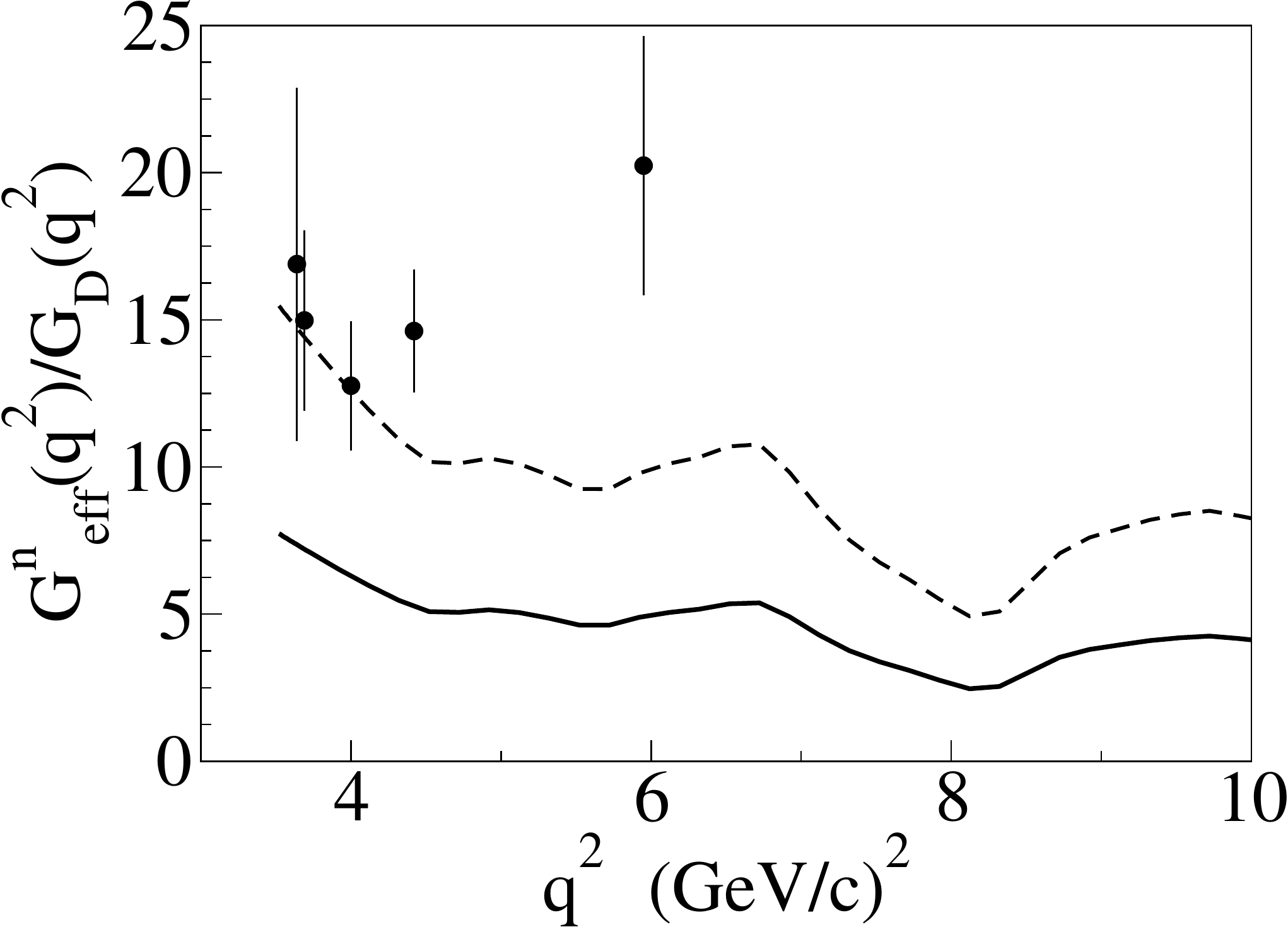}}
\caption{Experimental nucleon {\em  effective} form factors, Eq.~(\ref{geffa}), in the TL region compared to
the predictions of 
 the relativistic constituent quark model of Ref.
\cite{Demelo}, evaluated according to Eq.~(\ref{geff}). Left panel:
proton FF (data from~\cite{BABAR} and references quoted therein); notice the logarithmic scale for $q^2$.
 Right panel: neutron FF (data from Ref. 
\cite{Fenice}). Solid line: RCQM full calculation. 
Dashed line: neutron full calculation multiplied by an arbitrary factor of
two. 
Figure taken from Ref.~\cite{Demelo}.
}
\label{Nfftl}
\end{figure}

The free parameters determined through a fit to the available SL experimental data, $G^p_M(Q^2)$, 
 $G^n_M(Q^2)$, $\mu_p G^p_E(Q^2)/G^p_M(Q^2)$ and $G^n_E(Q^2)$, are four: i)
 two weights out three present in  the pair production term of  Eq.~(\ref{VMD2}), i.e.  
$Z_B = Z_{VM}^{IV} = 2.283 $  and 
$Z_{VM}^{IS} / Z_{VM}^{IV} = 1.12$; ii)
the power $p = 0.13$  in the valence amplitude, Eq.~(\ref{val});
iii) the power $r = 0.17$ in the SL and TL non valence vertexes, see e.g. Eqs.~(\ref{nval1})
and (\ref{offtl}). The minimization yields
$\chi ^2$ = 1.7. The overall quality of the SL fit can be appreciated from the comparisons 
 in Fig.~(\ref{fig:slffp}), and  from 
 both the calculated  proton charge radius 
$r_p = (0.903 \pm 0.004) ~ fm$ ($r_p^{exp} = (0.895 \pm 0.018) ~ fm$),  and 
the neutron charge slope at $q^2=0$,
 $-d G_E^n(q^2)/ d q^2  ~= (0.501 \pm 0.002) ~ 
{\rm (GeV/c)^{-2}} $ (exp value; $
 -d G_E^n(q^2)/ d q^2  = (0.512 
\pm 0.013) ~ {\rm (GeV/c)^{-2}}$). Surprisingly, the same quality in the fitting procedure can be
achieved by excluding the $\mu_p G^p_E(Q^2)/G^p_M(Q^2)$ data \cite{Demelo}. In Fig.~(\ref{Nfftl}), the experimental 
  {\em effective} FF's (see Eq.~(\ref{geffa})) for both proton and neutron are
compared with the RCQM results (evaluated according to Eq.~(\ref{geff})), that 
are true predictions of the model.
It is worth noting 
that a microscopic calculation could lead to the extraction of valuable information on the dynamical
content of a realistic RCQM. For instance, missing strengths at 
$q^2 = 4.5 ~ {\rm (GeV/c)^2}$ (nicely, the same feature appears in the  TL pion
FF~\cite{CQMpion})  and  
$q^2 = 8 ~ {\rm (GeV/c)^2}$, could point to an enrichment of the present VM data-base, and therefore to a strong
request of 
refining the RCQM approach in the VM sector. This simple example sheds
light on the potentially relevant results that one could achieve from the 
forthcoming accurate measurements
in the TL region and more and more improved    microscopic calculations.

\clearpage

\section{Conclusions sand  Perspectives}
\label{concl}

This review has illustrated  
the advancements  achieved so far, both experimentally and theoretically, in the 
investigation of the timelike electromagnetic form
factors of the nucleon. We have also presented the forthcoming experimental facilities, 
from which in the near future results are to be expected.

The development of the field, started with the pioneering experimental work 
carried out in Frascati~\cite{Adone73,Fenice} up 
to the forthcoming exciting experiments with high intensity $e^+e^-$ beams~\cite{vepp2000,bes3pb} as
well as -- hopefully polarized -- antiproton beams~\cite{PANDA,PAX} at FAIR, is demonstrating the
continuous interest in these challenging studies.
However, to measure separately the electric and magnetic  nucleon  FF's in the TL region is 
not an easy task, and  the present-day database is still given 
 in terms of an {\em effective } TL nucleon
FF, that for the electron-positron annihilation process reads (see Eq.~(\ref{geff})
and the corresponding remarks):
$$
|G_{\rm eff}^N (q^2)|=\sqrt{\sigma^{exp}_{e^+e^-\rightarrow N \bar{N}}(q^2)\over {4\pi\alpha^2 
\beta \over 3q^2} ~C_N(q^2)   \left [1
+\frac{2M^2_N}{q^2}\right ]}=\sqrt{q^2~|G_{M}^N(q^2)  |^2+ 2 M^2_N|G_{E}^N(q^2)|^2
\over q^2 +2M^2_N }
$$
where the rightmost-hand side shows the relations with the Sachs FF's in 
one-photon approximation. 

\begin{figure}[t]
\centering

\vspace{-1cm}
\includegraphics[width=12cm]{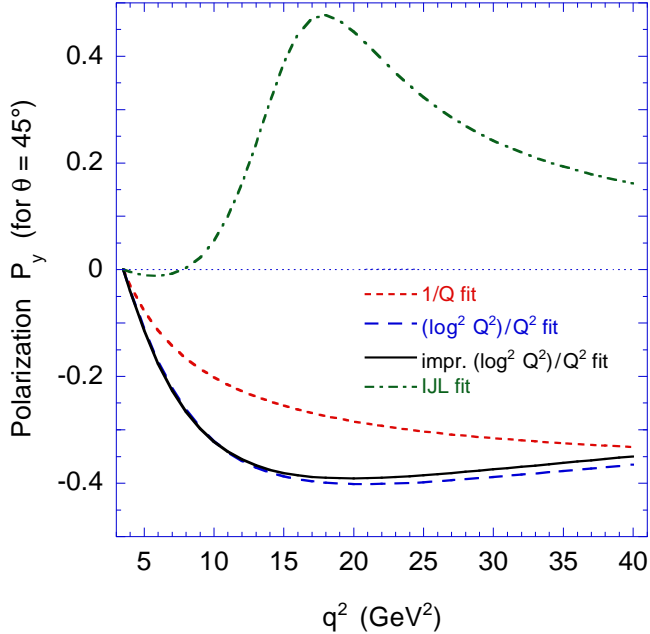} 

\vspace{-2cm}\caption{ Proton polarization, ${\cal P}_y$ (see Eq.~(\ref{polartl})), normal to the scattering plane 
calculated by using  different models. Dot-dashed line: the analytic
continuation of the SL proton FF of Iachello et al.~\cite{Iac73,Iac04}. The other
curves represent the analytic continuation of fits to the SL ratio
$F_1^p/F_p^p$, as indicated in the legend. Figure reprinted with 
permission from 
Ref.~\cite{Brod04}, \copyright 2004 American Physical Society.}
\label{Py_Brod04}
\end{figure}

The first success in the experimental investigations was the study of the total
cross section for the
reaction $e^+e^-\rightarrow N \bar{N}$ by exploiting the energy scan technique at
electron-positron colliders. 
In this respect, a major improvement was achieved
by introducing the initial state radiation technique, i.e.\ by using events like
$e^+e^-  \rightarrow N \bar{N} \gamma$ (cf.\ Subsect.~\ref{isr}). It was possible via ISR to
obtain a measurement of the proton form factor with so far unprecedented precision and
covering a very wide $q^2$ range.
It should be pointed out that the ISR technique is of great appeal 
also for the planned high-luminosity $e^+e^-$ B-factories, which are mainly devoted to the search of
New-Physics signatures in B-decays. The study of the process
$\sigma^{exp}_{e^+e^-\rightarrow N \bar{N}}$, lead to single out 
 three energy regions: i) the region around the production 
threshold, $q^2 \sim 4M^2_N$, that yields valuable information on the transition
from unphysical to physical regions (cf. Subsects. \ref{thresh} and
\ref{tlextre}); ii) the region where the Vector Meson Dominance and the 
interplay 
with the multi-cut analytic structure of the FF's fully display their 
 phenomenological strength in the studies of the non-perturbative regime of
QCD; iii) the asymptotic region where the realm of the perturbative QCD 
settles and the application of the Phragm\'en-Lindel\"off theorem is
representing a fundamental constraint for calculations based on microscopic models. 
As shown in the theoretical interpretation of the 
experimental  {\em effective} FF, (let us recall that such an   interpretation 
is driven by
the  one-photon approximation and that the  
two-photon   corrections seem to be small in TL region 
\cite{Tomasi08,Chen08,Adam07}), the electric and magnetic Sachs
FF's are both present in a {\em proper combination} (cf.\ the rightmost side of
the above expression). As previously mentioned, experimental 
efforts already have been carried out~\cite{PS170_94,BABAR} and will be
pursued in near future with the goal
to fully disentangle the electric and the magnetic FF's. To this end,
extending the measurements of the angular distributions (see~\cite{BABAR} for 
recent experimental achievements for the proton case) could provide 
a tool for determining the
ratio
$|G_E^N/ G_M^N| $. The {\em holy grail} in the field of TL nucleon FF's, 
however, is the investigation of the phases of the complex-valued FF's. This is 
 heuristically illustrated in Fig.~(\ref{Py_Brod04}), where the 
discrimination power of an accurate measurement
of the proton polarization perpendicular to the scattering plane, is shown.
Notice that $P_y$ does not depend upon
the polarization of the incoming electron beam, and once a polarimeter for the
measurement of the proton polarization will be included into the experimental
apparatus at an electron-positron collider, this measurement of $P_y$ could be
performed.
The experimental difficulties for achieving such a goal are known and are related to the
  measurement of the polarization degrees of
freedom, but the strong motivations to pursue such a goal were already 
contained 
 in the seminal work of
Ref.~\cite{Feder58}, where the bridge between the TL FF's , in particular
$\Imf\{F_{1(2)}^N\}$, and the hadronic amplitudes of hadronic {\em physical processes}
 was
established. Therefore, when measurements of $\Imf\{F_{1(2)}^N\}$ will be
available, the possibility to extract accurate information on both the em decays
of vector mesons and the hadron production (with given quantum numbers) via $N
\bar N$ should allow a carefully study of the hadronic states.   
 From the theoretical side, the starting point of this program has been 
 the analytic behavior of the complex-valued nucleon FF's, as imposed by
the causality, and the consequent dispersion relations (cf.\ Subsect.
\ref{DR}), that link the real and imaginary parts of the
nucleon FF's. 
The  
Lehmann-Symanzik-Zimmermann reduction formula made it possible, to express
$\Imf\{F_{1(2)}^N\}$ 
   in terms of the above mentioned hadronic amplitudes, that
 one can study in other experiments, since they are involved in physical (not
 virtual) processes. The first phenomenological attempts to model 
 $\Imf\{F_{1(2)}^N\}$, as shown in Subsects. \ref{TLDR}, \ref{tlextre} and 
 \ref{tlcont},   were based on 
 exploiting experimental inputs, like masses and widths of
the vector mesons, but also i) introducing adjusted parameters for  implementing
 a fine
tuning and for describing the
hadronic free propagation (i.e. the continuum associated to the multi-cut structure of the 
TL FF's) and ii) assuming theoretical constraints, particularly for the
 asymptotic behavior dictated by pQCD. The next step should be a refinement of
 our phenomenological tools,  extending  and enriching the
 microscopic calculations, like e.g.\ the one of Refs.~\cite{CQMpion,Demelo}, which are
  based on a LF Hamiltonian for 
describing the VM spectra and successfully applied to both the pion and the
nucleon. Another appealing approach is the one based on the Anti de Sitter/Conformal Field
Theory framework, where a  very nice description of  both SL and TL 
pion FF has been achieved~\cite{AdS}.
 In this way, by developing models with less and less phenomenological inputs,
we could link the measurements of $\Imf\{F_{1(2)}^N\}$ to QCD-based calculations.

In summary, a detailed comparison between
 accurate experimental data, in particular when both absolute values and phases
of the TL nucleon  FF's will become available,  and microscopic approaches 
will  allow to gain  deep insights
into the non-perturbative regime of QCD.

\section{Acknowledgments}
 We gratefully acknowledge illuminating 
discussions and helpful remarks from Henryk Czy{\.z}, 
Vladimir Druzhinin, Tobias Frederico, Frank Maas, Simone Pacetti, Carlo Presilla, and Marc Vanderhaeghen.
This work is supported by the Deutsche Forschungsgemeinschaft through the Collaborative
Research Center ``The Low-Energy Frontier of the Standard Model'' (SFB 1044).

\appendix 
\section{Matrix element of the conjugated current}
\label{matrix}
In this Appendix, the steps for obtaining the equivalent expression given in Eq.
(\ref{matrix0}), but for 
the matrix elements
  of $\langle 0|
\left[J_{em}^{\mu}(0)\right]^\dagger| {\cal N} (p,s)\bar{\cal N} (p',s')\rangle=
 \langle  {\cal N} (p,s)\bar{\cal N} (p',s')|
J_{em}^{\mu}(0)| 0\rangle^*$ are shown. One gets
\be 
\sqrt{{EE'\over M^2_N} } (2\pi)^3~\langle  {\cal N} (p,s)\bar{\cal N} (p',s')|
J_{em}^{\mu}(0)| 0\rangle^*=\nonu=
\sqrt{{E\over M_N } } (2 \pi)^{3/2} \left[i ~ \int d^4x~e^{i p'\cdot
x}~\theta(t)~ { 1\over  (2\pi)^4}\sum_{I\ne nucleon}  
 \langle {\cal N} (p,s)| \bar S_N(x)|I,B=0 \rangle
\right. \nonu \left.
 \times~\langle I,B=0| J_{em}^{\mu}(0)| 0\rangle   ~v_{s'}(p') 
\phantom{{ 1\over  2\pi}}\!\!\!\!\!\right]^*
=\nonu=
\sqrt{{E\over M_N } } (2 \pi)^{3/2} (-i) ~ \int d^4x~e^{-i (p'+p-p_I)\cdot
x}~\theta(t)~ { 1\over  (2\pi)^4}\sum_{I\ne nucleon}  
 ~\nonu
 \times~\bar v_{s'}(p')~
\langle 0| J_{em}^{\mu}(0)| I,B=0\rangle  \langle I,B=0| S_N(0)| {\cal N}
(p,s)\rangle
=\nonu=
\sqrt{{E\over M_N } } (2 \pi)^{3/2} (-i) ~{i\over 2 \pi}\int d\omega
{e^{-i\omega t}\over \omega +i\epsilon}~ \int d^4x~e^{-i (p'+p-p_I)\cdot
x}~
 { 1\over  (2\pi)^4}\sum_{I\ne nucleon}  
 ~\nonu
 \times~\bar v_{s'}(p')~
\langle 0| J_{em}^{\mu}(0)| I,B=0\rangle  \langle I,B=0| S_N(0)| {\cal N}
(p,s)\rangle
=
\nonu=
\sqrt{{E\over M_N } } (2 \pi)^{3/2} { 1\over  2\pi}\sum_{I\ne nucleon}  
 \delta^3({\bf p}_I-{\bf p}'-{\bf p} )~\left[{ {\cal P} \over E_I-E'-E} - i\pi 
 \delta(E_I-E'-E)\right]~\nonu
\bar v_{s'}(p')~
\langle 0| J_{em}^{\mu}(0)| I,B=0\rangle  \langle I,B=0| S_N(0)| {\cal N}
(p,s)\rangle 
\ee


\begin{thebibliography}{99}
\itemsep -2pt 
\bibitem{Hofst} R.W. McAllister and  R. Hofstadter, {\em Phys. Rev.} {\bf 102}, 851 (1956)
\bibitem{SLrev} C. E. Hyde-Wright and  K. de Jager, {\em Ann. Rev.  Nucl. and Part. Sci.} {\bf  54}, 217 (2004); 
C.F. Perdrisat, V. Punjabi, and M. Vanderhaeghen, {\em Progr. Part. Nucl. Phys.} {\bf  59}, 694 (2007); 
C.E. Carlson, {\em  Eur. Phys. J. Special Topics} {\bf  198}, 65 (2011); 
J. Arrington, K. de Jager, and C. F. Perdrisat, {\em Jou. Phys. Conf. Ser.} {\bf  299}, 012002 (2011).
\bibitem{VMD} D. Schildknecht, {\em Acta Phys.Polon.} B {\bf  37},595 (2006).

\bibitem{Feder58} P. Federbush, M. L. Goldberger, and S. B. Treiman, {\em  Phys. Rev.}
{\bf  112}, 642 (1958).
\bibitem{AdS} S. J. Brodsky, and G. F. de Teramond, {\em Phys. Rev.} D {\bf  77},
 056007 (2008); arXiv:1203.4025.
 \bibitem{CQMpion} J.P.B.C. de Melo, T. Frederico, E. Pace, and G. Salm\`e, 
{\em Phys. Lett.} B {\bf  581}, 75 (2004); {\em Phys. Rev.} D {\bf  73},
074013 (2006).
\bibitem{Demelo} J.P.B.C. de Melo, T. Frederico,  E. Pace, 
S. Pisano, and G. Salm\`e,   
{\em  Phys. Lett. } B {\bf  671}, 153 (2009).
\bibitem{LQCD} C. Alexandrou et al, {\em Phys. Rev.} D {\bf 83}, 114513 (2011).

{\em Phys. Rep.} {\bf  301}, 299 (1998).
\bibitem{BABAR} B. Aubert et al. (BaBar Collaboration), {\em  Phys. Rev.} D 
{\bf  73}, 012005 (2006).
\bibitem{PS170_94} G. Bardin et al. (PS170 Collaboration), {\em  Nucl. Phys.} B {\bf 411}, 3 (1994).





\bibitem{Adone73} M. Castellano et al., {\em Nuovo Cimento } {\bf 14} A, 1 (1973).

\bibitem{Fenice1} A. Antonelli et al. (Fenice Collaboration),  {\em Phys. Lett.} B {\bf  334}, 431 (1994).
\bibitem{Fenice} A. Antonelli et al. (Fenice Collaboration), {\em  Nucl. Phys. } B {\bf  517}, 3 (1998).

\bibitem{DM179} B. Delcourt et al. (DM1 Collaboration), {\em   Phys. Lett. } B {\bf  86}, 395 (1979).
\bibitem{DM283} D. Bisello et al. (DM2 Collaboration), {\em  Nucl. Phys.} B {\bf  224}, 379 (1983).
\bibitem{DM290} D. Bisello et al. (DM2 Collaboration), {\em  Z. Phys. } C {\bf  48}, 23 (1990).

\bibitem{BES05} M. Ablikim et al. (BES Collaboration), 
{\em  Phys. Lett. } B {\bf  630}, 14 (2005).
\bibitem{CLEO05} T.K. Pedlar et al. (CLEO Collaboration), {\em Phys. Rev. Lett.} {\bf  95}, 261803 (2005).

\bibitem{Fenice2} A. Antonelli et al. (Fenice Collaboration), {\em Phys. Lett.} B {\bf  313}, 283 (1993).

\bibitem{BNL65} M. Conversi et al., {\em Nuovo
Cimento} {\bf  XL}, 8926 (1965).

\bibitem{PS69} D. L. Hartill et al., {\em  Phys. Rev.} 
{\bf  184}, 1415 (1969).

\bibitem {PS7677} S. Bassompierre et al. (Mulhouse-Strasbourg-Torino Collaboration), 
{\em  Phys. Lett.} B {\bf  64}, 475 (1976); {\em  Phys. Lett.} B {\bf  68}, 477 (1977).

\bibitem{PS170_91} G. Bardin et al. (PS170 Collaboration), {\em  Phys. Lett.} B {\bf 255}, 149 (1991).
\bibitem{PS170_91b} G. Bardin et al. (PS170 Collaboration), {\em  Phys. Lett.} B {\bf 257}, 514 (1991).

\bibitem{E760_93} T. A. Armstrong et al. (E760 Collaboration), {\em  Phys. Rev. Lett.} {\bf  70}, 1212 (1993).

\bibitem {E835_99} M. Ambrogiani et al. (E835 Collaboration), {\em  Phys. Rev.} D {\bf  60}, 032002 (1999). 
\bibitem {E835_03} M. Andreotti et al. (E835 Collaboration), {\em  Phys. Lett.} B {\bf  559}, 20 (2003).

\bibitem{BABAR2} B. Aubert et al. (BaBar Collaboration), {\em Phys. Rev.} D {\bf  76}, 092006 (2007).

\bibitem{BELLE} G. Pakhlova et al. (BELLE Collaboration), {\em Phys. Rev. Lett.} {\bf  101}, 172001 (2008).

\bibitem{DM2plot} {\rm http://www.bo.infn.it/$\sim$castro/research/DM2/DM2.html}

\bibitem{zerwas} M.-S. Chen, and P.M. Zerwas, {\em Phys. Rev.} D {\bf  11}, 187 (1975).

\bibitem{eva} S. Binner, J.H. K\"uhn, and K. Melnikov, {\em Phys. Lett.} B {\bf  459}, 279
(1999).

\bibitem{isrbfactories}
M. Benayoun, S.I. Eidelman, V.N. Ivanchenko, and Z.K. Silagadze, 
{\em Mod. Phys. Lett.} A {\bf  14} 2605 (1999).

\bibitem{actis} 
S. Actis et al., {\em Eur. Phys. J.} C {\bf  66}, 585 (2010).

\bibitem{isrreview} V.P.Druzhinin, S.I. Eidelman, S.I. Serednyakov, and E.P. Solodov, 
{\em Rev. Mod. Phys.} {\bf  83}, 1545 (2011).

\bibitem{babar_2pi} B. Aubert et al. (BaBar Collaboration), {\em  Phys. Rev. Lett.}  
{\bf  103}, 231801 (2009).

\bibitem{babar_6pi} B. Aubert et al. (BaBar Collaboration), {\em  Phys. Rev.} D 
{\bf  73}, 052003 (2006).

\bibitem{babar_y4260} B. Aubert et al. (BaBar Collaboration), {\em  Phys. Rev. Lett.}  
{\bf  95}, 142001 (2005).

\bibitem{babar_y2175} B. Aubert et al. (BaBar Collaboration), {\em  Phys. Rev.} D 
{\em Rap. Comm.}
{\bf  74}, 091103 (2006).

\bibitem{baier} V.N. Baier, V.S. Fadin, {\em Phys. Lett.} B {\bf  27}, 223 (1968);
A.B. Arbuzov et al., {\em J. High Energy Phys.} {\bf 12} 009 (1998).

\bibitem{egle_fsr} E.A. Kuraev, V.V. Bytev, E. Tomasi-Gustafsson, and S. Pacetti,
{\em Phys. Rev.} D {\bf   84} 017301 (2011).

\bibitem{phokhara_ppbar}
H. Czy{\.z}, J. H. K\"uhn, E. Nowak, and G. Rodrigo,
{\em Eur. Phys. J.} C {\bf  35} 527 (2004).

\bibitem{phokhara_LLbar}
H. Czy{\.z}, A. Grzelinska, and J. H. K\"uhn,
{\em Phys. Rev.} D {\bf  75} 074026 (2007).

\bibitem{phokhara}
H. Czy{\.z}, A. Grzelinska, J. H. K\"uhn, and G. Rodrigo,
{\em Eur. Phys. J.} C {\bf  27} 563 (2003).

\bibitem{photos}
G. Navana, and Z. Was, 
{\em Eur. Phys. J.} C {\bf  51} 569 (2007).

\bibitem{eva4pi}
H. Czy{\.z}, and J. H. K\"uhn,
{\em Eur. Phys. J.} C {\bf  18} 497 (2001).

\bibitem{BELLE_B} K. Abe et al. (BELLE collaboration),
{\em Phys. Rev. Lett} {\bf  88} 181803 (2002); 
{\em Phys. Rev. Lett} {\bf  89} 151802 (2003).

\bibitem{BABAR_B} B. Aubert et al. (BaBar collaboration),
{\em Phys. Rev.} D {\bf  72}, 051101 (2005);
{\em Phys. Rev.} D {\bf  74}, 051101 (2006). 

\bibitem{BES_Psi} J.Z. Bai et al. (BES collaboration),
{\em Phys. Rev. Lett.} {\bf  91} 022001 (2003).

\bibitem{franzini} P.J. Franzini, and F.J. Gilman, 
{\em Phys. Rev.} D {\bf  32} 237 (1985).

\bibitem{6pi} P.L. Frabetti et al. (E687 collaboration),
{\em Phys. Lett.} B {\bf  514} 240 (2001);
M. Grilli et al. (DM2 collaboration), {\it Nuovo Cim.} 
{\bf  13}A, 593 (1973).



\bibitem{rosner} J.L. Rosner, {\em Phys.Rev.} D {\bf  74}, 076006 (2006).

\bibitem{vepp2000} E. Solodov, Proceedings of the XIV International Conference on Hadron 
Spectroscopy, Munich, June 13-17, 2011, edited by B. Grube, S. Paul, and N. Brambilla, 
arXiv: 1108:6174.

\bibitem{sff} Super KEK-B: Physics at Super B Factory, arXiv: 1003.5012;  
SuperB: A High-Luminosity Asymmetric $e^+e^-$ Super Flavor Factory. 
Conceptual Design Report, arXiv: 0709.0451.

\bibitem{bes3pb}  D.M. Asner et al., {\em Int. J. of Mod. Phys.} A {\bf  24}, Issue: 1 supp (2009). 

\bibitem{jpsinnbar} M. Ablikim et al. (BES-III collaboration),
accepted for  publication on {\em Phys. Rev.} D; arXiv: 1205.1036.

\bibitem{cristina} C. Morales-Morales, and F. Maas, Mainz, private communication.

\bibitem{KLOE}
A. Aloisio et al. (KLOE collaboration), {\em Phys. Lett.} B {\bf  606}, 12 (2005); 
F. Ambrosino et al. (KLOE collaboration), {\em Phys. Lett.} B {\bf  670}, 285 (2009). 

\bibitem{bes_tagger} 
R. Baldini Ferroli, S. Pacetti, and A. Zallo, 
{\em Nucl. Phys. Proc. Suppl.} {\bf  219-220}, 32 (2011).

\bibitem{fair} http://www.gsi.de/portrait/fair.

\bibitem{panda_pb} Physics performance report for PANDA, W. Erni et al. 
(PANDA collaboration), arXiv:0903:3905

\bibitem{PANDA} M. Sudol et al.,  {\em  Eur. Phys. J.} A {\bf  44}, 373 (2010).

\bibitem{panda_unphysical} 
C. Adamuscin, E.A. Kuraev, E. Tomasi-Gustafsson, and F. Maas, {\em Phys. Rev.} C {\bf  75}, 045205 (2007);
E.A. Kuraev, C. Adamuscin, E. Tomasi-Gustafsson, and F. Maas, {\em Phys. Lett.} B {\bf  649}, 400 (2007).

\bibitem{panda_tp} 	
G.I. Gakh, E. Tomasi-Gustafsson, {Nucl. Phys.} A {\bf  799} 127 (2008).

\bibitem{BF} S.J. Brodsky, and G.R. Farrar, {\em   Phys. Rev. } D {\bf  11}, 1309 (1975).

\bibitem{Iac73}
 F. Iachello, A.D. Jackson, and A. Lande, {\em Phys. Lett. } B {\bf  43}, 191
(1973).

\bibitem{Lomon02} E.L. Lomon, {\em  Phys. Rev.} C {\bf  66}, 045501 (2002).

\bibitem{Brod04} S.J. Brodsky, C.E. Carlson J.R. Hiller, and D.S. Hwang,
{\em Phys. Rev. } D {\bf  69}, 054022 (2004).
\bibitem{Bilenki93} S.M. Bilenkii, C. Giunti, and V. Wataghin, {\em Z. Phys.} C 
{\bf  59}, 475 (1993).

\bibitem{Tomasi05}E. Tomasi-Gustafsson, F. Lacroix, C. Duterte, and 
G. I. Gakh, 
{\em Eur. Phys. J.} A {\bf  24}, 419 (2005).
\bibitem{feher} M. Feher, F. Maas, Mainz, private communication.

\bibitem{PAX} H. Str\"oher, P. Lenisa, and F. Rathmann, {\em PoS} (STORI11) 030.

\bibitem{spinfilt} M.A. Leonova et al., {Phys. Rev. Lett.} {\bf  108}, 074801 (2012).  


\bibitem{Yennie} D. R. Yennie, M. M. L\'evy, and D. G. Ravenhall, {\em  Rev. Mod. Phys.} {\bf  29}, 144 (1957).

\bibitem{Hand} L.N. Hand, D. G. Miller, and Richard Wilson,  
{\em   Rev. Mod. Phys.} {\bf  35}, 335 (1963).

\bibitem{BD} J. D. Bjorken and S. D. Drell, "Relativistic Quantum Mechanics"
and "Relativistic Quantum Fields", Mac-Graw Hill Book Company, 91965).

\bibitem{esw}F. J. Ernst, R. G. Sachs and K. C. Wali,
{\em    Phys. Rev.} {\bf  119}, 1105 (1960).

\bibitem{Kelly02} J J. Kelly, {\em Phys. Rev.} C {\bf  66}, 065203 (2002).

\bibitem{Krutov09} A. F. Krutov and V. E. Troitsky, {\em Phys. Part.  Nuc.} {\bf  40}, 136 (2009).

\bibitem{Miller07} G. A. Miller, {\em   Phys. Rev. Lett. }{\bf  99}, 112001 (2007);
 {\em   Phys. Rev.} C {\bf  80}, 045210 (2009).

\bibitem{Diehl03} M. Diehl, {\em   Phys. Rep.} {\bf  388}, 41 (2003).

\bibitem{BL} S.J. Brodsky, and G.P. Lepage, {\em   Phys. Rev.} D {\bf  22}, 2157 (1980).

\bibitem{Akhiez58} A.I. Akhiezer, L.N. Rozentsveig,  and I.M. Shmushkevich, {\em
Sov. Phys. JETP} {\bf  6},588 (1958).
\bibitem{GW04} M. L. Goldberger, and K.M. Watson, 
 ''Collision Theory '', Dover Pub. Inc, 2004.
 
\bibitem{rekalo02} M.P. Rekalo, and Egle Tomasi-Gustafsson, 
arXiv:0202025v1 and references quoted therein.


 




\bibitem{Cabibbo61} N. Cabibbo and R. Gatto, {\em Phys. Rev. }{\bf  124},  1557 (1961).

\bibitem{Zich62} A. Zichichi, S.M. Berman, N. Cabibbo and  R. Gatto, {\em Nuovo
Cimento} {\bf  XXIV}, 170 (1962).
\bibitem{Druz11} V. Druzhinin, Novosibirsk, private communication. 
\bibitem{Brod95} S.J. Brodsky, A.H. Hoang , J.H. K{\"u}hn, T. Teubner,
{\em  Phys. Lett. } {B \bf   359}, 355 (1995).

\bibitem{Hoan97} A. H. Hoang, {\em  Phys. Rev.} D {\bf  56}, 7276 (1997).



\bibitem{Dubni96}
A. Z. Dubnickova, S. Dubnicka and M. P. Rekalo, {\em Nuovo Cimento} A  
{\bf  109}, 241 (1996).

\bibitem{Adam07} C. Adamuscin, G. I. Gakh, E. Tomasi-Gustafsson, arXiv:0704.3375v1 and references quoted
therein.

\bibitem{Arring11} J. Arrington a, P.G. Blundenb, W. Melnitchouk, {\em  Progr.  Part.  Nucl. Phys.} 
{\bf  66}, 782 (2011)

\bibitem{Tomasi08} E. Tomasi-Gustafsson, E.A. Kuraev, S. Bakmaev, S. Pacetti, {\em Phys. Lett.} B
 {\bf  659}, 197 (2008).

\bibitem{Chen08} D. Y. Chen, H. Q. Zhou, and Y. B. Dong, {\em  Phys. Rev.} C {\bf 78}, 045208 (2008).

\bibitem{Pire} T. Gousset and B. Pire, {\em Phys. Rev.} D {\bf 51}, 15 (1995).

\bibitem{Dalka10} O.D. Dalkarov, P.A. Khakhulin, and A.Yu. Voronin, {\em  Nucl. Phys.} A {\bf
 833}, 104 (2010).

\bibitem{Bruno09} B. El-Bennich, M. Lacombe, B. Loiseau, and S. Wycech, {\em   Phys. Rev. }
C {\bf  79}, 054001 (2009).

\bibitem{Haiden06} J. Haidenbauer, H.-W. Hammer, and Ulf-G. Mei{\ss}ner, A. Sibirtsev,
{\em Phys. Lett.} B {\bf  643}, 29 (2006). 

\bibitem{Juli}  T. Hippchen, J. Haidenbauer, K. Holinde, and V. Mull,
{\em Phys. Rev.} C {\bf  44}, 1323 (1991); {\em Phys. Rev.} C {\bf 44}, 1337
(1991).

\bibitem{Fonv09} H. Fonvieille, and V.A. Karmanov, {\em  Eur. Phys. J.} A 
{\bf  42}, 287 (2009).

\bibitem{Ellis} J. Ellis, and M. Karliner, {\em  New J. Phys. } {\bf 4}, 18 (2002).

\bibitem{KK} R. Kr\"onig, {\em Jou. Opt. Soc. Am.} {\bf 12}, 547 (1926); H. A. Kramers,
{\it Atti Congr. Intern. Fisici Como} {\bf 2}, 545 (1927).

\bibitem{LSZ} H. Lehmann, K. Symanzik, and W. Zimmerman, {\em  Nuovo Cimento }
{\bf  1}, 205 (1955).

\bibitem{Tit50} E.C. Titchmarsh, 
''The Theory of Functions'', 2nd ed., Oxford University
Press, 1950; R. P. Boas and H.P. Boas, ''Invitation to complex analysis'', 
MAA Textbook series,
    Washington D.C., 2010.
    
\bibitem{Chew58}G. F. Chew, R. Karplus, S. Gasiorowicz, and F. Zachariasen, {\em  Phys. Rev.}
{\bf  110}, 265 (1958).

\bibitem{Drell60}
S.D. Drell and F. Zachariasen, ''Electromagnetic Structure of
Nucleons'', Oxford University Press, New York, 1960.

\bibitem{Hammer07} M.A. Belushkin, H.-W. Hammer, and U.-G. Mei{\ss}ner,
{\em Phys. Rev. } C {\bf  75}, 035202 (2007).
\bibitem{Mergel96} P. Mergell, U.-G. Mei{\ss}ner and D. Drechsel, {\em  
Nucl. Phys.} A {\bf  596}, 367 (1996).

\bibitem{Hammer96}H.-W. Hammer, U.-G. Mei{\ss}ner, and D. Drechsel, 
{\em  Phys. Lett. } B {\bf  385}, 343 (1996).



\bibitem{Hammer06} H.-W. Hammer, {\em  Eur.Phys. J.} A {\bf  28}, 49 (2006).



\bibitem{Furuichi10}   S. Furuichi, H. Ishikawa, and K. Watanabe,
   {\em  Phys. Rev.} C {\bf   81},045209 (2010).
   \bibitem{Hohl} G. H\"ohler and H. H. Schopper, Landolt B\"ornstein, 
   New Series
Group I, Vol. 9b, p. 405. Pion Nucleon Scattering Part 2. Methods
and Results of Phenomenological Analyses, edited by K.-H.
Hellwege (Springer-Verlag, Berlin, Heidelberg, New York,
1983).

\bibitem{Bald99} R. Baldini, S. Dubni\v{c}ka, P. Gauzzi, S. Pacetti, E. Pasqualucci, and Y. Srivastava,
{\em Eur. Phys. J.} C {\bf  11}, 709 (1999).
\bibitem{Gourdin} M. Gourdin, {\em Phys. Rep.} {\bf  11}, 29 (1974).
\bibitem{Bald06} R. Baldini, C. Bini, P. Gauzzi, M. Mirazita, M. Negrini, and
S. Pacetti, {\em Eur. Phys. Jou.} C {\bf  46}, 421 (2006).

\bibitem{Pacetti07} S. Pacetti, {\em Eur. Phys. Jou.} A {\bf  32}, 421 (2007).



\bibitem{Iac04} F. Iachello, and Q. Wan, {\em Phys. Rev. }  C {\bf  69}, 055204 (2004)

\bibitem{Bij04} R. Bijker, and F. Iachello, {\em Phys. Rev. } C {\bf  69}, 068201 (2004),


\bibitem{CPSS95} F. Cardarelli, E. Pace, G. Salm\`e, and S. Simula,  
{\em Phys. Lett.}  B {\bf   357}, 267  (1995).

\bibitem{LomonTL} E. L. Lomon and S. Pacetti, {\em Phys. Rev.} {\bf  D 85}, 113004 
(2012).

\bibitem{Mandel55} S. Mandelstam, {\em Proc. Royal Soc.} A {\bf  233}, 248 (1955).

\bibitem{TobVM}T. Frederico, H.-C. Pauli, and S.-G. Zhou, 
{\em Phys. Rev.} D {\bf   66}, 054007 (2002);
{\em  Phys. Rev.} D {\bf  66},  116011 (2002).


\bibitem{PDG06} W.-M. Yao et al., {\em Jou. of Phys.} {\bf  33}, 1 (2006).
\bibitem{Brodsky98} 
S.J. Brodsky, H.C. Pauli, and S.S. Pinsky,
{\em Phys. Rep.} {\bf 301}, 299 (1998).  

\end{thebibliography}
\end{document}